%% file: articledef.tex
\documentclass[11pt,twoside]{article} 
\usepackage{epsfig,graphics,subfigure}
\usepackage{amssymb}
\usepackage[centertags]{amsmath}

\usepackage{endnotes}

\usepackage{stmaryrd}
\makeatletter
\def\@biblabel#1{#1.}
\makeatother
\newcommand{\citeyear}[2][\null]{\ifx#1\null \cite{#2} \else \cite[#1]{#2}\fi}
\newcommand{\citeaffixed}[3][\null]{\ifx#1\null \cite{#2} \else \cite[#1]{#2}\fi}
\newcommand{\citeasnoun}[2][\null]{\ifx#1\null \cite{#2} \else \cite[#1]{#2}\fi}

\newfont{\ensmathquatorze}{msbm10 scaled 1400}
\newfont{\ensmathonze}{msbm10 scaled 1100}
\newfont{\ensmathdix}{msbm10}
\newfont{\ensmathneuf}{msbm10 scaled 833}
\newfont{\ensmathhuit}{msbm10 scaled 694}
\newfam\ensmathfam                        
\textfont\ensmathfam=\ensmathonze        
\scriptfont\ensmathfam=\ensmathdix       
\scriptscriptfont\ensmathfam=\ensmathhuit
\def\ensmf{\fam\ensmathfam\ensmathonze}         

\newfont{\bboldquatorze}{bbold10 scaled 1400}
\newfont{\bbolddouze}{bbold12}
\newfont{\bboldonze}{bbold10 scaled 1100}
\newfont{\bbolddix}{bbold10}
\newfont{\bboldneuf}{bbold9}
\newfont{\bboldhuit}{bbold8}
\newfam\bboldfam                        
\textfont\bboldfam=\bboldonze        
\scriptfont\bboldfam=\bbolddix       
\scriptscriptfont\bboldfam=\bboldhuit
\def\bbld{\fam\bboldfam\bboldonze}              

\newfont{\calligravarieteonze}{rsfs10 scaled 1100}
\newfont{\calligravarietedix}{rsfs10}
\newfont{\calligravarieteneuf}{rsfs10 scaled 833}
\newfont{\calligravarietehuit}{rsfs10 scaled 694}
\newfam\clgvfam
\textfont\clgvfam=\calligravarieteonze
\scriptfont\clgvfam=\calligravarieteneuf
\scriptscriptfont\clgvfam=\calligravarietehuit
\def\clgv{\fam\clgvfam\calligravarieteonze}     

\newfont{\frakquatorze}{eufm10 scaled 1400}
\newfont{\frakonze}{eufm10 scaled 1100}
\newfont{\frakdix}{eufm10}
\newfont{\frakneuf}{eufm10 scaled 833}
\newfont{\frakhuit}{eufm10 scaled 694}
\newfam\fraktur
\textfont\fraktur=\frakonze
\scriptfont\fraktur=\frakdix
\scriptscriptfont\fraktur=\frakhuit
\def\frak{\fam\fraktur\frakonze}                

\newfont{\eurmquatorze}{eurm10 scaled 1400}
\newfont{\eurmonze}{eurm10 scaled 1100}
\newfont{\eurmdix}{eurm10}
\newfont{\eurmneuf}{eurm10 scaled 833}
\newfont{\eurmhuit}{eurm10 scaled 694}
\newfam\eurmfam
\textfont\eurmfam=\eurmonze
\scriptfont\eurmfam=\eurmdix
\scriptscriptfont\eurmfam=\eurmhuit
\def\eurm{\fam\eurmfam\eurmonze}                

\newfont{\eusmonze}{eusm10 scaled 1100}
\newfont{\eusmdix}{eusm10}
\newfont{\eusmneuf}{eusm10 scaled 833}
\newfont{\eusmhuit}{eusm10 scaled 694}
\newfam\eusm
\textfont\eusm=\eusmonze
\scriptfont\eusm=\eusmdix
\scriptscriptfont\eusm=\eusmhuit
\def\eusmft{\fam\eusm\eusmonze}                 

\newfont{\microfontecinq}{cmr5}
\newfont{\microfontesix}{cmr6}
\newfont{\microfontesept}{cmr7}
\newfont{\microfontehuit}{cmr8}
\newfam\microfonte
\textfont\microfonte=\microfontesept
\scriptfont\microfonte=\microfontesix
\scriptscriptfont\microfonte=\microfontecinq

\newfont{\fontegrecquegrasseonze}{cmmib10 scaled 1100}
\newfont{\fontegrecquegrassedix}{cmmib10}
\newfont{\fontegrecquegrasseneuf}{cmmib10 scaled 833}
\newfont{\fontegrecquegrassehuit}{cmmib10 scaled 694}

\newfont{\vieuxchiffres}{cmmi10 scaled 1100}

\newcommand{\NN}{{\bbld N}}                     

\newcommand{\ZZ}{{\bbld Z }}                    

\newcommand{\QQ}{{\bbld Q }}                    

\newcommand{\RR}{{\bbld R }}                    

\newcommand{\CC}{{\bbld C }}                    

\newcommand{\nonnull}{\backslash\{0\}}

\newcommand{\ea}{{\eurm a}}
\newcommand{\eb}{{\eurm b}}
\newcommand{\ec}{{\eurm c}}
\newcommand{\ed}{{\eurm d}}
\newcommand{\ee}{{\eurm e}}
\newcommand{\eh}{{\eurm h}}

\newcommand{\Lbbld}{{\ensmf L}}
\newcommand{\Mbbld}{{\ensmf M}}

\newcommand{\Lambdabbld}{{\mbox{\bbolddouze{\char'003}}}}
\newcommand{\Lambdabblds}{{\mbox{\bbolddix{\char'003}}}}
\newcommand{\Id}{{\bbld 1 }}                    
\newcommand{\Zero}{{\bbld 0 }}

\newcommand{\JJ}{{\ensmf J }}   

\newcommand{\intext}[1]                          
{\smash{#1}}                                      

\newcommand{\transpose}[1]{{}^{\mathrm{t}}#1}

\newcommand{\EXP}[1]{{ \mbox{\large e}^{\raisebox{.3ex}{$\sst #1$}}}}

\newcommand{\DEF}{\stackrel{\mbox{\rm\scriptsize def}}{=}}
\newcommand{\DEFt}{\stackrel{\mbox{\rm\tiny def}}{=}}

\newcommand{\arctg}{\mathop{\mathrm{arctg}}\nolimits}
\newcommand{\sgn}{\mathop{\mathrm{sgn}}\nolimits}
\newcommand{\Hess}{\mathop{\mathrm{Hess}}}
\newcommand{\spectre}{\mathop{\mathrm{sp}}}
\newcommand{\im}{\mathop{\mathrm{Im}}}

\newcommand{\textvector}[2]{\smash{\big(\begin{smallmatrix}
                                   \!#1\!\\\!#2\!
                                  \end{smallmatrix}
                            \big)}}

\newcommand{\eval}[1]{_{\left|#1\right.}}               

\newcommand{\tr}{\mathop{\mathrm{tr}}}

\newcommand{\imat}{\mathrm{i}}                          

\newcommand{\dst}{\displaystyle}
\newcommand{\tst}{\displaystyle}
\newcommand{\sst}{\scriptstyle}
\newcommand{\ssst}{\scriptscriptstyle}

\newcommand{\fatop}[2]{\genfrac{}{}{0pt}{0}{#1}{#2}}   
\newcommand{\fatops}[2]{\genfrac{}{}{0pt}{1}{#1}{#2}}  

\newcommand{\scl}{\fatops{ \raisebox{-.3cm}{$\textstyle\sim$} }
                         { \scriptstyle\hbar\to 0                             }
                 }

\newcommand{\kete}[1]{|\kern.3ex#1\kern.3ex\rangle}

\newcommand{\brae}[1]{\langle\kern.3ex #1 \kern.3ex|}

\newcommand{\cmt}[1]                                   
{\left\llbracket#1\right\rrbracket}                    

\newcommand{\poisson}[1]                                
{\{\kern-1.15mm|#1|\kern-1.19mm\}}                              
\newcommand{\spoisson}[1]                                       
{\left\{\kern-.9mm\left|#1\right|\kern-.9mm\right\}}            %

\newcommand{\finiteset}[2]{\left\{#1,\ldots\kern -0.15ex,#2\right\}}

\newcommand{\opp}{\hat{p}}
\newcommand{\opq}{\hat{q}}

\newcommand{\origin}{{\frak o}}                                       %
\newcommand{\impitau}{$\imat m\,\frac{\sst2\pi}{\sst\tau}$}               %
\newcommand{\sigmaz}{$\sigma_{{\!\ssst m0k}}$}
\newcommand{\sigmau}{$\sigma_{{\!\ssst m1(k-1)}}$}
\newcommand{\sigmaa}{$\sigma_{{\!\ssst m\alpha(k-\alpha)}}$}
\newcommand{\sigmak}{$\sigma_{{\!\ssst mk0}}$}
\newcommand{\hz}{${\eurm h}_{{\ssst\kern.12ex m0k}}$}
\newcommand{\hu}{${\eurm h}_{{\ssst\kern.12ex m1(k-1)}}$}
\newcommand{\ha}{${\eurm h}_{{\ssst\kern.12ex m\alpha(k-\alpha)}}$}
\newcommand{\hk}{${\eurm h}_{{\ssst\kern.12ex mk0}}$}
\newcommand{\ve}{\varepsilon}
\newcommand{\fpz}{${\frak p}^0$} 
\newcommand{\fqz}{${\frak p}_0$} 
\newcommand{\fpu}{${\frak p}^1$} 
\newcommand{\fpd}{${\frak p}^2$} 
\newcommand{\fpt}{${\frak p}^3$} 
\newcommand{\fpqu}{${\frak p}^4$} 
\newcommand{\fpc}{${\frak p}^5$} 
\newcommand{\fps}{${\frak p}^6$} 
\newcommand{\fpse}{${\frak p}^7$} 
\newcommand{\fph}{${\frak p}^8$} 
\newcommand{\fpn}{${\frak p}^9$} 
\newcommand{\fpm}{${\frak p}^-$}
\newcommand{\fpp}{${\frak p}^+$}
\newcommand{\fqm}{${\frak p}_-$}
\newcommand{\fqp}{${\frak p}_+$}
\newcommand{\fppqp}{${\frak p}^+_+$}
\newcommand{\fppqm}{${\frak p}^+_-$}      
\newcommand{\fpmqp}{${\frak p}^-_+$}      
\newcommand{\fpmqm}{${\frak p}^-_-$}

\newcommand{\pve}{$\propto\ve$}   
\newcommand{\psve}{$\propto\sqrt{|\ve|}$} 
 \newcommand{\pvetq}{$\propto|\ve|^{3/4}$} 
 \newcommand{\pvek}{$\propto|\ve|^{\ell/4-1/2}$}  

\newcommand{\vextr}{$v_{\textrm{extr}}$}   
\newcommand{\vtran}{$-(\sgn{a})v_{\textrm{trans}}$}

\newcommand{\ora}[1]{
\vbox {\ialign {##\crcr \scriptsize  \rightarrowfill \crcr \noalign 
{\kern 1pt \nointerlineskip} $\hfil \displaystyle {#1}\hfil $\crcr }}}
                                                               
\numberwithin{equation}{subsection}
\renewcommand{\theenumi}{(\kern -0.15ex{\em\roman{enumi}\/})}

\begin{document}

\thispagestyle{empty}
\vspace{1cm}
			\begin{center}
\renewcommand{\thefootnote}{\fnsymbol{footnote}}
{\Huge  Normal forms and complex periodic orbits in semiclassical
expansions of Hamiltonian systems}

\vspace{1cm}

{\Large\sc P. Leb\oe uf${}^1$ and A. Mouchet${}^2$ }\\

${}^1$ Laboratoire de Physique Th{\'e}orique et Mod\`eles 
Statistiques\footnote{Unit{\'e} de recherche de
                        		l'Universit{\'e} Paris XI 
					associ{\'e}e au CNRS
                   } \\ 91406 Orsay Cedex, France.
\\[1ex]
${}^2$ Laboratoire de Math\'ematiques 
			et de Physique Th\'eorique,\footnote{CNRS UPRES-A 6083.} \\
			Avenue Monge, 
			Parc de Grandmont 37200 Tours, France.
\\
 mouchet@celfi.phys.univ-tours.fr
\\ http:$\backslash\backslash$www.phys.univ-tours.fr/\~{}mouchet
\renewcommand{\thefootnote}{\arabic{footnote}}

\end{center}

\begin{abstract}
Bifurcations of periodic orbits as an external parameter is varied are a
characteristic feature of generic Hamiltonian systems. Meyer's
classification of normal forms provides a powerful tool to understand the
structure of phase space dynamics in their neighborhood. We provide a
pedestrian presentation of this classical theory and extend it by
including systematically the periodic orbits lying in the complex plane on
each side of the bifurcation. This allows for a more coherent and unified
treatment of contributions of periodic orbits in semiclassical expansions.
The contribution of complex fixed points is find to be exponentially
small only for a particular type of bifurcation (the extremal one). In all
other cases complex orbits give rise to corrections in powers of $\hbar$
and, unlike the former one, their contribution is hidden in the ``shadow''
of a real periodic orbit.
\end{abstract}

\vfill

71 pages, 4 tables, 20 figures

\newpage

\underline{Proposed running head:} Semiclassics in the complex plane.

\vspace{3\baselineskip}

\raisebox{5.5cm}{
\parbox[b]{6cm}{
		\begin{center}

	     \textsf{\bfseries Amaury MOUCHET} \\
	     \textrm{\footnotesize Ma\^\i tre de conf\'erences de physique}\\
	      \rule[1mm]{1cm}{.1mm}\\
	     {\itshape\normalsize Laboratoire de math\'ematiques et de 
	                          physique th\'eorique}\\[1ex]
	     \textbf{\footnotesize UPRES-A 6083}\\
	      \rule[1mm]{1cm}{.1mm}\\
	     \textbf{\normalsize Facult\'e des sciences}\\[1ex]
	     \textrm{\footnotesize Avenue Monge, Parc de Grandmont}\\[1ex]
	     \textrm{ 37200\ \  \textsc{Tours - France}}

	         \end{center}
	       } 
	      }
\hspace{3.5cm}
\raisebox{6.66cm}{
\parbox[b]{8.2cm}{
             \mbox{\hspace{0cm}\epsfig{file=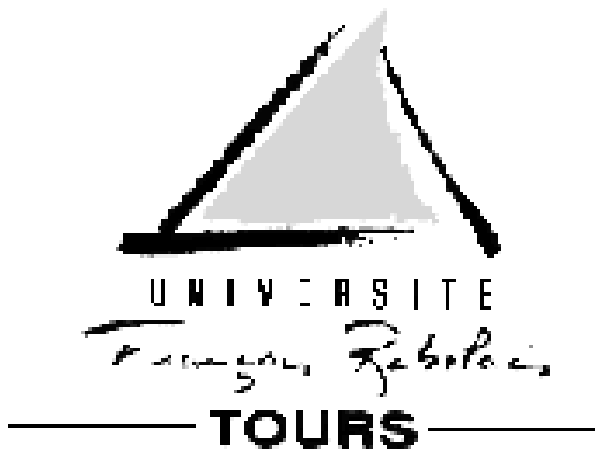,scale=.75}}\\[1.5em]

	     \hspace{-2.5cm}	
             \parbox{8.2cm}{\footnotesize
			    \underline{Tel} : (33).(0)2.47.36.69.28\\
		            \underline{Fax} : (33).(0)2.47.36.69.56\\
		            \underline{e-mail} :
			mouchet@celfi.phys.univ-tours.fr\\
			     \underline{www} : \ 
	                     {\tt http://bbrother.phys.univ-tours.fr/\~{}mouchet
                             } 
			    }
                  }
		  }


\begin{tabular}{p{2cm}p{5cm}@{\hspace{2ex}}}
$\DEF$&is equal by definition to \\

$\sgn(x)$&sign of $x$  \\

${\eusmft A}\backslash {\eusmft B}$
&  elements of set ${\eusmft A}$ not in ${\eusmft B}$ \\

${\eusmft A}^{\pm}$&positive or negative numbers of ${\eusmft A}$\\

${\eusmft A}\times {\eusmft B}$&cartesian product of two sets \\

$\Re(z), \Im(z)$& real, imaginary part of~$z\in\CC$\\

${\ensmf A}\circeq{\ensmf B}$&${\ensmf A}$ and ${\ensmf B}$ are two equivalent matrices\\



$O_{\finiteset{x_1}{x_n}}(k)$& (set of) smooth function(s) of $x\mathop{=}\finiteset{x_1}{x_n}$  whose derivatives up to 
order~$k-1$ vanish at~$\finiteset{0}{0}$. \\

${\frak o}$, ${\frak a}$, ${\frak p}$, etc. & fixed points or periodic orbits\\

${\ensmf A, B, R}, \Lambdabbld$& $2\times 2$-matrices\\

$\spectre(U)$& spectrum of matrix $U$ \\

$\ve^\star$& $|\ve|$ if $\ell=3$, $\sqrt{|\ve|}$ otherwise 

\end{tabular}


\newpage

\section{Introduction}
\label{sec:introduction}

Semiclassical methods have had many applications in the description of the
quantum mechanical behavior of particles whose classical analog is chaotic.
These methods are essentially based on the knowledge of the classical
periodic orbits. Aside from these classical contributions, there are other
ones, usually associated to non-classical effects such as diffraction or
tunneling. The main motivation of this work is to better understand a
particular type of contribution to trace formulae, i.e. the complex periodic
orbits. We are thus interested in semiclassical expansions which generically
take the form of an asymptotic series
\begin{equation}\label{traceformulae}
 	Q(\hbar)\ \scl\ \sum_{{\frak p}}A[{\frak p}]\,
			\EXP{\imat W[{\frak p}]/\hbar}\;
			\bigg(
				1+\sum_{k\geqslant1} a_k[{\frak p}]\,\hbar^k
			\bigg) \ .
\end{equation}
Here, the first sum concerns (periodic) orbits and fixed points of the
classical system. Both~$A[{\frak p}]$ and~$W[{\frak p}]$ are complex
functions of geometrical quantities attached to each~{\frak p}. The
prefactor~$A[{\frak p}]$ is well defined if the periodic orbits are assumed
to be isolated. This is typically the case for fully chaotic systems, as for
example the Sinai billiard or the Bunimovich stadium. However, no smooth
potential is known to be fully chaotic and the generic scenario is a
coexistence of regular and irregular motion in phase space (the so called
mixed dynamics). In this case the phase space structure is much more
complicated than in the fully chaotic regime because the frontier between
regular and irregular motion is not smooth, but fractal, and its details are
very sensitive to changes of external parameters controlling the dynamics.

At the level of periodic orbits, this complexity is reflected on the one
hand in a rich structure of coexisting stable and unstable orbits and on the
other hand by the occurrence of many bifurcations as some parameter is
varied. Close to a bifurcation several periodic orbits can lie very close to
each other, and exactly at the bifurcation they coalesce. Semiclassically,
the distance between periodic orbits is measured in terms of a phase space
length-scale fixed by the value of $\hbar$. For distances smaller than this
scale, the prefactor 
$A[{\frak p}]$
in~\eqref{traceformulae} 
diverges, and some uniform
approximation treating simultaneously the contribution of several orbits has
to be implemented. For two-freedom systems, for which a complete
classification of generic bifurcations exists 
\cite{Meyer70a}
 (see also table I
below), this was done in 
Refs.~\cite{Ozorio/Hannay87a,Sieber96a}.
 Because most
bifurcations imply the creation of new or the disappearance of existent
periodic orbits as a parameter is varied, uniform approximations are not
only a good way to solve the problem of the divergence of the prefactor,
they also take care of the (unphysical) discontinuities that would have
occurred in 
~\eqref{traceformulae} 
 if suddenly a term is added (or subtracted) to the
interferent sum. In fact, the most natural way to obtain a continuous
physical picture through the bifurcation is to extend the classical
dynamics (i.e. the coordinates and momenta) into the complex plane. Then it
is easy to see, as we will show below, that classically there is no
discontinuity (i.e., no creation or destruction) at the bifurcation but
rather an exchange of periodic orbits flowing from the complex to the real
plane and vice versa as a parameter is varied. (Complex periodic orbits and
more generally complex classical trajectories are solutions of the classical
equations of motion whose coordinate and/or momenta are complex variables.
Time is always considered as a real variable). Exactly at the bifurcation
point several orbits coming from the complex and/or the real plane coalesce.

The inclusion of complex orbits in semiclassical formulae is thus an
important issue and may be considered as part of a general program on {\sl
exact semiclassical mechanics}. The aim of this program is to compute exact
quantum results from semiclassical analysis (see for example the review by
A. Voros 
\cite{Voros94a}
. As Balian and Bloch have pointed out 
\cite{Balian/Bloch74b}
, quantum mechanics can be fully reconstructed from
classical trajectories if real as well as complex trajectories are included.
This program, which for generic mixed systems remains for the moment at a
formal level, has however been pursued successfully for one-dimensional
autonomous systems 
\cite{Voros83a, Voros94b}
as for the free motion of a particle on a compact surface of constant
negative curvature 
\cite{Cartier/Voros88a}
. In spite of the fact that the latter is a
prototype of fully chaotic motion, the accomplishment of the program in this
case heavily relies in the fact that the semiclassical expressions are exact.

The appearance of complex trajectories in exact semiclassical expressions is
of course easy to understand. Consider for example the tunneling effect. By
definition this effect is classically forbidden and hence is not taken into
account in expressions such as 
Eq.~\eqref{traceformulae} 
where only real classical
trajectories are included. It is however remarkable that only complex {\sl
classical} paths (and not arbitrary complex paths) are needed in order to
correctly describe it 
\cite{Balian/Bloch74b}.
Ref.~\cite{Leboeuf/Mouchet94a},
where complex periodic orbits
are used in order to include tunneling effects in the computation of the
density of states of a mixed system, provides an illustration of this point.
Other interesting recent works exploring different aspects of tunneling in
the mixed or fully chaotic regimes can be found in 
Refs.\cite{Lin/Ballentine90a,Bohigas+93a,Tomsovic/Ullmo94a,Doron/Frischat95a,Creagh/Whelan96a,Frischat/Doron98a}.

As mentioned before, the amplitude~$A[{\frak p}]$ in
Eq.(\ref{traceformulae}) depends on the geometry of the classical flow and
is connected to the linearized motion around~{\frak p}. If~{\frak p} is
isolated, it can be expressed in terms of the eigenvalues of the monodromy
matrix~$\Mbbld$ at~{\frak p} which is, by definition, the linearized part of
the transversal map~$\Phi$ after one return in the neighborhood of any
point~{\frak o} of~{\frak p}. In order to find these eigenvalues, one can,
for example, diagonalise the quadratic part of the classical Hamiltonian by
constructing a suitable symplectic chart in which the later is made up of
only the relevant geometric monomials. This  procedure 
can be generalized to higher orders: it
is the object of normal form theory. In hamiltonian physics, it has been
introduced by~\citeasnoun{Poincare57a} and~\citeasnoun{Birkhoff27a}.

In order to obtain geometrical semiclassical~$\hbar$ expansions, one must
therefore classify the generic situations which may happen in the
neighborhood of periodic orbits as an external parameter is varied. In
\citeasnoun{Meyer70a} \nocite{Meyer71a} \cite{Meyer86a}
such a classification was performed for one-parameter hamiltonian
systems with two degrees of freedom. In the original
paper, the aim was to prove the genericity in the mathematical sense. In a
more recent work~\cite{Meyer/Hall92a}, the authors use the Lie
transformation theory developed in the context of many body classical
problems~\citeaffixed[and references therein] {Henrard70a,Cushman+83a}{see
for example}, which is a powerful method for obtaining normal forms. It has
also the great advantage of giving a systematic method that can be
numerically implemented. However, because our final aim are semiclassical
expansions, the classical objects we are interested in are the generating
functions -- the phases in Feynman integrals -- rather than the explicit
maps themselves. Especially when the bifurcation is of an extremal or
transitional type~(see section~\ref{sec:nofounfham} below), the explicit
expression for the generating functions is not trivially obtained from
Meyer's work. We will give a precise expression for them in
section~\ref{sec:nofogenfun}.

Before, the aim of section~\ref{sec:reduction} is to introduce the one-parameter
hamiltonian systems with two degrees of freedom, and of
sections~\ref{sec:nofoindham} and~\ref{sec:nofounfham} is to give an original
presentation of Meyer's classification. We will neither use the language of singularity theory nor that of
Lie transformations; we will only need some elementary algebraic operations and
refer to a physicists' intuitive approach of genericity. We will systematically
use Taylor expansions and make generic statements about the coefficients.
 
In section~\ref{sec:geomprop}, we make a detailed study of the
structure of the geometrical classical skeleton based on the previous results
and include systematically the complex fixed points in the analysis of each
type of bifurcation. The reader familiar with the theory of normal forms or
not interested in the proofs can directly go to this section were the
essential ingredients are summarized. 

The quantum expansions based on these classical structures are finally
introduced in section~\ref{sec:quant}. We go through all the different types
of bifurcations encountered in two-dimensional systems and systematically
evaluate the contribution of the real as well as the complex periodic orbits
involved. The main difficulty is associated with the complex orbits. When
one uses the steepest descent method in semiclassical approximations one
must take into account global properties of the integration domain to find
the possible deformation of the integration path that determines the
contribution of the orbit. In particular, the direction in which one crosses
the saddle points must be known in order to correctly compute the Maslov
indices of the orbits. When the dynamics is embedded in a smooth family, the
choice of critical points may change discontinuously as the external
parameters are varied. This produces the well known Stokes'phenomenon
extensively studied by {\sc
Berry~\&~Howls}~\citeyear{Berry/Howls91a,Berry/Howls94a} (see
also~\cite{Howls91a}). The difficulties when dealing with this highly
non-local properties may be attenuated if one considers only complex orbits
near a bifurcation. Aside from these considerations, our interest is to
compute the relative weight of the contribution of complex periodic orbits.
Several previous studies in the mixed regime~\cite{Kus+93a,Leboeuf/Mouchet94a}
have shown that some complex orbits give an exponentially small
contribution. Nevertheless we will argue in section~\ref{sec:quant} that
these exponentially small contributions occur in only one type of
bifurcation: the extremal case. For all the others the contribution of
complex periodic orbits -- if they exist -- is proportional to a power law
in~$\hbar$. In other words, these complex periodic orbits contribute to 
$a_k\hbar^k$-like terms in Eq.(\ref{traceformulae}).

\newpage

\section{Poincar{\'e} reduction}
\label{sec:reduction}

Let us consider an autonomous hamiltonian system with two degrees of
freedom. Denote $G$ its hamiltonian, defined on a four-dimensional phase
space. Let~{\frak p} be a periodic orbit of period~$T>0$ lying in the
three-dimensional energy shell~$\Sigma(E)$. Let us construct another
three-dimensional submanifold~{\frak S} which is sufficiently small to
intersect~{\frak p} only once. Let~${\frak o}\DEFt{\frak S}\cap{\frak p}$
and ${\frak s}(E)\DEFt{\frak S}\cap\Sigma(E)$ (see
figure~\ref{fig:reduction}). \vfill

\begin{figure}[!ht]
\begin{center}
\input{reduction.pstex_t}
\caption{\sl\baselineskip=0.25in\label{fig:reduction}
         Poincar{\'e} reduction in the neighborhood of a periodic
         orbit~{\frak p}.
        }
    	
\end{center}
\end{figure}

\vfill
Let us call~$\vartheta$ the angle coordinate defined in a tubular
neighborhood of~{\frak p}. On~{\frak p}, it represents the arclength
coordinate modulo the total length~$\tau$ of~{\frak p} counted in a suitable
unit system. Moreover there is, by construction, a one-to-one relation
between~$\vartheta$ and the time~$t$ since everywhere locally around~{\frak
p},~$d\vartheta/dt\neq0$. One can thus consider~$t$ to be a smooth function
of~$\vartheta$, $t(\vartheta)$. Let~$I$ be the coordinate canonically
conjugate to~$\vartheta$ and~$(p,q)$ the other two symplectic coordinates.
The hamiltonian~$G$ is locally a function of these
variables:~$G(I,p,\vartheta,q)$. Since~$\partial G/\partial I=d\vartheta/dt\neq0$, one can
make use of the implicit function theorem to define a
function~$H(p,q;\vartheta;E)$ by 
\begin{equation}
G(I,p,\vartheta,q)=E\iff I=H(p,q;\vartheta;E)\;.
\end{equation}
From the expression of the partial derivative of the implicit functions we
get

\vspace{-.9\baselineskip}
\noindent
\begin{subequations}
        \hfill
        \raisebox{0cm}[0cm][0cm]
	{\makebox[0cm][l]{\rule{5.1cm}{0cm}$\left\{\rule{0cm}{1.2cm}\right.$}
	}
	\parbox{16.9cm}{
            		 \begin{align}
                                      \frac{dp}{d\vartheta}&
				      =\frac{dp/dt}{d\vartheta/dt}
				      =-\frac{\partial G/\partial q}
				             {\partial G/\partial I}
				      =+\frac{\partial H}{\partial q}\;;\\[1ex]
               			      \frac{dq}{d\vartheta}&
				      =\frac{dq/dt}{d\vartheta/dt}
				      =+\frac{\partial G/\partial p}
				      	     {\partial G/\partial I}
				      =-\frac{\partial H}{\partial p}  \ .  
         		 \end{align}
	                }        
\end{subequations}
\smallskip

\noindent Therefore, this construction -- due
to~\citeasnoun[chapter XII, \S141]{Whittaker64a}, see
also~\cite[chapter VI, \S3]{Birkhoff27a} --
reduces locally the original two freedom system to an equivalent
one-dimensional system. The price to pay is that the new system is
nonautonomous since the new hamiltonian~$H$ depends on the new
time~$\vartheta$. Nevertheless, the reduced dynamics, now living
on~${\frak s}(E)$, is \mbox{$\tau$-periodic} due to the periodicity
of~$\vartheta$.
{\frak o}~is an equilibrium point of~$H$ obeying, $\forall\,\vartheta$,
\begin{equation}\label{def:generaleq1}
        \frac{\partial H}{\partial p}\mbox{$\eval{\fatops{p=0}{q=0}}$}=0
        \qquad\mathrm{and}\qquad
        \frac{\partial H}{\partial q}\mbox{$\eval{\fatops{p=0}{q=0}}$}=0 \ ,
\end{equation}
if~$(p,q)$ is centered on~{\frak o}. The reduced dynamics mimics the
transversal motion stroboscopically with period $\tau$. The 
corresponding one-step iteration map~$\Phi(E;\tau)$ defined
on~${{\frak s}(E)}$ is the so-called Poincar{\'e} map.
Exploring what happens around~{\frak p} when we go outside~$\Sigma(E)$
is equivalent to considering that the reduced dynamics is embedded in a
one-parameter~($E$) unfolding. Since the symplectic chart on~${{\frak s}(E)}$
is defined in a neighborhood of it, we will adopt a passive point of view
 by considering that we stay for all~$E$ in the same bidimensional phase
 space~~$\clgv{P}$.
 
\section{Normal form for an individual hamiltonian} \label{sec:nofoindham}

From the preceding arguments, changing slightly the notation, let us 
start from a
smooth Hamiltonian~$H(p,q;t)$ defined
on a bidimensional phase space~$\clgv{P}$ and which is 
\mbox{$\tau$-periodic} in time ($\tau>0$).      
We will assume that the origin is an isolated fixed point. 

If we denote~\intext{$w\DEF\textvector{p}{q}$}~
and~\intext{$\nabla_{\!w}\DEF\textvector{\partial_p}{\partial_q}$},
the equations of motion can be written in the compact form 
\begin{equation}\label{eqmo}
        \dot{w}(t)=\JJ\,\nabla_{\!w}H\big(w(t);t\big)
\end{equation}
where ``$\dot{\phantom{w}}$''$\,\DEF d/dt$~and~\intext{$
      \JJ\DEF\bigl(
              \begin{smallmatrix}
                        0&-1\\ 1&\phantom{-}0
              \end{smallmatrix}
              \bigr)
      $}.
      
\subsection{Elimination of the time-dependence in the linearized motion}
\label{subsec:elimtime}
The classification of the dynamics in the neighborhood
of an isolated fixed point is based on the linearization of the motion around
it.

Linearizing  equation~(\ref{eqmo}) around the origin it follows
\begin{equation}\label{leqmo1}
        \dot{w}(t)=\Lambdabbld(t)\,w(t)
\end{equation}
where
\begin{equation}\label{def:Lambda}
        \Lambdabbld(t)\DEF\JJ\Hess\!\big[H(w;t)\big]
        \eval{w=\binom{0}{0}}
        =\JJ
        \begin{pmatrix}
        \partial^{\,2}_{p,p}H(w;t)&\partial^{\,2}_{p,q}H(w;t)\\[2ex]
        \partial^{\,2}_{q,p}H(w;t)&\partial^{\,2}_{q,q}H(w;t)
        \end{pmatrix}
        \eval{w=\binom{0}{0}}.
\end{equation}
It is natural to expect that the linearized dynamics~(\ref{leqmo1})
is roughly akin to the true dynamics provided we are close enough to the 
origin.
We will come back to this point at the end of the subsection.

Recall that a symplectic (complex) matrix~${\ensmf S}$
is defined by the condition 
\begin{equation}
        \transpose{{\ensmf S}}\,\JJ\,{\ensmf S}=\JJ
        \qquad\mbox{($\transpose{{\ensmf S}}$
                     denotes the transpose of ${\ensmf S}$)
                    }
        \label{def:symp1}
\end{equation}
In two dimensions, this is equivalent to
\begin{equation}\label{def:symp2}
        \det{\ensmf S}=1 \ .
\end{equation}
   
By definition \citeaffixed[definition~3.1.13]{Abraham/Marsden85a}{cf.} a
complex matrix~${\ensmf K}$
is an infinitesimal symplectic matrix if it verifies 
\begin{equation}\label{def:infsymp}
        \JJ\,{\ensmf K}=-\transpose{{\ensmf K}}\,\JJ\;.
\end{equation}
Then it can easily  be checked that~$\Lambdabbld(t)$ is
an infinitesimal symplectic matrix.
As its name suggests, the exponential of an infinitesimal symplectic matrix
is symplectic.

Let~${\ensmf U}(t)$ be the evolution matrix associated
with~$\Lambdabbld(t)$. In other words, $\,w(t)={\ensmf U}(t)\,w_0\;$~is
the unique solution of equation~(\ref{leqmo1}) such that~$w(0)=w_0$. We have 
\begin{equation}\label{diffeqU}
        \dot{\ensmf U}(t)=\Lambdabbld(t)\,{\ensmf U}(t)
        \quad\text{with}\quad
        {\ensmf U}(0)=\Id\DEF\bigl(
                                   \begin{smallmatrix}
                                        1&\phantom{-}0\\ 0&\phantom{-}1
                                   \end{smallmatrix}
                             \bigr)\; \ .
\end{equation}
One often writes the formal solution of the latter equation as a time
 ordered product 
\begin{equation}\label{chroprod}
     {\ensmf U}(t)={\sst\eusmft T}\,\EXP{\int_0^t\Lambdabblds(t')dt'}\;.
\end{equation}
It is straightforward to show that~${\ensmf U}(t)$ is symplectic for
all $t$.
Let~$\Lbbld$
be any time-independent in\-fi\-ni\-te\-si\-mal symplectic matrix. Then, being the
product of two symplectic 
matrices, \intext{\hbox{$
         {\ensmf F}(t)\DEFt\EXP{t\,\Lbbld}\bigl[{\ensmf U}(t)\bigr]^{-1}\
         $}} 
is also symplectic.
If we make the symplectic, time-dependent change of coordinates
$\widetilde{w}={\ensmf F}(t)\,w$, then equation (\ref{leqmo1}) becomes
$\dot{\widetilde{w}}(t)=\Lbbld\,\widetilde{w}(t)$. All the information
of the original linearized dynamics has been hidden in the change of
variables. Let us keep in mind that~$H(p,q;t)$ is a reduced hamiltonian and
hence that~$\clgv{P}$ is a (local)~Poincar{\'e} surface of section
centered at one point of an isolated periodic orbit~{\frak p} of a
higher dimensional system. Then, the time-dependence of the flow
defined by equation~(\ref{eqmo}) can be seen as a transversal projection
of the non-reduced autonomous flow as we move along~{\frak p}. The
time periodicity of the reduced flow comes from the fact that
after one cycle along~{\frak p} we are back on~$\clgv{P}$.
Hence we should consider only real \mbox{$\tau$-periodic} symplectic
changes of coordinates,  ${\ensmf F}(\tau)=\Id$. $\Lbbld$~must then be
chosen such that~\intext{$\EXP{\tau\,\Lbbld}=\Mbbld\DEFt{\ensmf U}(\tau)$},
the so-called monodromy matrix.
$\tau\,\Lbbld$~exists and can be any logarithm of~$\Mbbld$
since~$\Mbbld$ has no zero eigenvalues as a consequence of 
equation~(\ref{def:symp2}).
Nevertheless, it may happen that no {\em real} logarithm of~$\Mbbld$
exists. In this case one possibility is to relax the constraint
on~${\ensmf F}(t)$ by retaining only the intersection
with the Poincar{\'e} surface of section every second crossing. That means
that~${\ensmf F}(t)$ must only be~\mbox{$2\tau$-periodic}. 
Because~${\ensmf U}(2\tau)=[{\ensmf U}(\tau)]^2$
and using the fact that every real matrix that has a square root 
has a real logarithm \cite[theorem~2 of~II.E]{Meyer/Hall92a}, we can 
indeed choose the infinitesimal symplectic matrix~$\Lbbld$ to be real
without loss of generality.
This is the statement of the Floquet 
theorem~\cite{Floquet1883a}~(see also the work of {\sc
Cherry}~\citeyear{Cherry27a,Cherry27b,Cherry28a}). There exists 
a symplectic coordinate system in which the linearized motion~(\ref{leqmo1})
is governed by a constant real infinitesimal symplectic 
matrix~$\Lbbld$. 
Equivalently, in a well chosen symplectic chart, the hamiltonian can be written
\begin{equation}
        H(p,q;t)=\underbrace{\frac{1}{2}\,\ec \,p^2+\ed \,pq
                             +\frac{1}{2}\,\ee \,q^2
                            }_{\dst\DEF h^{(2)}(p,q)}
                +\,h^{({\sst\geqslant3})}(p,q;t)    
\end{equation}
in a neighborhood of the origin after having subtracted the irrelevant
value of~$H(0,0;t)$. $({\eurm c,d,e})\in\RR^3$~and $h^{({\sst\geqslant3})}$~is
a \mbox{$\tau$-periodic} function whose first and second 
derivatives with respect to~$p$ or~$q$ vanish at the origin for all~$t$.
In this case~$\Lambdabbld(t)$ is a constant matrix~$\Lambdabbld$ and,
 from definition~(\ref{def:Lambda}) 
\begin{equation}
        \Lbbld=\Lambdabbld={\begin{pmatrix}
                                \eurm-d           & \eurm-e\            \\
                                \eurm\phantom{-} c & \eurm\phantom{-}d\
                     \end{pmatrix} \ ,
                    }
\end{equation}
and the monodromy matrix~$\Mbbld$ is such that 
\begin{equation}\label{monoL}
        \Mbbld^{\,|\varsigma|}=\EXP{\varsigma\tau\Lbbld}
        \qquad\mathrm{with}\qquad
        \left\{
        \begin{array}{ll}
        \varsigma=\pm1&\mbox{if $\Mbbld$ has a real logarithm;}\\[1ex]
        \varsigma=\pm2&\mbox{if $\Mbbld$ has no real logarithm.}
        \end{array}\right.
\end{equation}
The choice of sign of~$\varsigma$ depends on the choice of the arrow of time.
The linearized motion, even though it 
gives an idea of the nature of the true dynamics, cannot reproduce its fine 
structure. The time elimination makes the quadratic
part of~$H$ integrable even though we have started from an a priori non-integrable one.
The chaotic complexity is hidden in the higher-order terms and the aim of 
normal form theory is to retain only the main actors of this rich structure.

\subsection{Normal form for the quadratic part of the hamiltonian}
\label{subsec:nofoquadham}


After having eliminated the time-dependence in the quadratic part of
the hamiltonian, we are now ready to carry on the simplification
of~$h^{(2)}(p,q)$ by constructing a more suitable symplectic
chart~$(p',q')$.
The general classification of the normal forms of quadratic hamiltonians
in any number of dimensions has been done a long time ago by~\citeasnoun{Williamson36a}.
In our case, since we are dealing with a bidimensional phase space, we will
give a more hand-waving demonstration which however illustrates
the main guidelines for deriving Meyer's classification in a 
pedestrian way.

Let us characterize the corresponding symplectic transformation 
by a generating smooth function~$S(p',q)$. The change of 
coordinates~$(p,q)\mapsto(p',q')$ is implicitly defined by 

\noindent
\begin{subequations}\label{sympmap0}
        \hfill
        \raisebox{-0.1cm}[0cm][0cm]{\makebox[0cm][l]{\rule{6.75cm}{0cm}$\left\{\rule{0cm}{.7cm}\right.$}}\parbox{16.9cm}{
         \begin{align}
              p\phantom{'}&=\partial_{q}S(p',q)\;;\\[1ex]
               q'&=\partial_{p'}S(p',q)  
	 \end{align}}        
\end{subequations}

\noindent
in the neighborhood of the origin, provided that 
\begin{equation}\label{ddSnonnull}
        \partial^{\,2}_{p'\!,q}S(0,0)\neq0\;.
\end{equation}
Since without loss 
of generality the transformation
can be assumed to leave the origin invariant,  $S(p',q)$~can be written as 
\begin{equation}\label{genfunctquad}
        S(p',q)=\frac{1}{2}\,\gamma\,p^{\prime\,2}+\delta\,p'q
                +\frac{1}{2}\,\eta\,q^2
                +O_{p'\!,\,q}(3)
\end{equation}
where~$(\gamma,\eta)\in\RR^2$~and~$\delta\in\RR\nonnull$, by virtue of
condition~(\ref{ddSnonnull}). It is thus straightforward from
equations~(\ref{sympmap0}) to obtain the linear part of the symplectic
change of coordinates and deduce the new expression of the quadratic
hamiltonian 
\begin{multline}\label{newhamquad}
        h^{(2)}(p',q')=\frac{1}{2}\underbrace{\left[
                                  \frac{\gamma^2}{\delta^2}
                                  (
                                 \ec \,\eta^2+2\ed \,\eta+\ee 
                                  )
                                  -2\gamma(\ed +\ec \,\eta)
                                  +\ec \,\delta^2     
                                  \right]
                               }
                              _{\dst\DEF\ec '}\,p^{\prime\,2}
                          \\[2ex]
                          +
                          \underbrace{\left[
                                   -\frac{\gamma}{\delta^2}
                                   (
                                  \ec \,\eta^2+2\ed \,\eta+\ee 
                                   )
                                   +\ed 
                                   +\ec \,\eta
                                 \right]
                                }                  
                               _{\dst\DEF\ed '}\,p'q'
                   \\[2ex]
                   +\ \frac{1}{2}\underbrace{\left[
                                  \frac{1}{\delta^2}
                                  (
                                 \ec \,\eta^2+2\ed \,\eta+\ee 
                                  )        
                                \right]
                               }
                              _{\dst\DEF\ee '}\,q^{\prime\,2}
                   +\,O_{p'\!,\,q'}(3)   
\end{multline}

Generically none of the coefficients~$({\eurm c,d,e})$ vanish. Nevertheless,
we will also deal with situations where one real scalar like~${\eurm c,d,e}$
or~$\tr(\Mbbld^2)-2$ can vanish a priori, since we are going to study not
only individual hamiltonians but one-parameter unfoldings. Actually, we can
choose~$(\delta,\gamma,\eta)$ in order to cancel some of the
coefficients~$({\eurm c',d',e'})$. For example, the cancellation of~${\eurm e'}$
implies solving an algebraic equation of degree at most two in~$\eta$
whose discriminant is 
\begin{equation}
        \eurm d^\mathrm{2}-c\kern .15ex e\;=-\det(\Lbbld).
\end{equation}
The simplified form we want to obtain depends essentially on the nature of
the eigenvalues of the monodromy matrix which governs the qualitative regime
about the fixed point.

\noindent
Let us then briefly review the spectrum~$\spectre(\Mbbld)$ of~$\Mbbld$.
Since the monodromy matrix is symplectic, from condition~(\ref{def:symp2})
we have
\begin{equation}\label{detMeq1}
        \det(\Mbbld)=1
\end{equation}
Moreover~$\Mbbld$ is real and
hence~$\spectre(\Mbbld)$ is symmetric with respect
to the real axis:~$\upsilon\in\spectre(\Mbbld)
          \Rightarrow\upsilon^*\in\spectre(\Mbbld)
         $.
         
Let us now introduce the usual equivalence relation between matrices.
Two matrices ${\ensmf A}$~and~${\ensmf B}$ will be equivalent -- we will
denote~${\ensmf A}\circeq{\ensmf B}\,$ -- if by definition they represent the
same linear map but expressed in two different bases; in other words there
exists an invertible matrix~${\ensmf P}$ such 
that~${\ensmf A}={\ensmf P}^{-1}{\ensmf B}{\ensmf P}$. Without loss of
generality ${\ensmf P}$~can be chosen with determinant one and thus
symplectic.
The Jordan decomposition of~$\Mbbld$ asserts that \citeaffixed[chapter VI]{Turnbull/Aitken61a}{see, for example,}
\begin{equation}
        \Mbbld\circeq
                     \begin{pmatrix}
                                 \,\mu  & 0\, \\[1ex]
                                 \,\tilde\mu    & \tst\frac{1}{\mu}\,   
                     \end{pmatrix}
        \qquad \mathrm{where}\qquad \mu\in\RR\cup U(1)\nonnull
        \qquad \mathrm{and}\qquad \tilde\mu\in\{0,1\}\;.
\end{equation}

It is possible to distinguish three different cases:  
\begin{enumerate}
        \item  $\spectre(\Mbbld)\subset\RR\backslash\{-1,0,1\}
                \hspace{1.1ex}
                \iff
                \bigl|\tr(\Mbbld)\bigr|>2$ 
                and {\frak o} is said to be an {\em unstable} fixed point;
               \label{unstable}
        \item  $\spectre(\Mbbld)\subset\{-1,1\}
                \hspace{1.2ex}\qquad
                \iff
                \bigl|\tr(\Mbbld)\bigr|=2$
                and {\frak o} is said to be a 
                {\em marginal }
                fixed point\endnote{\  In the literature
                             \citeaffixed[App.~27]{Arnold/Avez67a}{e.g.},
                             one can
                             find {\em parabolic} fixed point.};
                \label{marginal}
        \item  $\spectre(\Mbbld)\subset U(1)\backslash\{-1,1\}
                \iff
                \bigl|\tr(\Mbbld)\bigr|<2$
                and {\frak o} is said to be a {\em stable} fixed 
		point\endnote{\  In the literature
                                 \citeaffixed[App.~27]{Arnold/Avez67a}{e.g.}
                                 and \cite[\S~7]{Berry78a},
                                 one can find
                                 {\em elliptic} fixed point.}.
                \label{stable}
\end{enumerate}
 Generically, in the
absence of any symmetry but if the dynamics is embedded in a one 
parameter family, the two eigenvalues of~$\Mbbld$ may be degenerate
(the marginal case). It would require at least two parameters to
get~$\Mbbld\in\{\Id,-\Id\}$
and, following our philosophy, we discard from the discussion
all sets of codimension greater than two. In other words we will 
have~$\tilde\mu=1$ in the generic marginal case, which
can therefore be divided in two sub cases 

\ref{marginal}
\parbox{10cm}{
   \begin{enumerate}
        \item[{\rm\ref{marginal}'\phantom{'}}]\label{marginal1}
                $\tr(\Mbbld)=+2\quad\mbox{and generically}\quad
                \Mbbld\circeq
                             \bigl(
                                   \begin{smallmatrix}
                                        1&\phantom{-}0\\ 1&\phantom{-}1
                                   \end{smallmatrix}
                             \bigr)\;;$
        \item[{\rm\ref{marginal}''}]\label{marginal-1}
               $\tr(\Mbbld)=-2\quad\mbox{and generically}\quad
                \Mbbld\circeq
                             \bigl(
                                   \begin{smallmatrix}
                                        -1&\phantom{-}0\\ \phantom{-}1&-1
                                   \end{smallmatrix}
                             \bigr)\;.$
   \end{enumerate}
               }

\noindent
\begin{figure}[!hb]
\begin{center}
\input{RU1.pstex_t}
\caption{\sl\baselineskip=0.25in\label{RU1}
        Geometrical representation of the locus of~$\spectre{\ensmf M}$
        in the~$\CC$ plane.
        }
\end{center}
\end{figure}

Moreover, case~\ref{unstable} is usually also subdivided in the following
way 

\ref{unstable}
\parbox{13cm}{
   \begin{enumerate}
        \item[{\rm\ref{unstable}'\phantom{'}}]\label{hyperbolic1}
                $\tr(\Mbbld)>+2$
                and {\frak o} is said to be a {\em hyperbolic} fixed point;
        \item[{\rm\ref{unstable}''}]\label{hyperbolic-1}
               $\tr(\Mbbld)<-2$ 
                and {\frak o} is said to be an {\em inverse hyperbolic}
                fixed point.
   \end{enumerate}
               }

We now want to transpose this classification onto the
spectrum~$\spectre(\Lbbld)$ of~$\Lbbld$.
Recall that the relation between~$\Lbbld$~and~$\Mbbld$ is given
by~(\ref{monoL}). It can be easily verified that  in cases~\ref{unstable}''
and~\ref{marginal}'' the monodromy matrix
has no real logarithm and hence~$\varsigma\mathop{=}\pm2$ while in all others 
cases\hbox{ -- \ref{unstable}',~\ref{marginal}',~\ref{stable} -- }we 
can take~$\varsigma\mathop{=}\pm1$.

Since~$\Lbbld$ is infinitesimal symplectic, property~(\ref{def:infsymp})
implies that its spectrum~$\spectre(\Lbbld)$ is symmetric with respect
to zero:~$\upsilon\in\spectre(\Lbbld)
          \Rightarrow-\upsilon\in\spectre(\Lbbld)
         $.
Moreover, because~$\Lbbld$ is real, $\spectre(\Lbbld)$~is symmetric
with respect to the real
axis:~$\upsilon\in\spectre(\Lbbld)
          \Rightarrow\upsilon^*\in\spectre(\Lbbld)
     $.
Let
\begin{equation}
        \Lbbld\circeq
                    \begin{pmatrix}
                                \lambda  & \phantom{-}0\\
                                 \tilde{\lambda}& -\lambda
                    \end{pmatrix}
        \qquad \mathrm{where}\qquad \lambda\in\RR\cup(\imat\,\RR)
        \qquad \mathrm{and}  \qquad\tilde\lambda\in\{0,1\}\;,
\end{equation}
be the Jordan decomposition of~$\Lbbld$. Then, relation~(\ref{monoL})
yields 
\begin{equation}
        \Mbbld^{|\varsigma|}\circeq
                     \begin{pmatrix}
                                \,\mu^{|\varsigma|}  
                                & 0
                         \\[2ex]
                                \tst\tilde\mu\bigl[
                                                2-|\varsigma|+(|\varsigma|-1)
                                                (\mu+\frac{1}{\mu})
                                                \bigr]
                                & \;\tst\frac{1}{\dst\mu^{|\varsigma|}}\,
                     \end{pmatrix}
                     \circeq
                     \begin{pmatrix}
                          \EXP{\varsigma\tau\lambda}  
                        & 0
                        \\[1ex]
                          \tst\frac{\tilde{\lambda}}{\lambda}
                          \sinh(\varsigma\tau\lambda)          
                        &   \EXP{-\varsigma\tau\lambda}
                      \end{pmatrix} \ ,
\end{equation}
which implies that~$ \EXP{2\tau\lambda}=\mu^2$
with~$\lambda\in\RR\cup(\imat\,\RR)$. Therefore, in the
unstable case, we have~$\lambda\in\RR\nonnull$ and
then \mbox{$-\det(\Lbbld)>0$}. In the stable case, we
have~$\lambda\tau\in\imat(\RR\backslash\pi\ZZ)$ and
then~$-\det(\Lbbld)<0$. In both
cases~$\tilde\mu=0~$~and~$\tilde\lambda=0$.
In the marginal case, we have~$\Mbbld^2=\Id+{\ensmf X}$ where, generically,
\begin{equation}\label{marginal:X}
        {\ensmf X}\in\Big\{ \bigl(
                                   \begin{smallmatrix}
                                        0&\phantom{-}0\\ 
                                        2&\phantom{-}0
                                   \end{smallmatrix}
                             \bigr),
                             \bigl(
                                   \begin{smallmatrix}
                                        0&\phantom{-}0\\
                                        -2&\phantom{-}0
                                   \end{smallmatrix}
                             \bigr)
                     \Big\}
                     \qquad\mathrm{modulo\ }\circeq\;.
\end{equation}
Therefore, possibly changing the arrow of time,~$\tau\mapsto-\tau$,
we can choose
\begin{equation}\label{marginal:L}
        \Lbbld\circeq\frac{1}{2\tau}\,
                                      \begin{pmatrix}
                                          \phantom{-}0&0\,\\
                                          -2&0\,
                                      \end{pmatrix}
                                      \;;
\end{equation}  
then~$\lambda=0$~and~$\det(\Lbbld)=0$. Other choices, 
say~$\tilde{\,\Lbbld}$, of real
infinitesimal symplectic matrices  obeying relation~(\ref{monoL})
are possible and not necessarily with~$\det(\!\tilde{\,\Lbbld})=0$. But
following what we have done in subsection~\ref{subsec:elimtime}, we
can make a (linear and \mbox{$|\varsigma|\tau$-periodic} ) symplectic change 
of variables corresponding
to~${\ensmf F}(t)\DEF
    \EXP{-t\tilde{\kern 0.5ex\Lbbld}}\,\EXP{-t{\,\Lbbld}}$
in order to deal with an~$\Lbbld$ whose determinant vanishes.
Summarizing
\begin{enumerate}\label{discussionL}
        \item unstable case 
              $\lambda\in\RR\nonnull\hspace{2ex}
              \quad\mathrm{and}\quad -\det(\Lbbld)>0\;$;
        \item marginal case 
              $\lambda=0
              \hspace{6.2ex} 
              \quad\mathrm{and}\quad -\det(\Lbbld)=0\;$;
        \item stable case 
              \intext{$\tst\lambda\in\tau^{-1}\imat\,(\RR\backslash\pi\ZZ)
              \hspace{0.4ex} 
              \quad\mathrm{and}\quad -\det(\Lbbld)<0\;$}.
\end{enumerate}

\noindent
We can now  go back to the reduction of the 
quadratic part of the 
hamiltonian~(\ref{newhamquad}). 
\smallskip



%
%
                \textbf{\ref{unstable}\ Unstable case: }\label{strobunstable}
We can choose at least one real value of~$\eta$ which
makes~$\ee '$ vanish and, provided we
take $\gamma=\pm\ec \,\delta^2/[2\sqrt{-\det(\Lbbld)}]$,
we can also cancel~$\ec '$. At last, we
get \mbox{$h^{(2)}(p',q')=\pm\sqrt{-\det(\Lbbld)}\,p'q'$}. The global sign of
the hamiltonian becomes irrelevant by a time reversal transformation and 
we finally find a symplectic chart in which the quadratic 
Hamiltonian can be written as 
\begin{equation}\label{hamquad:unstable}
        h^{(2)}(p,q)=\lambda\,pq
        \qquad\mathrm{with}\qquad \lambda\in\RR^{\ssst+}\!\nonnull
\end{equation}
after having relabeled the new coordinates.
$h^{(2)}(p,q)$~corresponds to an integrable dynamics whose trajectories lie
on hyperbolae of center~{\frak o} and whose  asymptotes, the~$p$ and $q$~axes,
are the two stable and unstable manifolds 
respectively~(see the right column of~table~\ref{tab:quadham}).
Recall (cf Eq.(\ref{monoL})) that in case~\ref{unstable}'' the
hamiltonian~$h^{(2)}(p,q)$ describes the stroboscopic view of the real motion
with a period twice the original physical period~$\tau$.
Technically such a prescription misses a $-\Id$~factor in the
equations of motion~(\ref{leqmo1}) and
what effectively happens  is that the system goes from one
branch of a hyperbola to the other branch between~$t=0$~and~$t=\tau$.
Then, between~$t=\tau$~and~$t=2\tau$, it goes back to the original branch.
The $2\tau$-stroboscopic view does not see this flipping.

%
%
                \textbf{\ref{marginal} \ Marginal case: }           
If we have~$\ed \neq0$ and hence~$\ec \neq0$ we can
choose~$\eta=-\ed /\ec $ to
get~$\ec \,\eta^2+2\ed \,\eta+\ee =0$.
Then both~$\ed '$~and~$\ee '$ vanish.
Moreover, we take~$\delta=|\tau\ec |^{-1/2}$ to
 obtain~$|\ec '|=1/\tau$.
By changing possibly~$t$ to~$-t$ we get  
\begin{equation}\label{hamquad:marginal}
        h^{(2)}(p,q)=\frac{1}{2\tau}p^2
\end{equation}
after having relabeled the new coordinates.
$h^{(2)}(p,q)$~corresponds to an integrable dynamics whose trajectories
are straight lines parallel to the $q$~axis
(see the middle column of table~\ref{tab:quadham}).
The distinction between cases~\ref{marginal}' and~\ref{marginal}'' is
analogous to that of the unstable cases. In~\ref{marginal}'', during one
period~$\tau$, the system flips back and forth between a couple of lines 
symmetric with respect to~{\frak o}.

%
%
                \textbf{\ref{stable} \ Stable case: }
 We
have $\ec \,\eta^2+2\ed \,\eta+\ee \neq0$ for every real~$\eta$ and
then~$\ee '$
can never be cancelled. Nevertheless we can
choose~$\gamma=\delta^2(\ec \,\eta+\ed )/
                    (\ec \,\eta^2+2\ed \,\eta+\ee )$
in order to have $\ed '=0$.
Besides, we can make $\ee '=\ec '$ by
choosing $\delta=|(\ec \,\eta^2+2\ed \,\eta+\ee )|^{1/2}\det{(\Lbbld)}^{-1/4}$.
If we define~\intext{$\omega\DEFt|\lambda|$} and possibly change the
arrow of time we find the following normal form 
\begin{equation}\label{hamquad:stable}
        h^{(2)}(p,q)=\frac{1}{2}\,\omega\,(p^2+q^2)
\qquad\mathrm{with}\qquad 
        \omega\in\frac{1}{\tau}\,(\RR^{\ssst+}\!\backslash\pi\ZZ)
\end{equation}
after having relabeled the new coordinates.
$h^{(2)}(p,q)$~corresponds to an integrable dynamics whose trajectories
are circles centered on~{\frak o} and whose angular speed is ~$\omega$ 
(see the right column of table~\ref{tab:quadham}).

\noindent
\begin{table}[Ht]
\begin{tabular}{||l||c|c||c|c||c||}\hline
        &\multicolumn{2}{c||}{\bf\ref{unstable}\ unstable case}
        &\multicolumn{2}{c||}{\bf\ref{marginal}\  marginal case}
        &{\bf\ref{stable}\ stable case }\\
        \cline{2-5}
        &{\bf\ref{unstable}'\phantom{'}}
        &{\bf\ref{unstable}''}
        &{\bf\ \ \ \ \ref{marginal}'\phantom{'}}
        &{\bf\ref{marginal}''}
        &\\
        \hline\hline
        $\Mbbld$:\kern2pt monodromy
        &\multicolumn{2}{c||}
                        {$\spectre(\Mbbld)\subset\RR\backslash\{-1,0,1\}$}
        &\raisebox{-1.5ex}[0ex][0ex]{$\Mbbld\mathop{\circeq}\bigl(
                        \begin{smallmatrix}
                                1&\phantom{-}0\\1&\phantom{-}1
                        \end{smallmatrix}
                                            \bigr)$}\!\!
        &\raisebox{-1.5ex}[0ex][0ex]
                        {$\Mbbld\mathop{\circeq}\bigl(\!
                        \begin{smallmatrix}
                       -1&\phantom{-}0\\\phantom{-} 1&-1
                   \end{smallmatrix}
          \bigr)$}\!\!
        &$\spectre(\Mbbld)\mathop{\subset} U(1)\backslash\{-1,1\}$
	\rule[-1.5ex]{0ex}{4ex}\\
        \cline{2-3}\cline{6-6}
        matrix
        &$\tr(\Mbbld)\mathop{>}2$
        &$\tr(\Mbbld)\mathop{<}-2$
        &       
        &
        &$-2<\tr(\Mbbld)<2$
	\rule[-1.5ex]{0ex}{4ex}\\
        \hline
        $\varsigma\in\{1,\pm2\}$
        &$\varsigma=1$
        &$|\varsigma|=2$
        &$\varsigma=1$
        &$\varsigma=-2$
        &$\varsigma=1$\\
        \hline
         $\Mbbld^{|\varsigma|}=\EXP{\varsigma\tau\Lbbld}$
        &\multicolumn{2}{c||}{$-\det(\Lbbld)>0$}
        &\multicolumn{2}{c||}{\raisebox{-1.5ex}[0ex][0ex]{
        $\Lbbld\circeq\frac{1}{2\tau}\bigl(
                                   \begin{smallmatrix}
                                        0&\phantom{-}0\\ 2&\phantom{-}0
                                   \end{smallmatrix}
                             \bigr)$
                     }}
        &$-\det(\Lbbld)<0$
	\rule[-1.5ex]{0ex}{4ex}\\
         \cline{2-3}\cline{6-6}
         $\spectre(\Lbbld)\mathop{=}\{\lambda,-\lambda\}\!\!$
        &\multicolumn{2}{c||}{$\lambda\in\RR^+\nonnull$}
        &\multicolumn{2}{c||}{}
        &$\omega=-\imat\lambda\in\frac{1}{\tau}\,(\RR^+\backslash\pi\ZZ)$
	\rule[-1.5ex]{0ex}{4ex}\\
        \hline
         $h^{(2)}(p,q)=$
        &\multicolumn{2}{c||}{$\lambda\,pq$}
        &\multicolumn{2}{c||}{$\frac{1}{2\tau}p^2$}
        &$\frac{1}{2}\,\omega\,(p^2+q^2)$
	\rule[-1.5ex]{0ex}{4ex}\\
        \hline
        \parbox[b]{2cm}{\begin{flushleft} 
                        flow around\\  
                        the origin\\
                        $(\omega=|\lambda|)$\\
                        \rule{0ex}{1ex}\\
                        \end{flushleft}}
        &\multicolumn{2}{c||}{\input{fig_hyperbolic.pstex_t}}
        &\multicolumn{2}{c||}{\input{fig_parabolic.pstex_t}}
        &\input{fig_elliptic.pstex_t}\\
        \hline
\end{tabular}
\caption{\em Classification of generic quadratic hamiltonians and 
         normal forms in the neigh\-bor\-hood of
        an isolated fixed point~{\frak o}.
        \label{tab:quadham}}
\end{table}
Together with equations~(\ref{hamquad:unstable}), (\ref{hamquad:marginal})
and~(\ref{hamquad:stable}) it constitutes our first classification of normal
forms. In the next subsections, we will refine it by including higher
order terms in the hamiltonians and by considering one-parameter
families of hamiltonians and normal forms.


\subsection{Birkhoff normal form for the full hamiltonian: unstable
and stable cases}
\label{subsec:Birkhypell}

If we write the equations of motion in complex coordinates 

\begin{subequations}\label{def:complexcoord}
        \hfill\parbox{6cm}{\begin{equation}
\makebox[0cm][l]{\rule{.6cm}{0cm}$\left\{
\rule{0cm}{0.8cm}\right.$}
        \begin{align}
                z(p,q)= p+\imat q \\[1.5ex]
                \bar{z}(p,q)= p-\imat q 
        \end{align}\end{equation}
                  }
       $ \qquad\iff\qquad$
        \parbox{6cm}{\begin{equation}
                      \makebox[0cm][l]{\rule{.3cm}{0cm}$\left\{
\rule{0cm}{0.8cm}\right.$}
        \begin{align}        
                p(z,\bar{z})=&\frac{1}{2}(\bar{z}+z)\\[1ex]
                q(z,\bar{z})=&\frac{1}{2}\imat(\bar{z}-z) \ ,
        \end{align}\end{equation}
                    }\hfill
\end{subequations}

\noindent
then for the stable case~(\ref{hamquad:stable}) we get the differential
equations~$\dot{z}\mathop{=}\imat\omega z$
and~$\dot{\bar{z}}\mathop{=}-\imat\omega\bar{z}$ which correspond to the
complexified
hamiltonian~$h^{(2)}(z,\bar{z})\mathop{=}-\imat\omega\,z\bar{z}$
with~$(z,\bar{z})$ playing the role of complexified canonical variables.
Moreover, from a function~$F(z',\bar{z})$ holomorphic
with respect to its two complex variables in the neighborhood of the origin
and satisfying~$\partial_{z'\!,\,\bar{z}}F(0,0)\neq0$, one can construct a
{\em real} symplectic transformation of the form~(\ref{sympmap0}) for the
generating function~$S(p',q)={\dst\frac{1}{2}}
\Bigl(p'q'+pq-\Im\bigl[F(z',\bar{z})\bigr]\Bigr)$, where~$(p,q)$ and
$(z,\bar{z})$ are connected by the relations~(\ref{def:complexcoord}) and
similarly for~$(p',q')$, $(z',\bar{z}')$.

Then, we will study both unstable and stable cases from
a hamiltonian of the form 
\begin{equation}\label{ham1}
        H(u,v;t)=\lambda\,uv\ +\ \sum_{m\,\in\,\ZZ}\ 
                \sum_{\fatops{(\alpha,\beta)\,\in\,\NN^2}
                             {\alpha+\beta\,=\,k}
                     }
                \eh_{{\ssst\kern.12ex m\alpha\beta}}\;u^\alpha\,v^\beta\,
                \EXP{\imat m\frac{2\pi}{\tau}t}
                \ +\ O_{u,v}(k+1)       
\end{equation}
where we made use of the~\mbox{$\tau$-periodicity}.

From what precedes, if we are dealing with an unstable case
then~$\lambda\mathop{\in}\RR^{\ssst+}\!\nonnull$, $(u,v)$~are the
real coordinates~$(p,q)$ and
the coefficients~$\eh_{{\ssst\kern.12ex m\alpha\beta}}$ are
complex numbers with a priori only one 
constraint,~$\eh^*_{{\ssst\kern.12ex m\alpha\beta}}
     =\eh_{{\ssst\kern.12ex-m\alpha\beta}}$
(\mbox{$H(p,q;t)$} must be real for real $(p,q)$).
If the origin is
stable, $\lambda\mathop{\in}-\imat\,\RR^{\ssst+}\!\nonnull$~and
$(u,v)$~are the complexified coordinates~$(z,\bar{z})$ from which the
real ones are obtained via equations~(\ref{def:complexcoord}). In this case,
$\eh_{{\ssst\kern.12ex m\alpha\beta}}$~are complex numbers such 
that~$\eh_{{\ssst\kern.12ex m\alpha\beta}}^*=
             \eh_{{\ssst\kern.12ex
                                   -m\beta\alpha}}$
(\mbox{$H(z,\bar{z};t)$} must be real
for~$\bar{z}=z^*$).
In both cases, $k$~is the least integer greater or equal
to three such that~$\eh_{{\ssst\kern.12ex m\alpha\beta}}\neq0$
for some~$(\alpha,\beta)$ with~$\alpha+\beta=k$.

We now want to simplify the expression of~$H$ by eliminating as many terms
as possible in the series~(\ref{ham1}). The symplectic change of coordinates
is given by the generating function
\begin{equation}\label{genfunct1}
                S(u',v;t)=u'v\ +\ \sum_{m\,\in\,\ZZ}\ 
                \sum_{\fatops{(\alpha,\beta)\,\in\,\NN^2}
                             {\alpha+\beta\,=\,k}
                     }
                \sigma_{{\!\ssst m\alpha\beta}}\;u^{\prime\,\alpha}\,v^\beta\,
                \EXP{\imat m\frac{2\pi}{\tau}t}
                \ +\ O_{u'\!,v}(k+1)      
\end{equation}
where the transformation of degree less or equal to~$k-1$ corresponds to the
identity since we do not want to modify the terms of the same order in~$H$.
$\sigma_{{\!\ssst m\alpha\beta}}$~are complex coefficients which we now fix.
The symplectic transformation generated by the function~(\ref{genfunct1}) is 
\bigskip

\begin{subequations}\label{sympmap1}
        \hfill\makebox[0cm][l]{\rule{1.8cm}{0cm}$\left\{\rule{0cm}{1.4cm}\right.$}
        \parbox{16cm}{
        \begin{eqnarray}
                u(u',v';t)
                &=&\dst
                    u'+\frac{1}{v'}
                    \sum_{m\,\in\,\ZZ}\
                    \sum_{\fatops{(\alpha,\beta)\,\in\,\NN^2}
                                                  {\alpha+\beta\,=\,k}
                         }
                    \beta\,\sigma_{{\!\ssst m\alpha\beta}}\;
                    u^{\prime\,\alpha}\,v^{\prime\,\beta}\,
                    \EXP{\imat m\frac{2\pi}{\tau}t}
                    + O_{u'\!,v'}(k)      
                \\[2ex]
                v(u',v';t)
                &=&\dst
                    v'-\frac{1}{u'}
                    \sum_{m\,\in\,\ZZ}\
                    \smash{\sum_{\fatops{(\alpha,\beta)\,\in\,\NN^2}
                                        {\alpha+\beta\,=\,k}
                                }
                         }      
                    \alpha\,\sigma_{{\!\ssst m\alpha\beta}}\;
                    u^{\prime\,\alpha}\,v^{\prime\,\beta}\,
                    \EXP{\imat m\frac{2\pi}{\tau}t}
                    + O_{u'\!,v'}(k)              
        \end{eqnarray}}
\end{subequations}

\bigskip
\noindent and the
hamiltonian~$\dst H'(u',v';t)=H\big[u(u',v';t),v(u',v';t);t\big]
             +\frac{\partial S}{\partial t}\big[u',v(u',v';t);t\big]$
takes the form  
\begin{multline}\label{ham2}
        H'(u',v';t)=\lambda\,u'v'+\sum_{m\,\in\,\ZZ}\, 
                \sum_{\fatops{(\alpha,\beta)\,\in\,\NN^2}
                             {\alpha+\beta\,=\,k}
                     }\!
                \Bigl[\eh_{{\ssst\kern.12ex m\alpha\beta}}
                       +
                       \Bigl(\lambda(\beta-\alpha)
                             +\imat m\frac{2\pi}{\tau}
                       \Bigr)
                       \sigma_{{\!\ssst m\alpha\beta}}
                \Bigr]
                u^{\prime\,\alpha}\,v^{\prime\,\beta}\,
                \EXP{\imat m\,\frac{2\pi}{\tau}t}\\[2ex]
                +\ O_{u'\!,v'}(k+1)\;.
\end{multline}
By taking~$\sigma_{{\!\ssst m\alpha\beta}}$ so that the term between square
brackets vanishes we can cancel all the terms in the double sum except the ones
which correspond to~$m,\alpha,\beta$ such that the so-called resonant
condition is fulfilled 
\begin{equation}\label{resrel}
        \lambda(\beta-\alpha) +\imat m\,\frac{2\pi}{\tau}=0.
\end{equation}
Now the original hamiltonian may be simplified as follows.
If we assume that all the non-resonant terms have been cancelled
up to degree~$k-1$ in~$(p,q)$, we can get rid of those
of degree~$k$ with the help of the generating function~(\ref{genfunct1})
provided we choose 
\begin{equation}\label{genfracchoice}
\left\{\rule{-.1cm}{.7cm}\right.\raisebox{.8ex}{$
\begin{array}{ll}
        \dst
        \sigma_{{\!\ssst m\alpha\beta}}=
                  -\frac{  \eh_{{\ssst\kern.12ex m\alpha\beta}}
                        }{ \lambda(\beta-\alpha)+\imat m\,2\pi/\tau
                         }
        &\quad\mbox{for the non-resonant~$m,\alpha,\beta$,}\\[2ex]
        \dst
        \sigma_{{\!\ssst m\alpha\beta}}=0
        &\quad\mbox{for the resonant~$m,\alpha,\beta$.}
\end{array}$}
\end{equation}
Since we can repeat this procedure order by order, we are left with a
hamiltonian whose normal form is made up exclusively of resonant terms. This
obstruction to the linearization of differential equations is well known
from the works of~Poincar{\'e}, Dulac~and later on Sternberg, Grobman and
Hartman \cite{Arnold88a} among others. It is remarkable that the resonant relations which
prevent the reduction via diffeomorphism are the same resonant relations as
those we obtained above, even though they lie in a more restricted class of
coordinate transformations (the symplectic ones). The convergence of the
normalization procedure of a dynamical system is an old and delicate problem
because of the possibly small denominators in
expression~(\ref{genfracchoice}). 
                                 
Let us now discuss the nature of the resonant terms depending on whether
the origin is unstable or stable.
\paragraph{\ref{unstable} \ Unstable case: }
Since~$\lambda\mathop{\in}\RR^{\ssst+}\!\nonnull$ the resonant relation is
fulfilled only if~$m=0$~and~$\alpha=\beta$. Thus, 
for~$n\in\NN\backslash\{0,1\}$ there always exists a real symplectic coordinate
chart in which the hamiltonian has the form   
\begin{equation}\label{ham:Birkhyper1}
        H(p,q;t)=\lambda\,pq +\sum_{k=2}^n\,\eh_k(pq)^k+O_{p,q}(2n+1)
\end{equation}
for some~$\eh_k\in\RR$. This normal form has been derived for the
first time by \citeasnoun[\S\hspace{.5ex}III.8]{Birkhoff27a}
\citeaffixed{Gustavson66a}{see also}.
$n\geqslant2$~is
a cut off integer whose value fixes the accuracy with which we want to
reproduce the full dynamics and $2n$~will be called the order of the normal
form. The limit~$n\to\infty$ has been shown to be locally
convergent~\cite[\S4]{Moser56a} and hence the dynamics is equivalent to an
integrable~one.
\paragraph{\ref{stable} \ Stable case: }
We have~$\dst\lambda=\imat\omega
\in\frac{1}{\tau}\,\imat(\RR^{\ssst+}\!\backslash\pi\ZZ)$.

\noindent
$\rightarrowtriangle\boxed{\dst\frac{\omega\tau}{2\pi}\notin\QQ}\,$: 
The hamiltonian can also
be reduced to a Birkhoff normal form, which in real coordinates, is
\begin{equation}\label{ham:Birkelli}
        H(p,q;t)=\frac{1}{2}\,\omega\,(p^2+q^2) 
                 \ +\ \sum_{k=2}^n\,\tilde\ea _k(p^2+q^2)^k
                 \ +\ O_{p,q}(2n+1).
\end{equation}
for some~$\tilde\ea _k\in\RR$ and for every integer~$n\geqslant2$. In
action-angle variables,

\begin{subequations}\label{def:aavar}
        \hfill\parbox{6cm}{\begin{equation}
\makebox[0cm][l]{\rule{.6cm}{0cm}$\left\{
\rule{0cm}{0.9cm}\right.$}
        \begin{align}
                 I=\frac{1}{2}\,(p^2+q^2)\\[1.5ex]
               \theta=\arctg\!\left(\frac{q}{p}\right) 
        \end{align}\end{equation}
                  }
       $ \qquad\qquad\iff\qquad$
        \parbox{6cm}{\begin{equation}
                      \raisebox{-0.08cm}[0cm][0cm]{\makebox[0cm][l]{\rule{.7cm}{0cm}$\left\{
\rule{0cm}{0.74cm}\right.$}}
        \begin{align}        
                 p=\sqrt{2I}\cos\theta\\[1ex]
                q=\sqrt{2I}\sin\theta
        \end{align}\end{equation}
                    }\hfill
\end{subequations}

\noindent
and after having rescaled the coefficients,
$\ea _k\DEF2^k\tilde\ea _k$, it takes the form,
\begin{equation}\label{ham3}
        H(I,\theta;t)=\omega I 
                 \ +\ \sum_{k=2}^n\,\ea _kI^k
                 \ +\ O_I(2n+1)\;.
\end{equation}
The problem of small denominators appears now clearly. The
non-convergence of the normalization procedure as~$n\to\infty$
is due to the fact that the rational numbers~$\QQ$ are dense in~$\RR$:
$\sigma_{{\!\ssst m\alpha\beta}}$~in equation~(\ref{genfracchoice})
can be arbitrarily large
if we take~$\beta-\alpha$ small enough.

\noindent
$\rightarrowtriangle\boxed{\dst\frac{\omega\tau}{\pi}\in\QQ\backslash\NN}\,$: 
Let~$\ell\in\NN\backslash{\{0,1,2\}}$ be the smallest integer such that it
exists~$r\in\NN\nonnull$ with 
\begin{equation}
        \omega=\frac{r}{\ell}\,\frac{2\pi}{\tau}\;.
\end{equation}
Then, the resonant terms correspond to~$(m,\alpha,\beta)\in\ZZ\times\NN^2$
obeying~$(\beta-\alpha)r=-\ell m$. They are of degree~$s\DEFt\alpha+\beta$
in~$(p,q)$. We can get rid of the time-dependence of the normal form
if we describe the dynamics in a frame uniformly rotating around the origin
at angular speed~$\omega$. This corresponds to the symplectic
transformation 
\begin{equation}\label{rotframe}
  \Bigg\{
        \begin{array}{ccc}
                z      &\longmapsto&z'(z,\bar{z})= z\,\EXP{+\imat\omega t}
		\\[2ex]
                \bar{z}&\longmapsto&\bar{z}^{\,\prime}(z,\bar{z})
                                    =\bar{z}\,\EXP{-\imat\omega t}
        \end{array}\quad\mbox{generated by}\quad
        S(z',\bar{z})=z'\bar{z}\,\EXP{-\imat\omega t},
\end{equation}
leading to
\begin{equation} 
        H'(z',\bar{z}^{\,\prime};t)=
        H\bigl(z'\EXP{-\imat\omega t},
               \bar{z}^{\,\prime}\EXP{\imat\omega t}
               ;t
         \bigr)
        +\frac{\partial S}{\partial t}\bigl(z',
                                       \bar{z}^{\,\prime}\EXP{\imat\omega t}
                                       \bigr)
\end{equation}
which has no longer any quadratic terms nor any time-dependence in all the
resonant ones. 

The latter can be classified in three sets~$R_-$, $R_0$~and~$R_+$
defined by
\begin{equation}
        R_\pm\,\Bigg\{\begin{array}{l}
                        \beta=\alpha\pm\kappa\ell\\[2ex]
                        s=2\alpha\pm\kappa\ell\geqslant3
                      \end{array}
        ,\qquad 
        R_0\,\Bigg\{\begin{array}{l}
                        \beta=\alpha\\[2ex]
                        s=2\alpha\geqslant3
                    \end{array}
        \qquad \mathrm{for}\ 
        \alpha=0,1,2,\ldots\ \mathrm{and}\  
        \kappa=0,1,2,\ldots\;.
\end{equation}
$R_0$~corresponds to the Birkhoff like terms. We obtain~$R_-$ from~$R_+$
 exchanging~$\beta$ with~$\alpha$.
\begin{figure}[!ht]
\begin{center}
\input{RRR.pstex_t}
\caption{\sl\baselineskip=0.25in\label{fig:RRR}
         Graphical representation of the resonant terms in a  hamiltonian
         in the neighborhood of a stable fixed point whose
         linearized motion is a rotation of rational angular speed.
        }
\end{center}
\end{figure}
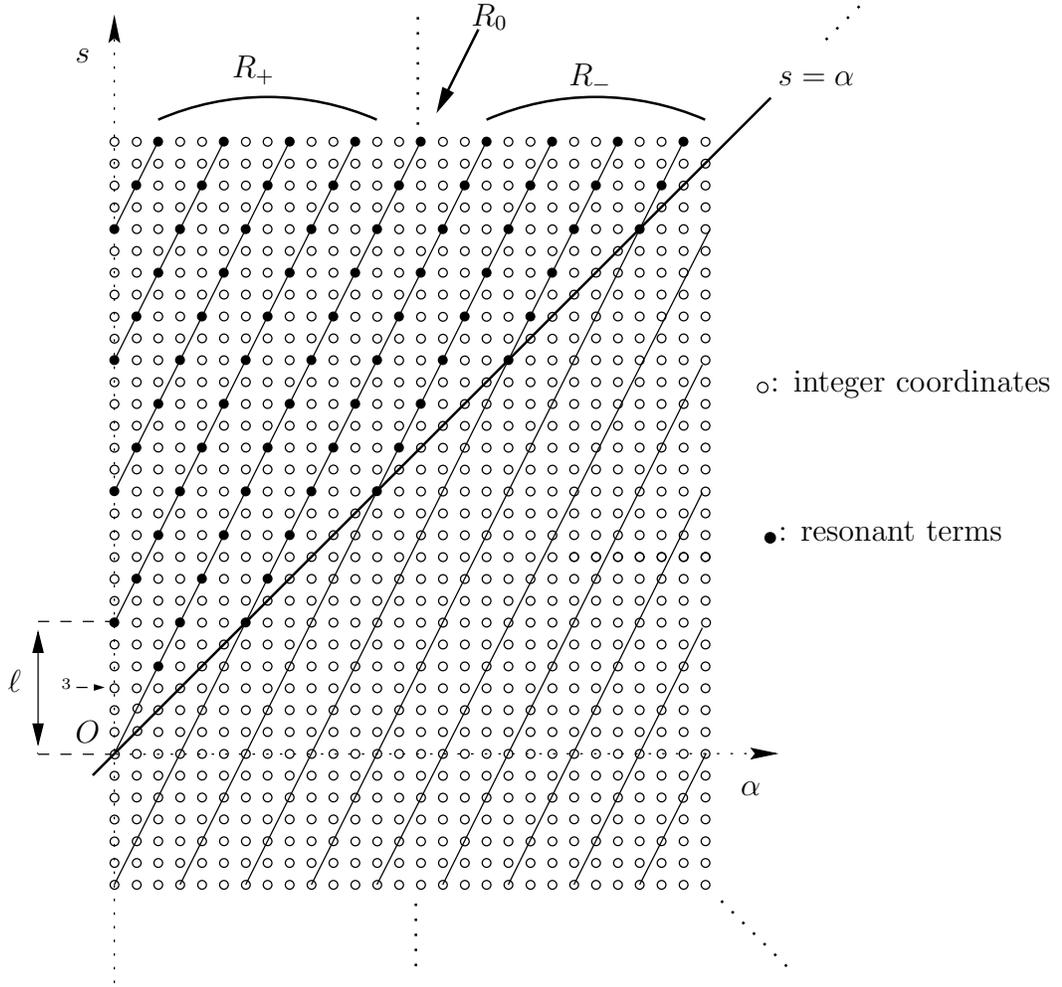

These sets have a graphical interpretation
in the \mbox{$(s,\alpha)$-plane} (see figure~\ref{fig:RRR} and
Ref.\cite{Bruno70a}]). The resonant terms are associated with the points of
integer coordinates~$s\geqslant\alpha\geqslant0$~and~$s\geqslant3$ which lie on
straight lines
of slope two, whose intercept are strictly positive, zero and strictly negative
for~$R_-$, $R_0$~and~$R_+$ respectively. 

For~$\ell\geqslant3$, in addition of the Birkhoff~like terms, 
the resonant terms of lowest degree in~$(p,q)$ are
$(\alpha=0,\beta=\ell,m=-r)$~and~$(\alpha=\ell,\beta=0,m=r)$. Hence,
dropping the primes of the rotating coordinates, we have 
\begin{equation}\label{ham4}
        H(z,\bar{z};t)=\!\!\sum_{\fatops{k\,\in\,\NN}
                                        {2\,\leqslant\,k\,\leqslant\,\ell/2}
                                }
                            \!\!\tilde\ea _k(z\bar{z})^k
                       \ +\ \eh_{{\ssst\kern.12ex r\ell0}}^*
                            \;\bar{z}^{\,\ell}
                       \ +\ \eh_{{\ssst\kern.12ex r\ell0}}
                            \;z^\ell
                       \ +\ O_{z,\bar{z}}(\ell+1)\;,
\end{equation}
for some~$\tilde\ea _k\in\RR$
and~$\eh_{{\ssst\kern.12ex r\ell0}}\in\CC$.
In real coordinates~$(p,q)$, we get 
\begin{equation}\label{ham5}
        H(p,q;t)=\!\!\sum_{\fatops{k\,\in\,\NN}
                                  {2\,\leqslant\,k\,\leqslant\,\ell/2}
                          }
                            \!\!\tilde\ea _k(p^2+q^2)^k 
                       \ +\ |\eh_{{\ssst\kern.12ex r\ell0}}
                                        |\;\Re\big[(p+\imat q)^\ell
                                \big]   
                       \ +\ O_{p,q}(\ell+1)\;.
\end{equation}
In action-angle variables 
\begin{equation}\label{ham6}
        H(I,\theta;t)=\!\!\sum_{\fatops{k\,\in\,\NN}
                                       {2\,\leqslant\,k\,\leqslant\,\ell/2}
                               }
                            \!\!\ea _kI^k
                       \ +\ \eb _\ell I^{\ell/2}
                            \cos(\ell\theta)
                       \ +\ O_{\!\sqrt{I}}(\ell+1)
\end{equation}
where~$\ea _k\DEF2^k\tilde\ea _k$,
      $\eb _\ell\DEF2^{\ell/2}|\eh_{{\ssst\kern.12ex r\ell0}}|$.
In order  to  cancel the  irrelevant 
phase~$\theta_0\DEF\arg(\eh_{{\ssst\kern.12ex r\ell0}})$,
we have made a final rotation:~$\theta\mapsto\theta-\theta_0$. 
        
\subsection{Normal form for the full hamiltonian: marginal case}
\label{subsec:nofofullmarg}

 We start from the real hamiltonian
\begin{equation}\label{ham7}
        H(p,q;t)=\frac{1}{2\tau}p^2+\ \sum_{m\,\in\,\ZZ}\
                \sum_{\fatops{(\alpha,\beta)\,\in\,\NN^2}
                             {\alpha+\beta\,=\,k}
                     }
                \eh_{{\ssst\kern.12ex m\alpha\beta}}\;p^\alpha\,q^\beta\,
                \EXP{\imat m\frac{2\pi}{\tau}t}
                \ +\ O_{p,q}(k+1),
\end{equation}

\enlargethispage{-2\baselineskip}

\noindent$k$ being a fixed integer greater or equal to three
and~$\eh^*_{{\ssst\kern.12ex m\alpha\beta}}
     =\eh_{{\ssst\kern.12ex\rule[0.25ex]{.5ex}{.02ex}m\alpha\beta}}$.
We are looking for a suitable real generating function of the form 
\begin{equation}\label{genfunct2}
                S(p',q;t)=p'q\ +\ \sum_{m\,\in\,\ZZ}\ 
                \sum_{\fatops{(\alpha,\beta)\,\in\,\NN^2}
                             {\alpha+\beta\,=\,k}
                     }
                \sigma_{{\!\ssst m\alpha\beta}}\;p^{\prime\,\alpha}\,q^\beta\,
                \EXP{\imat m\frac{2\pi}{\tau}t}
                \ +\ O_{p'\!,\,q}(k+1) \ .
\end{equation}
Similarly to transformation~(\ref{sympmap1}) we get 

\begin{subequations}\label{sympmap2}
        \hfill\makebox[0cm][l]{\rule{1.8cm}{0cm}$\left\{\rule{0cm}{1.4cm}\right.$}
      \parbox{16cm}{
        \begin{eqnarray}
                p(p',q';t)
                &=&\dst
                    p'+
                    \sum_{m\,\in\,\ZZ}\
                    \sum_{\fatops{(\alpha,\beta)\,\in\,\NN^2}
                                                  {\alpha+\beta\,=\,k}
                         }
                    \beta\,\sigma_{{\!\ssst m\alpha\beta}}\;
                    p^{\prime\,\alpha}\,q^{\prime\,\beta-1}\,
                    \EXP{\imat m\frac{2\pi}{\tau}t}
                    + O_{p'\!,\,q'}(k)      
                \\[2ex]
                q(p',q';t)
                &=&\dst
                    q'-
                    \sum_{m\,\in\,\ZZ}\
                    \smash{\sum_{\fatops{(\alpha,\beta)\,\in\,\NN^2}
                                        {\alpha+\beta\,=\,k}
                                }
                         }      
                    \alpha\,\sigma_{{\!\ssst m\alpha\beta}}\;
                    p^{\prime\,\alpha-1}\,q^{\prime\,\beta}\,
                    \EXP{\imat m\frac{2\pi}{\tau}t}
                    + O_{p'\!,\,q'}(k) \ .
        \end{eqnarray}}
\end{subequations}

\bigskip\noindent
For all integers~$m$~and~$(\alpha,\beta)$ such that~$\alpha+\beta=k\geqslant3$,
the coefficient
of~$p^{\prime\,\alpha}\,q^{\prime\,\beta}\,\EXP{\imat m\frac{2\pi}{\tau}t}$
in the expansion of the
hamiltonian \mbox{$\dst H'(p',q';t)=H\big[p(p',q';t),q(p',q';t);t\big]
             +\frac{\partial S}{\partial t}\big[p',q(p',q';t);t\big]$}
is
\begin{equation}
   \left\{
   \begin{array}{ll}
        \dst
        \eh_{{\ssst\kern.12ex m\alpha\beta}}
        +\imat  m\,\frac{2\pi}{\tau}\,\sigma_{{\!\ssst m\alpha\beta}}
        -2(\alpha+1)\,\sigma_{{\!\ssst m(\alpha+1)(\beta-1)}}&\quad
        \forall\,\alpha\geqslant0,\forall\,\beta>1\ \mbox{such that}\ 
                                                    \alpha+\beta=k\ ;
        \\[1.5ex]
        \dst
        \eh_{{\ssst\kern.12ex mk0}}
        +\imat  m\,\frac{2\pi}{\tau}\,\sigma_{{\!\ssst mk0}}
        &\quad\mathrm{for}\ \alpha=0\ \mathrm{and}\ \beta=k .   
   \end{array}
   \right.
\end{equation}

If we want to cancel them for fixed~$k$~and~$m$, we must solve the matrix
equation \stepcounter{equation}

\noindent
\input{matrix.pstex_t}
\vspace{-.8\baselineskip}
\nopagebreak\begin{equation}\label{eqmat1}\end{equation}
\noindent  which we denote by 
\begin{equation}\label{eqmat2}
        A_0\,\sigma=\eh\;.
\end{equation}
Since $\det A_0=\big(\imat m\,\frac{2\pi}{\tau}\big)^{k+1}$, the above equation
has always
a solution for every~$m\neq0$~and~$\eh$, thus all the time-dependent
terms can be
cancelled. When~$m=0$, the kernel of~$A_0$ is one-dimensional and
generated by~${\eurm u}_1\DEF(1,0,\dots,0)$ and the image~$\im A_0$ is a
hyper plane of equation~${\eurm u}_{k+1}.\eh=0$
with~${\eurm u}_{k+1}\DEF(0,\dots,0,1)$ and where
we have introduced the canonical scalar product. Let us now
decompose the right hand side of~(\ref{eqmat1}) in the following manner 
\begin{equation}
        \eh=\eh_{\ssst\parallel}
                 +\eh_{\ssst\perp}
\end{equation}
where~$\eh_{\ssst\parallel}\in\im A_0$
and~$\eh_{\ssst\perp}.\im A_0=0$.
We can choose~$\sigma=A_0^{-1}\eh_{\ssst\parallel}$, and the only
coefficients which cannot be eliminated correspond to the elements orthogonal
to~$\im A_0$, that is  those which are parallel to~${\eurm u}_{k+1}$ 
\begin{equation}\label{hhhzeroh}
        \eh_{{\ssst\kern.12ex 00k}}=
        \eh_{{\ssst\kern.12ex 01(k-1)}}= 
        \cdots=\eh_{{\ssst\kern.12ex 0\alpha(k-\alpha)}}=\cdots=
        \eh_{{\ssst\kern.12ex 0(k-1)0}}=0\quad\mathrm{and}\quad  
        \eh_{{\ssst\kern.12ex 0k0}}\neq0\;.
\end{equation}
The latter are therefore the resonant terms and the form of the
hamiltonian is, after relabeling the coordinates 
\begin{equation}\label{ham8.1}
        H(p,q;t)\ =\ \frac{1}{2\tau}p^2\ +\ \sum_{k=3}^n\eh_kq^k\ +\ O_{p,q}(n+1),
\end{equation}
for some~$\eh_k\in\RR$ and for any integer~$n\geqslant3$.

\section{Normal forms for one-dimensional parameter families of hamiltonians}
  \label{sec:nofounfham}      
Following what we have planned in section~\ref{sec:reduction}, we will now consider
that the hamiltonian~$H(p,q;t)$ 
is embedded in a smooth family depending on one real parameter, say~$\ve$. A
typical example of the control parameter is the
energy of a non-reduced two degrees of freedom system. Without loss of
generality, we assume that the origin~{\frak o} is an isolated
equilibrium point of~$H(p,q;t;\ve)$, obeying equations~(\ref{def:generaleq1})
at~$\ve=0$. We can follow step by step the discussion of the previous section
for~$H(p,q;t;\ve=0)$ and classify its generic normal form around~{\frak o}.
Now the question is to study what happens to the dynamics if we vary~$\ve$ 
in a neighborhood of zero. We must first wonder about the behavior of the
fixed point. By virtue of the implicit function theorem applied to the equations

\begin{subequations}\label{def:generaleq2}
        \hfill\makebox[0cm][l]{\rule{2.8cm}{0cm}$\forall\,t\in \RR,\qquad$}
        \makebox[0cm][l]{\rule{4.8cm}{0cm}$\left\{\rule{0cm}{1.2cm}\right.$}\parbox{14cm}{
         \begin{align}
                \frac{\partial H}{\partial p}(p,q;t;\ve)
                    &=0\;;
                \\[2ex]
                \frac{\partial H}{\partial q}(p,q;t;\ve)
                    &=0   
         \end{align}}        
\end{subequations}

\noindent
in a neighborhood of~$(p,q;\ve)=(0,0;0)$, we can follow as a function of
$\ve$  the isolated solutions if
\begin{equation}\label{implicit:H}
        0\neq
        \det\big(\Hess\!\big[H(w;t;\ve)\big]\big)\eval{w=\binom{0}{0},\,\ve=0}
        =\det
        \begin{pmatrix}
        \partial^{\,2}_{p,p}H(w;t;\ve)&\partial^{\,2}_{p,q}H(w;t;\ve)\\[2ex]
        \partial^{\,2}_{q,p}H(w;t;\ve)&\partial^{\,2}_{q,q}H(w;t;\ve)
        \end{pmatrix}
        \eval{w=\binom{0}{0},\,\ve=0}
        \;.
\end{equation}
An equivalent condition may be written for the monodromy matrix~$\Mbbld$.
If~$w\mapsto\Phi(w;\ve)$ is the (local) Poincar{\'e} map, the implicit 
function theorem applied to 
\begin{equation}
        \Phi(w;\ve)=w
\end{equation}
states that we can follow smoothly the fixed point if
\begin{equation}\label{implicit:M}
        0\ \neq\ 
        \det\bigl[\partial_w\Phi(w;\ve)-\Id\bigr]\eval{w=\binom{0}{0},\,\ve=0}
        =\ \det\bigl[\Mbbld(\ve)-\Id\bigr]\eval{\ve=0}
        =\ 2-\tr\bigl[\Mbbld(\ve)\bigr]\eval{\ve=0}.
\end{equation}
The last equality comes from property~(\ref{detMeq1}).
From table~\ref{tab:quadham} we see 
that the latter condition is fulfilled in both stable and
unstable cases as well as in case~\ref{marginal}'', because
there exists a neighborhood~${\clgv I}$
with~$0\mathop{\in} {\clgv I}\mathop{\subset}\RR$
and a smooth family of points~${\frak p}(\ve)$ such
that~${\frak p}(0)={\frak o}$, and~${\frak p}(\ve)$ is a fixed point
of~$\Phi(\kern.2ex\cdot\,;\ve)$ for all~$\ve$ in~${\clgv I}$. If we
now make an appropriate
smooth \hbox{$\ve$-dependent} change of
coordinates in order to
translate back~${\frak p}(\ve)$ to~{\frak o}, then we deal with
a hamiltonian family~$H(w;t;\ve)$ which admits~{\frak o} as 
equilibrium point for all~$\ve$ in~${\clgv I}$.
Elimination of the time dependence of the quadratic part of the hamiltonian
around~{\frak o} can be done as described in subsection~\ref{subsec:elimtime}
in the whole~${\clgv I}$.
The \mbox{$\tau$ or $2\tau$-periodic} moving frame in which we follow
the dynamics is smooth with respect to~$\ve$.
At this stage, we must distinguish cases~\ref{unstable} and~\ref{stable}
 from case~\ref{marginal}''. On the one hand, in the former two cases
the stability of the fixed point will not change if~${\clgv I}$ is small
enough because from a topological point of view
property~$\tr\bigr[\Mbbld(\ve)\bigr]\mathop{\neq}2$ defines an open
set for~$\ve$. On the other hand, case~\ref{marginal}'' corresponds
to~$\tr\bigr[\Mbbld(\ve\mathop{=}0)\bigr]\mathop{=}-2$ which 
 breaks down as soon as~$\ve\mathop{\neq}0$. More precisely, generically
~$\tr\bigr[\Mbbld(\ve)\bigr]$ will cross transversally
the value~$-2$~and~{\frak o} will change from stable to unstable
being marginal only at~$\ve=0$. 
We will study first the stable and unstable unfoldings, then the
marginal unfolding will be considered and we will finish
with the study of case~\ref{marginal}' where even the presence of
a fixed point cannot be guaranteed  as soon as~$\ve\neq0$. When embedded in a
one-parameter unfolding,  case~\ref{marginal}' and~\ref{marginal}'' are
respectively called {\em extremal} and {\em transitional} 
by~\citeasnoun[resp.~definition~1.9 and definition~1.12]{Meyer70a}.

\subsection{Normal forms for unstable and stable unfoldings}
\label{subsec:nofounfoldus}
From what precedes, we can deduce that
if~{\frak o} is an unstable (resp.~stable) fixed point
of~$H(w;t;\ve\mathop{=}0)$ it will remain a fixed point
of~$H(w;t;\ve)$ for all~$\ve$ in a small enough neighborhood~${\clgv I}$ 
of zero. 
Indeed, in both cases all the symplectic changes of coordinates we have
made in subsection~\ref{subsec:nofoquadham} and~\ref{subsec:Birkhypell} vary 
smoothly with~$\ve$.

In the unstable case, we are led to the Birkhoff normal
form~(\ref{ham:Birkhyper1}) with now~$\lambda$ 
and~\intext{$\big\{\eh_k\big\}_{k\,\in\,\finiteset{2}{n}}$} being smooth real
functions of~$\ve$ 
\begin{equation}\label{def:Hu}
        H_u(p,q;t;\ve)
        =\lambda(\ve)\,pq +\sum_{k=2}^n\,\eh_k(\ve)\,(pq)^k
        +O_{p,q}(2n+1)\;.
\end{equation}

In the stable case, we cannot prevent~$\omega$ from
crossing~\intext{$\dst\frac{\pi}{\tau}\,\QQ\backslash\NN$} infinitely many 
times. Let us then assume that
\begin{equation}\label{omega0}
        \dst\omega(\ve\mathop{=}0)\,=\,\omega_0\,
        \DEF\,\frac{r}{\ell}\,\frac{2\pi}{\tau}\;
\end{equation}
with~$r\in\NN\nonnull$~and~$\ell\in\NN\backslash{\{0,1,2\}}$ coprimes. Choose
the neighborhood~$I\supset0$ so that it contains no
rational~$\frac{\tau}{\pi}\,\omega(\ve)$
whose denominator is lower than~$\ell$. Moreover since
generically~$\frac{d\omega}{d\ve}\eval{\ve=0}\mathop{\neq}0$,
we can always choose the control parameter in order to 
satisfy~$\omega(\ve)-\omega_0=\ve$.
Thus, in a rotating frame defined by~(\ref{rotframe}) 
with~$\omega=\omega_0$ we get, from 
expressions~(\ref{ham5}) and~(\ref{ham6}),
\begin{subequations}\label{def:Hell}
\begin{equation}\label{def:Hellcart}
        H_\ell(p,q;t;\ve)=\frac{1}{2}\,\ve\,(p^2+q^2)+
                 \!\!\sum_{\fatops{k\,\in\,\NN}
                                  {2\,\leqslant\,k\,\leqslant\,\ell/2}
                          }
                            \!\!\tilde\ea _k(\ve)\,(p^2+q^2)^k 
                       \ +\ \big|\eh_{{\ssst\kern.12ex\ell0\ell}}
                                              (\ve)\big|
                                        \;\Re\big[(p+\imat q)^\ell
                                \big]   
                       \ +\ O_{p,q}(\ell+1)
\end{equation}
and
\begin{equation}\label{def:Hellaa}
        H_\ell(I,\theta;t;\ve)=\ve I+
                      \!\!\sum_{\fatops{k\,\in\,\NN}
                                       {2\,\leqslant\,k\,\leqslant\,\ell/2}
                               }
                            \!\!\ea _k(\ve)\,I^k
                       \ +\ \eb _\ell(\ve)\,I^{\ell/2}
                            \cos(\ell\theta)
                       \ +\ O_{\!\sqrt{I}}(\ell+1)
\end{equation}
\end{subequations}
in momentum-position and action-angle variables, respectively.

\subsection{Normal form for the transitional fixed point}
\label{subsec:nofotrans}

Since the stability of the fixed point will change when
crossing~$\ve=0$, derivation of the normal form of the unfolding is not
as straightforward as in the previous subsection.
Let us then start from the following hamiltonian  
\begin{equation}\label{ham11}
        H(p,q;t;\ve)=\underbrace{\frac{1}{2}\,\ec (\ve)\,p^2
                                 +\ed (\ve)\,pq
                                 +\frac{1}{2}\,\ee (\ve)\,q^2
                                }_{\dst\DEF h^{(2)}(p,q;\ve)}
                     +\ O_{p,q}(3)\;.
\end{equation}
$\ec $, $\ed $ and $\ee $
are smooth functions of~$\ve$ such that 
\begin{equation}
        \qquad\ec (0)=\frac{1}{\tau}\;,\qquad\ed (0)=0\;,\qquad\ee (0)=0\; .
\end{equation}
In Eq.(\ref{ham11}) there are no linear
terms in~$(p,q)$ because we have already shown that
the origin {\frak o} is fixed  in a suitable
symplectic chart for all~$\ve$.
Recall that the hamiltonian~(\ref{ham11}) corresponds to a
\hbox{$2\tau$-periodic} stroboscopic view of the true dynamics.

We will proceed as in subsection \ref{subsec:nofoquadham}. We look for
a family of new symplectic charts obtained via a family of generating
functions of the form
\begin{equation}\label{genfunctquadfam}
        S(p',q;\ve)=\frac{1}{2}\,\gamma(\ve)\,p^{\prime\,2}
                +\delta(\ve)\,p'q
                +\frac{1}{2}\,\eta(\ve)\,q^2
                +O_{p'\!,\,q}(3)
\end{equation}
where~$(\delta,\gamma,\eta)$ are now smooth real functions of~$\ve$ such 
that~$\delta$ does not vanish in~${\clgv I}$. We are led to an expression
identical to
(\ref{newhamquad}) for~$h^{(2)}(p',q')$. The length of~${\clgv I}$
can possibly be reduced
in order to have, $\forall\,\ve\in I,\ec(\ve) >0$. Moreover since
generically~\intext{$\frac{ d({\ssst\eurm d^\mathrm{2}      -ce})}{ d\ve}
              \eval{\ve\mathop{=}0}\neq0$},
we can reparametrize the unfolding
by~$\ve'\DEFt-\tau({\eurm d^\mathrm{2}-ce})$. We can then choose
\begin{equation}\label{choicegde}
        \forall\,\ve\in I,\qquad
        \gamma(\ve)=0;\quad
        \delta(\ve)=1/\sqrt{\tau\ec(\ve)}; \quad
        \eta(\ve)=-\ed/\ec \ ,
\end{equation}
which implies that~$\ec'=1/\tau$,~$\ed'=0$, and~$\ee'=\ve$.

Dropping all the primes, the normal form of the 
quadratic terms of the unfolding of a transitional point is
\begin{equation}
 	h^{(2)}(p,q;\ve)=\frac{1}{2\tau}p^2+\ve\,q^2\;.
\end{equation}

\vspace{-1pt}
\noindent
We can now reproduce step by step the reasoning of 
subsection~\ref{subsec:nofofullmarg}. Starting from 
\begin{equation}
        H(p,q;t;\ve)=\frac{1}{2\tau}p^2+\ve\,q^2+\sum_{m\,\in\,\ZZ}\
                \sum_{\fatops{(\alpha,\beta)\,\in\,\NN^2}
                             {\alpha+\beta\,=\,k\geqslant3}
                     }
                \eh_{{\ssst\kern.12ex m\alpha\beta}}(\ve)
                \;p^\alpha\,q^\beta\,
                \EXP{\imat m\frac{2\pi}{\tau}t}
                \ +\ O_{p,q}(k+1),
\end{equation}
we construct a generating function~(\ref{genfunct2}) with 
the~$\sigma$'s being smooth functions of~$\ve$. We are led to a
 matrix equation~$A(\ve)\,\sigma(\ve)=\eh(\ve)$
with~$A(\ve\mathop{=}0)=A_0$ [see~(\ref{eqmat1})]. Possibly
reducing the length of~${\clgv I}$,
we can assume that~$\forall\,\ve\in {\clgv I},\det A(\ve)\neq0$ when~$m\neq0$,
and then the time dependence can be cancelled uniformly in~$\ve$.
In fact, we can also eliminate all the monomials of order greater
than three except those terms corresponding to~(\ref{hhhzeroh}) since we
cannot cancel them {\em uniformly} in~$\ve$. We thus obtain the
full normal form for the unfolding in the transitional case
\begin{equation}\label{nofotrans1}
        H(p,q;t;\ve)=\frac{1}{2\tau}\,p^2+\ve\, q^2
                     +\sum_{k=3}^n\eh_k(\ve)\,q^k\ +\ O_{p,q}(n+1),
\end{equation}
where~$\{\eh_k\}_{k\in\finiteset{3}{n}}$ are some real functions
of~$\ve$. 

\noindent 
There are certain constraints on the~$\eh$'s due to the fact that
the hamiltonian~(\ref{nofotrans1}) describes the
dynamics in a~\mbox{$(-2\tau)$-periodic} stroboscopic view.
After one period~$\tau$, the monodromy matrix verifies
\begin{equation}\label{monodromy2bis}
\Mbbld(\ve\mathop{=}0)\circeq
                         \begin{pmatrix}
                                        -1&\phantom{-}0\,\\ \phantom{-}1&-1\,
                         \end{pmatrix}\;.
\end{equation}

It can be shown using the Lie
transformation~\cite[\S VII. E. 5]{Meyer/Hall92a}
that we must have~$H(p,-q;t;\ve=0)=H(p,q;t;\ve=0)+\ O_{p,q}(n+1)$ which
implies~$\forall\,k\in\NN\nonnull,\ \eh_{2k+1}(0)=0$.
We are going to prove the same result for
the first terms of the normal form~(up to $n=6$) but without the help of
the Lie transformation theory.
We will show in this subsection that~$\eh_3(0)=0$. We will later on show in
subsection~\ref{subsec:genfuntrans}
that~\intext{$\frac{d{\ssst\eh}_3}{d\ve}\eval{\ve=0}=0$}  and~$\eh_5(0)=0$ (see
equations~(\ref{ab0})). 

Let us fix~$\ve=0$, define~$v_3(q)\DEFt\eh_3(0)q^3$ and anticipate a
little bit on the calculus of Poincar{\'e} stroboscopic maps of
section~\ref{sec:nofogenfun}.
We apply formula~(\ref{poissonserie1}) in order to calculate the Poincar{\'e}
map at time~$-2\tau$. The Poisson brackets are given by
\begin{align}
       &\poisson{q,H}=\frac{1}{\tau}\,p\ +\ O_{p,\,q}(3)\;;
       \\[1ex]
       &\poisson{p,H}=-v_3'(q)\ +\ O_{p,\,q}(3)\;;
       \\[1ex]
       &\poisson{\poisson{p,H},H}=-\frac{1}{\tau}\,p\,v_3''(q)
                                   \ +\ O_{p,\,q}(3)\;;\\[1ex]
       &\poisson{\poisson{\poisson{p,H},H},H}=-\frac{1}{\tau^2}\,p^2v_3'''(q)
                                   \ +\ O_{p,\,q}(3)\;;
       \\[1ex]
       &\,\{\kern-1.15mm|\ldots\{\kern-1.15mm|p,
                  \underbrace{
                              H|\kern-1.15mm\},\dots,H
                              |\kern-1.15mm\}
                             }_{n\ \mathrm{times,}\;n\geqslant4}
                         \ \in\ O_{p,\,q}(3)\;.
\end{align}

Then it is straightforward
to get~$w_2\DEFt w(-2\tau)$ up to order two in~$w_0\DEFt w(0)$ 

\noindent
\begin{subequations}\label{poinmap220}
        \hfill
         \raisebox{-0.08cm}[0cm][0cm]{
        \makebox[0cm][l]{\rule{1.5cm}{0cm}$\left\{\rule{0cm}{0.9cm}\right.$}}
        \parbox{16.4cm}{
         \begin{align}
                \hspace{0.2cm}p_2&=p_0+2\tau\,v_3'(q_0)-2\tau
\,p_0\,v_3''(q_0)+\frac{4}{3}\,\tau\,p_0^2\,v_3'''(q_0)+O_{p_0,\,q_0}(3)\;;
                \\[1ex]
                 \hspace{0.2cm}q_2&=q_0-2p_0-2\tau\,v_3'(q_0)
-\frac{4}{3}\,\tau\,p_0\,v_3''(q_0)-\frac{2}{3}\,\tau\,p_0^2\,v_3'''(q_0)+O_{p_0,\,q_0}(3)\;;
         \end{align}}        
\end{subequations} 

\noindent
Now we use the fact that this map is the square of another map
whose linear part is known (see equation~(\ref{monodromy2bis}))

\noindent
\begin{subequations}\label{poinmap22}
        \hfill
         \raisebox{-0.1cm}[0cm][0cm]{
        \makebox[0cm][l]{\rule{4.8cm}{0cm}$\left\{\rule{0cm}{0.75cm}\right.$}}
        \parbox{16.4cm}{
         \begin{align}
                \hspace{0.2cm}p_1&=-p_0+f_2(p_0,q_0)+O_{p_0,\,q_0}(3)\;;
                \\[2ex]
                 \hspace{0.2cm}q_1&=-q_0+p_0+g_2(p_0,q_0)+O_{p_0,\,q_0}(3)\;;
         \end{align}}        
\end{subequations} 

\noindent
($f_2$ and~$g_2$ are two quadratic forms in~$w$). After having
iterated the last map and identified the result
with~(\ref{poinmap220})
we get
\begin{equation}\label{v3g2}
 2\tau\,v_3'(q)-2\tau
\,p\,v_3''(q)+\frac{4}{3}\,\tau\,p^2\,v_3'''(q)=f_2(-p,p-q)-f_2(p,q)  
\end{equation}
for all~$(p,q)$ in a neighborhood of the origin. In particular on
the \hbox{$q$-axis} relation~(\ref{v3g2}) implies~$\eh_3(0)=0$.
We have then shown that the unfolding of a transitional point
has the form
\begin{equation}\label{nofotrans2}
        H(p,q;t;\ve)=\frac{1}{2\tau}\,p^2+\ve\, q^2+\ve\tilde\eh_3(\ve)q^3
                     +\sum_{k=4}^n\eh_k(\ve)\,q^k\ +\ O_{p,q}(n+1),
\end{equation}
were~\intext{$\dst\tilde\eh_3(\ve)\DEF\,\eh_3(\ve)/\ve$} is a smooth real
function.

\subsection{Normal form for an extremal fixed point}
In this case condition~(\ref{implicit:M}) is not fulfilled. Generically
we cannot guaranty that we can follow smoothly the fixed point
as soon as~$\ve\neq0$ and then we cannot even make use of the
arguments of subsection~\ref{subsec:elimtime} to remove smoothly
the time-dependence of the quadratic part of $H(p,q';t;\ve)$
uniformly with respect to~$\ve$ in a neighborhood of zero.
We must therefore reconsider our approach from its very beginning.

\subsubsection{Uniform elimination of the time-dependence in the 
quadratic part of the unfolding}

The most general form for a one-parameter linearized \mbox{$\tau$-periodic}
differential equation is
\begin{equation}\label{leqmo2}
        \dot{w}(t;\ve)=\Lambdabbld(t;\ve)\,w(t;\ve)+a(t;\ve)
\end{equation}
where~$\Lambdabbld(t;\ve)$ (resp.~$a(t;\ve)$ ) is a real one-parameter family
of \mbox{$\tau$-periodic} $2\times2$~matrices (resp. \mbox{$2$-vectors}).
If~${\ensmf U}(t;\ve)$ is the operator defined by equations~(\ref{diffeqU})
and~(\ref{chroprod}), then the extremal case
corresponds precisely
to~${\ensmf U}(t;\ve\mathop{=}0)=\Mbbld(\ve\mathop{=}0)=
                                \bigl(\begin{smallmatrix}
                                1&\phantom{-}0\\1&\phantom{-}1
                                \end{smallmatrix}\bigr)$.
From now on and until the end of this subsection we should
keep in mind that all the quantities
we are dealing with depend smoothly on~$\ve$ but
 for simplicity we will not write down explicitly this dependence.
 Let~$\Lbbld$ (resp.~$b$)
be any smooth family of time-independent in\-fi\-ni\-te\-si\-mal
symplectic matrices
(resp. \mbox{$2$-vectors}). In the new symplectic coordinate system defined 
by~$\widetilde{w}(t)=\EXP{t\,\Lbbld}\bigl[{\ensmf U}(t)\bigr]^{-1}\,w(t)+b(t)$,
equation~(\ref{leqmo2}) becomes 
\begin{equation}\label{leqmo3}
        \dot{\widetilde{w}}(t)
        =\Lbbld\,\widetilde{w}(t)+\dot b-\Lbbld\,b
	+\underbrace{\EXP{t\,\Lbbld}\bigl[{\ensmf
	U}(t)\bigr]^{-1}\,a(t)}_{\dst \DEF\ \breve a} \ .
\end{equation}
There will be no explicit time-dependence if
\begin{equation}\label{tieq}
	\dot b-\Lbbld\,b+\breve a(t)=c
\end{equation}
where
now~$c$ is a (family of) constant \mbox{$2$-vector(s)}. As we have seen
in subsection~\ref{subsec:elimtime}, the change of variables has to be real
and \mbox{$\tau$-periodic}. Actually it is an extremal case if we
choose a real \mbox{$\tau$-periodic}~$b$ 
and~$\Lbbld=\frac{1}{\tau}\bigl(\begin{smallmatrix}
                                0&\phantom{-}0\\ 1&\phantom{-}0
                                \end{smallmatrix}
                             \bigr)$.
Let us then expand~$a(t)$ and~$b(t)$ in Fourier series
\begin{equation}
        \breve{a}(t)=\sum_{n\,\in\,\ZZ}\breve{a}_n\,\EXP{\imat n\frac{2\pi}{\tau}t}
        \quad\mathrm{and}\quad
        b(t)=\sum_{n\,\in\,\ZZ}b_n\,\EXP{\imat n\frac{2\pi}{\tau}t}
        \quad\mathrm{where}\quad\forall\,n\in\ZZ,\ (\breve{a}_n,b_n)\in\CC^2 \ .
\end{equation}
Then equation~(\ref{tieq}) leads to
\begin{equation}\label{eqmat3}
        \forall\,n\in\ZZ,\quad
        \left(\imat n\,\frac{2\pi}{\tau}\Id-\Lbbld\right)b_n
        =\delta^{\ssst (K)}_{n,0}\,c-\breve{a}_n
\end{equation}
($\delta^{\ssst (K)}$ is the Kronecker symbol).
Since the eigenvalues of~$\Lbbld$ are both small for~$\ve$ near zero,
the matrix equation~(\ref{eqmat3}) can be solved for every~$n\in\ZZ\nonnull$.
It only remains~$\Lbbld\, b_0=\breve{a}_0-c$ which can also be solved provided
that~$\breve{a}_0-c$ belongs to~$\im\Lbbld=\RR\textvector{1}{0}+O_\ve(1)$.
If~$\breve{a}_0=\textvector{\alpha_1}{\alpha_2}$
we can therefore eliminate  the time dependence of~(\ref{leqmo2}) uniformly
in~$\ve$ by choosing~$c\DEFt\textvector{0}{\alpha_2}+O_\ve(1)$ 
and~$b_0\DEFt\Lbbld{-1}(\breve{a}_0-c)+O_\ve(1)$.
Moreover, since~$\Lbbld$ is real we have~$\forall\,n\in\ZZ,\;b_n^*=b_{-n}$ 
which ensures that~$b(t)$ is real. 

We have then shown that even in the one-parameter unfolding of an extremal
point, we can choose a suitable local symplectic 
coordinate system smooth in~$\ve$ in which the dynamics corresponds 
to a hamiltonian whose 
terms of degree one and two in~$(p,q)$ are constant.

\subsubsection{Normal form for the quadratic unfolding}
From what precedes, let us start from the most general form for a
time-independent hamiltonian  
\begin{equation}
        H(p,q;t;\ve)=\underbrace{
                             \ea (\ve)\,p  
                             +\eb (\ve)\,q
                             +\frac{1}{2}\,\ec (\ve)\,p^2
                             +\ed (\ve)\,pq
                             +\frac{1}{2}\,\ee (\ve)\,q^2
                            }_{\dst\DEF h^{(2)}(p,q;\ve)}
                        +\ O_{p,q}(3)
\end{equation}
with~$(\ea ,\eb ,\ec ,\ed ,\ee )$
being smooth functions of~$\ve$ such that 
\begin{equation}\label{abcd0e2}
        \ea (0)=0\;;\qquad\eb (0)=0\;;\qquad\ec (0)=\frac{1}{\tau}\;;
        \qquad\ed (0)=0\;;\qquad\ee (0)=0\;.
\end{equation}
 If~${\clgv I}$ is small enough we 
have~$\forall\,\ve\in {\clgv I},\ec >0$. Then,
if we make the smooth translation~\intext{$\dst p'=p+\frac{\ea }{\ec }$} and
if we relabel the coefficients of~$h^{(2)}$, we can write 
\begin{equation}
h^{(2)}(p,q;\ve)=\eb(\ve)\,q+\frac{1}{2}\,\ec (\ve)\,p^2
                        +\ed (\ve)\,pq
                        +\frac{1}{2}\,\ee (\ve)\,q^2\;.
\end{equation}
As was done in subsection~\ref{subsec:nofotrans}, the
symplectic change of variables corresponding to the generating 
function~(\ref{genfunctquadfam}) gives 
\begin{equation}\begin{split}
       h^{(2)}(p',q';\ve)=\frac{1}{\delta}\,\eb\,q'                 
                   +          
                                  \frac{1}{2}\left[
                                  \frac{\gamma^2}{\delta^2}
                                  (
                                 \ec \,\eta^2+2\ed \,\eta+\ee 
                                  )
                                  -2\gamma(\ed +\ec \,\eta)
                                  +\ec \,\delta^2     
                                  \right]\,p^{\prime\,2}
                   \\[1em]
                   +
                                 \left[
                                   -\frac{\gamma}{\delta^2}
                                   (
                                  \ec \,\eta^2+2\ed \,\eta+\ee 
                                   )
                                   +\ed 
                                   +\ec \,\eta
                                 \right]\,p'q'             
                   +           \frac{1}{2}
                                \left[
                                  \frac{1}{\delta^2}
                                  (
                                 \ec \,\eta^2+2\ed \,\eta+\ee 
                                  )        
                                \right]\,q^{\prime\,2}
                   +\,O_{p'\!,\,q'}(3) \ .
\end{split}\end{equation}
 Let
us make the same choice as in~(\ref{choicegde}).
Aside from Eq.~(\ref{abcd0e2}), generically we have no additional relations
between~$(\ea ,\eb ,\ec ,\ed ,\ee )$, in
particular~\smash{$\frac{d{\ssst \eb\sqrt{\tau\ec}}}
{d\ve}\eval{\ssst\ve=0}\neq0$}
and without loss of generality we can reparametrize
the unfolding to~\intext{$\ve'\DEFt\eb\sqrt{\tau\ec}$}.
Dropping all the primes, the normal 
form of the  quadratic terms of the unfolding of an extremal point is
\begin{equation}
 	h^{(2)}(p,q;\ve)=\frac{1}{2\tau}p^2+\ve\,q+\ve\eh_2(\ve)\,q^2 \ ,
\end{equation}
where~$\eh_2$ is a smooth generic real function of~$\ve$.
Getting the normal form for higher orders can be done
uniformly with respect to~$\ve$ exactly in the same way as described in 
subsection~\ref{subsec:nofotrans}. A
full normal form for the unfolding in the extremal case can then be written
\begin{equation}\label{nofoex1}
        H(p,q;t;\ve)=\frac{1}{2\tau}p^2+\ve\,q+\ve\tilde\eh_2(\ve)\,q^2
        +\sum_{k=3}^n\eh_k(\ve)q^k\ +\ O_{p,q}(n+1),
\end{equation}
where $\{\eh_k\}_{k\in\finiteset{3}{n}}$ are some generic real functions
of~$\ve$. Through the smooth 
translation \intext{$q'=q+\frac{\ve\tilde\eh_2(\ve)}{3\eh_3(\ve)}$}
we can moreover set~$\tilde\eh_2(\ve=0)=0$. 

\section{Geometrical structure of the trajectories}
\label{sec:geomprop}

We have now all the necessary elements to describe the unfolding of the
dynamics for each of the above mentioned hamiltonian
normal forms. In particular, staying in a neighborhood of the origin~{\frak o} 
we search for a characterization, as a function of~$\ve$, of the typical
phase-space scales of the classical structures involved in each bifurcation. We
also want to explicitly locate the real as well as the complex fixed points
involved. The following will be valid as long as~$\ve$ remains
small compared to all algebraic expressions that can be made from the
coefficients appearing in the normal form and having the same dimensions.

Recall that the dynamics generated by the normal forms coincides with the true
one if we  look at the trajectories stroboscopically with a period that
depends on  the nature of the equilibrium point at~$\ve=0$~(see
table~\ref{tab:quadham}). We must also keep in mind that, in the stable case, the
hamiltonian~(\ref{def:Hell}) describe the dynamics in a
 rotating frame~(see equations~(\ref{rotframe})).

\subsection{Unstable Point}
The unfolding of an unstable point is given by hamiltonian~(\ref{def:Hu}).
For each~$\ve$ in~${\clgv I}$, the origin~$\frak o$ is the only fixed point
and it remains unstable for~${\clgv I}$ small enough. 

\subsection{Extremal bifurcation}
\begin{figure}[!ht]
\begin{center}
\rule{.7cm}{0cm}\input{vextr.pstex_t}\vspace{-.5cm}	 
\caption{\sl\baselineskip=0.25in\label{fig:vextr}
         Potential $v_\mathrm{extr}(q)$.	 
       }\vspace{-.5cm}	 
\end{center}
\end{figure}
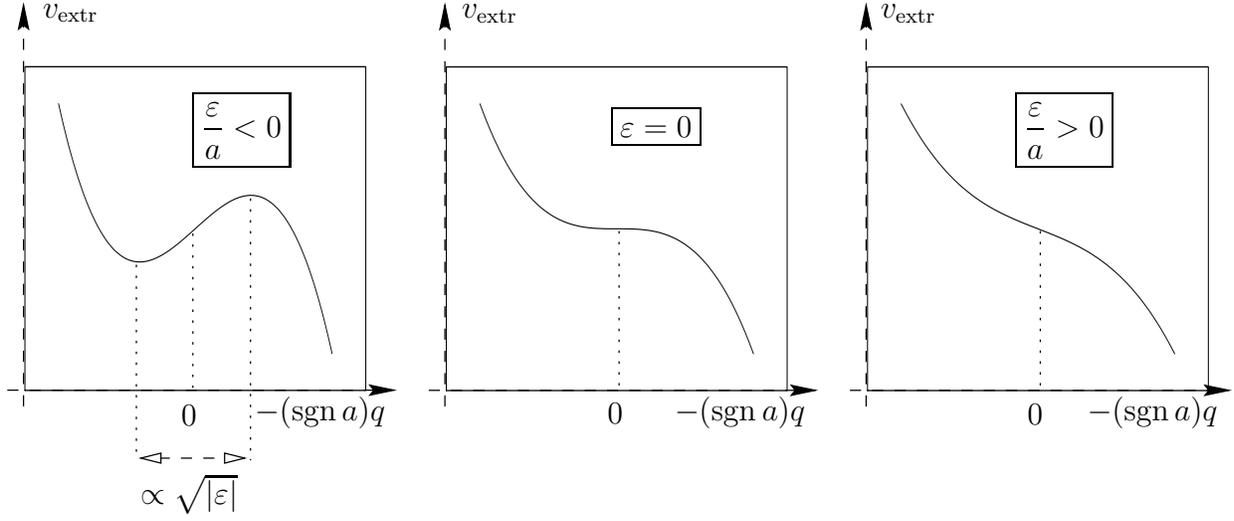

The normal form~(\ref{nofoex1}) can be rewritten  as follows (recall that
we can take~$\tilde\eh_2(\ve=0)=0$)
\begin{equation}\label{def:H1}
        H_1(p,q;t;\ve)\ =\ \frac{1}{2\tau}\,p^2\ +\ \ve\,q\ +\ \frac{1}{3}\,a\,q^3
                     \ +\ \frac{1}{4}\,b\,q^4\ +\ 
       O_{p,\,q,\,\sqrt{{\ssst|}\smash{\ve}{\ssst|}\vphantom{p}}}(5)
\end{equation}
where~$a,\ b$ are generic real numbers and~$\ve$ is such 
that~$|\ve|\ll| a^3/b^2|$.
In figure~\ref{fig:vextr} we have plotted the 
potential~$v_\mathrm{extr}(q)\DEFt\ve\,q\ +\ \frac{1}{3}\,a\,q^3$
whose behavior governs the dynamics when~$q\simeq0$. 
The fixed points of~$H_1(p,q;t;\ve)$ are given by~$p=0$~and~$q^2=-\ve/a $.

\begin{list}{$\rightarrowtriangle$}{}
\item$\boxed{\ve/a <0}$ There are two fixed points which are real 
\begin{equation}
        {\frak p}_\pm\ \DEF\ \left(\begin{array}{l}
                        \dst p=0
                        \\[1.5ex]
                        q=q_\pm\DEF
                        \pm\sqrt{\left|\ve/a \right|}
                        \end{array}
                 \right)\ +\ O_{\sqrt{{\ssst|}\smash{\ve}{\ssst|}}}(2)\;.
\end{equation}
The equations of motion are

\noindent
\begin{subequations}
        \hfill
         \raisebox{-0.08cm}[0cm][0cm]{
        \makebox[0cm][l]{\rule{4.8cm}{0cm}$\left\{\rule{0cm}{0.74cm}\right.$}}
        \parbox{15.4cm}{
         \begin{align}
                \dot p&=-\ve\ -\ a \,q^2\ +\ O_{p,\,q,\,\sqrt{{\ssst|}\smash{\ve}{\ssst|}\vphantom{p}}}(3) \;;
                \\[1ex]
                 \dot q&=\dst\frac{1}{\tau}\,p\ +\ O_{p,\,q,\,\sqrt{{\ssst|}\smash{\ve}{\ssst|}\vphantom{p}}}(3) \;.
         \end{align}}        
\end{subequations}

\noindent
Their linearization around~${\frak p}_\pm$ gives
\begin{equation}\label{leqmoex}
\biggl(\fatop{\dot u}{\dot v}\biggr)=\biggl(\fatop{\
0\quad-2aq_\pm}{1/\tau\hfill0\ \;}\biggr)\biggl(\fatop{u}{v}\biggr)
\end{equation}
Applying the results of subsection \ref{subsec:nofoquadham} we deduce
from the discussion of subsection 3.2 that~${\frak
p}_{\sgn{a}}$ is stable with a local angular speed given
by~$\omega=\sqrt{2/\tau}|\ve a |^{1/4}$. ${\frak p}_{\!-\sgn{a}}$~is an unstable
point whose stable and unstable manifolds have respectively the following
tangents:~$u=-\sqrt{2/\tau}|\ve a |^{1/4}v$ and~$u=\sqrt{2/\tau}|\ve a
|^{1/4}v$. The equation of the separatrix is 
\begin{equation}
       \ \frac{1}{2\tau}\,p^2\ +\ \ve\,q\ +\ \frac{1}{3}\,a\,q^3
        =H_1(0,q_{\!-\sgn{a }};t;\ve)
        =-\,\frac{2 }{3\sqrt{|a |}}\;|\ve|^{\frac{3}{2}} \ .
\end{equation}
It is the boundary of an area whose height and width vary with~$|\ve|$ 
as~$|\ve|^{1/2}$~and~$|\ve|^{3/4}$, 
respectively~(see figure~\ref{fig_1eposext}).
\vspace{-.5cm}
\begin{figure}[!ht]
\begin{center}
\input{fig_1eposext.pstex_t}
\caption{\sl\baselineskip=0.25in\label{fig_1eposext}
        Flow near the bifurcation of an extremal fixed point
        when~$\ve/a <0$.\vspace{-1cm}	 
       }
\end{center}
\end{figure}

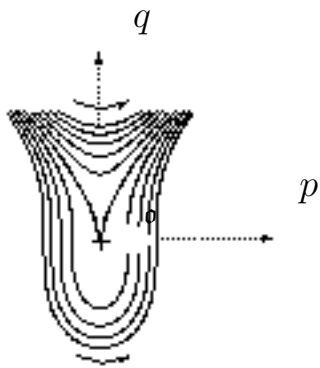
\begin{figure}[!ht]
\begin{center}
\input{fig_1enulext.pstex_t}
\caption{\sl\baselineskip=0.25in\label{fig_1enulext}
        Flow near the bifurcation of an extremal fixed point
        at~$\ve=0$.
       }
\end{center}
\end{figure}
\item$\boxed{\ve/a =0}$ By construction, we have one and only one
fixed point: the origin~{\frak o}. The two fixed points~${\frak p}_\pm$ which are
present when~$\ve/a \neq0$ now coalesce.
The linearized flow has been studied
in section \ref{subsec:nofoquadham} and looks like 
the marginal case in table~\ref{tab:quadham}.
If we take into account the first non-linear terms we get two families
of smooth trajectories separated by a curve
given by
\begin{equation}
        \frac{1}{2\tau}\,p^2+\frac{1}{3}\,a \,q^3=0\;,
\end{equation}
which has a vertical cusp at~{\frak o} where the stable and unstable
manifolds meet~(see figure~\ref{fig_1enulext}).

\item$\boxed{\ve/a >0}$
There are two fixed points which are now complex 
\begin{equation}
        {\frak p}_\pm\ \DEF\ \left(\begin{array}{l}
                        \dst p=0
                        \\[1.5ex]
                        q=\imat q_\pm=
                        \pm\imat\sqrt{\left|\ve/a \right|}
                        \end{array}
                 \right)\ +\ O_{\sqrt{{\ssst|}\smash{\ve}{\ssst|}}}(2)\;.
\end{equation}

In the Poincar{\'e} surface of section, the flow has no fixed point and the
linearized flow is given by 
equations~(\ref{leqmoex})~(see figure~\ref{fig_1enegext}).
\begin{figure}[!hb]
\begin{center}
\input{fig_1enegext.pstex_t}
\caption{\sl\baselineskip=0.25in\label{fig_1enegext}
        Flow near the bifurcation of an extremal fixed point
        when~$\ve/a>0$.
       }
\end{center}
\end{figure}

\end{list}

\subsection{Transitional point}
\begin{figure}[!ht]
\begin{center}
\rule{.7cm}{0cm}\input{vtran.pstex_t}
\caption{\sl\baselineskip=0.25in\label{fig:vtran}
         Potential $v_\mathrm{tran}(q)$.
       }
\end{center}
\end{figure}
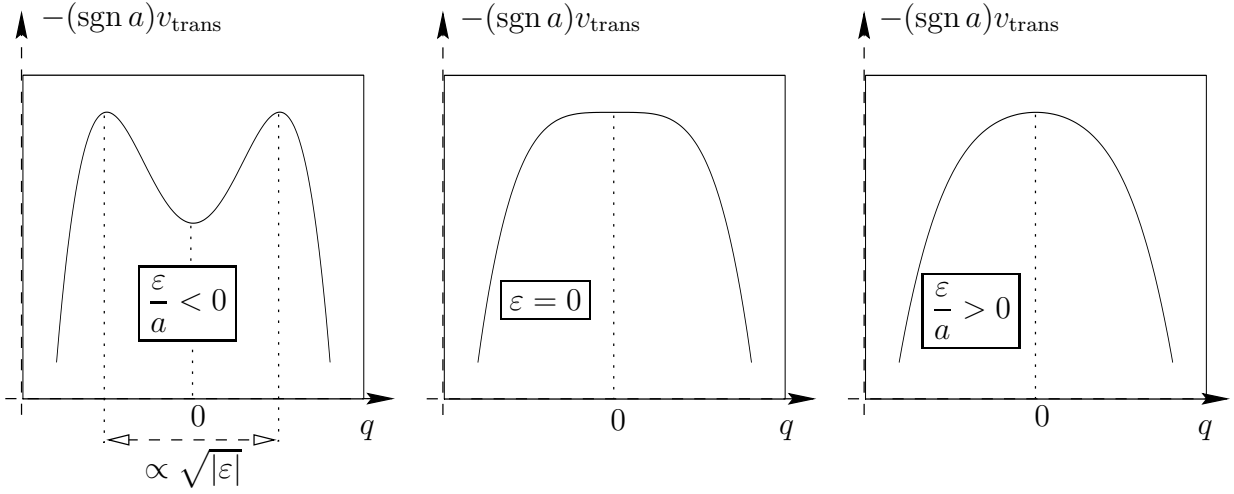

The normal form~(\ref{nofotrans2}) can be reordered as follows
\begin{equation}\label{def:H2}
        H_2(p,q;t;\ve)=
			\frac{1}{2\tau}\,p^2+\ve\,q^2+\frac{1}{4}\,a\,q^4
			+\frac{1}{3}\,\ve\,b\,q^3+\frac{1}{5}\,c\,q^5
			+\eh_4'(0)\ve\,q^4+\eh_6(0)q^6
        		+\ O_{\!\sqrt{\vphantom{\ssst|}|p|},\,q,
			      \,\sqrt{{\ssst|}\smash{\ve}{\ssst|}
			      	      \vphantom{p}}
			     }(7)
\end{equation}
where~$a\DEF4\eh_4(0)$, $b\DEF3\tilde\eh_3(0)=3\eh_3'(0)$,
and~$c\DEF5\eh_5(0)$. 
We are in a regime where~$|\ve|\ll| a/b^2|$ and~$|\ve|\ll| c/b^3|$.
In figure~\ref{fig:vtran} we have plotted the 
potential~$v_\mathrm{tran}(q)\DEFt\ve\,q^2+\frac{1}{4}\,a\,q^4$
whose behavior governs the dynamics.

The fixed points of~$H(p,q;t;\ve)$ are the origin~{\frak o} and the two
points defined by~$p\simeq0$, $q^2\simeq-2\ve/a $.

\begin{list}{$\rightarrowtriangle$}{}
\item$\boxed{\ve/a <0}$ Aside from the origin, there are two other real
fixed points 
\begin{equation}
        {\frak p}_\pm\ \DEF\ \left(\begin{array}{l}
                        \dst p=0
                        \\[1.5ex]
                        q=q_\pm\DEF
                        \pm\sqrt{2\left|\ve/a \right|}
                        \end{array}
                 \right)\
                        +\ O_{\sqrt{{\ssst|}\smash{\ve}{\ssst|}}}(2)\;.
\end{equation}

The equations of motion are

\noindent
\begin{subequations}\label{eqmotrans}
        \hfill
        \raisebox{-0.08cm}[0cm][0cm]
	{\makebox[0cm][l]{\rule{4.4cm}{0cm}$\left\{\rule{0cm}{0.74cm}\right.$}
	}
        \parbox{15.4cm}{
         		 \begin{align}
                		\dot p&
				 =-2\ve\,q-a\,q^3
				 +O_{\!\sqrt{\vphantom{\ssst|}p},
				     \,q,
				     \,\sqrt{{\ssst|}\smash{\ve}{\ssst|}
				     \vphantom{p}}
				    }(4)\;;
                		\\[1ex]
                 		\dot q&
				=\dst\frac{1}{\tau}\,p
				 +O_{\!\sqrt{\vphantom{\ssst|}|p|},
				     \,q,
				     \,\sqrt{{\ssst|}\smash{\ve}{\ssst|}
				     \vphantom{p}}
				     }(4)  \;.
         		 \end{align}
	 	        }        
\end{subequations} 

\noindent
Their linearization around~{\frak o} gives
\begin{equation}\label{leqmotranso}
\biggl(\fatop{\dot u}{\dot v}\biggr)=\biggl(\fatop{\
0\quad-2\ve}{1/\tau\hfill0\ \;}\biggr)\biggl(\fatop{u}{v}\biggr) \ .
\end{equation}
Therefore {\frak o}~is stable if~$a <0$ and the angular speed around it
in this case is~$\omega=2\sqrt{2|\ve|/\tau}$. If~$a >0$,
{\frak o}~is unstable and the stable and unstable manifolds have
tangents~$u=\mp\sqrt{2|\ve|/\tau}\,v$.
The linearization of equations~(\ref{eqmotrans}) about~${\frak p}_\pm$ gives
\begin{equation}
\biggl(\fatop{\dot u}{\dot v}\biggr)=\biggl(\fatop{\
0\quad4\ve}{1/\tau\hfill0\ \;}\biggr)\biggl(\fatop{u}{v}\biggr) \ .
\end{equation}
Then~${\frak p}_\pm$ have a stability opposite to~{\frak o}:
if~$a<0$, ${\frak p}_\pm$~are both unstable and in a chart centered on them,
the stable and unstable manifolds have 
tangents~$u=\pm2\sqrt{|\ve|\tau}\,v$.
If~$a>0$, ${\frak p}_\pm$~are both stable and the angular speed around them
is~$\omega=2\sqrt{|\ve|\tau}$.

The equation of the separatrix is
\begin{equation}
        \left\{
                \begin{array}{lr}
                        \dst
                        \frac{1}{2\tau}\,p^2+\ve\,q^2+\frac{1}{4}\,a \,q^4
                        + O_{\sqrt{{\ssst|}\smash{\ve}{\ssst|}}}(5)=H_2(0,0;t;\ve)=0
                        &\mathrm{if}\ a>0\\[3ex]
                        \dst
                        \frac{1}{2\tau}\,p^2+\ve\,q^2+\frac{1}{4}\,a \,q^4
                        + O_{\sqrt{{\ssst|}\smash{\ve}{\ssst|}}}(5)
                        =H_2(0,q_\pm;t;\ve)=\frac{1}{a}\,\ve^2
                        &\mathrm{if}\ a<0 \ .
                \end{array}
        \right.
\end{equation}
In both cases, the separatrix is the boundary of an area whose
height and width 
varies with~$|\ve|$ as~$|\ve|^{1/2}$~and~$|\ve|$,
respectively~(see figure~\ref{fig_2eposext}).

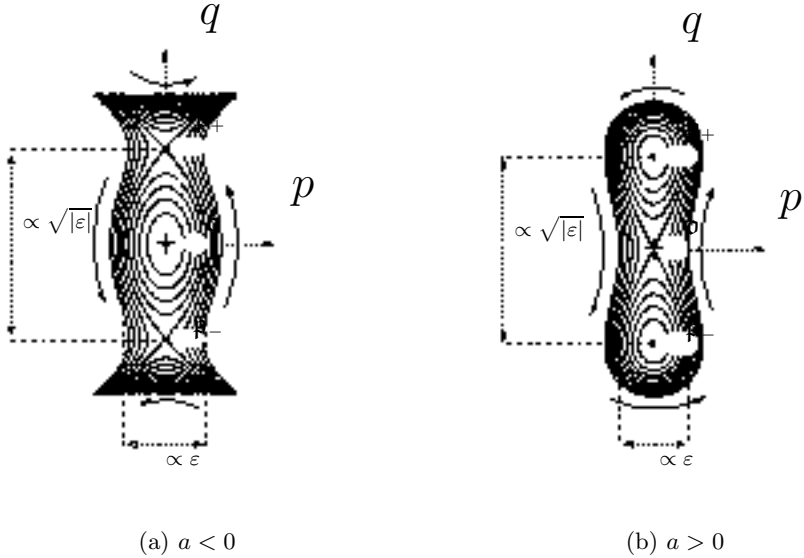
\begin{figure}[!ht]
\begin{center}
\vspace{-.5cm}
\mbox{\subfigure[$a<0$]{\input{fig_2eposext.pstex_t}}\quad
      \subfigure[$a>0$]{\input{fig_2eposbisext.pstex_t}}
     }
\vspace{-.5cm}
\caption{\sl\baselineskip=0.25in\label{fig_2eposext}
        Flow near the bifurcation of a transitional fixed point
        when~$\ve/a <0$.
       }
\end{center}
\end{figure}

\item$\boxed{\ve/a =0}$ By construction, the origin is the only
fixed point. The two fixed points~${\frak p}_\pm$ which are
present when~$\ve/a \neq0$ have merged at the origin.
The linearized flow has been studied
in section \ref{subsec:nofoquadham} and looks like the marginal case in
table~\ref{tab:quadham}.
If we take into account the non-linear terms we get families
of smooth trajectories separated by the 
curve~$\frac{1}{2\tau}p^2+\frac{1}{4}a q^4
+O_{\sqrt{{\ssst|}\smash{\ve}{\ssst|}}}(5)$ 
represented by
a couple of tangent parabolae if~$a <0$ and a single point if~$a >0$~(see figure~\ref{fig_2enulext}).

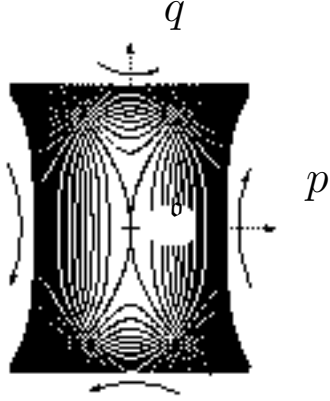
\begin{figure}[!ht]
\begin{center}
\vspace{-.5cm}
\input{fig_2enulext.pstex_t}
\vspace{-.5cm}
\caption{\sl\baselineskip=0.25in\label{fig_2enulext}
        Flow near the bifurcation of a transitional fixed point
        at~$\ve=0$.
       }
\vspace{-.5cm}       
\end{center}
\end{figure}

\item$\boxed{\ve/a >0}$ {\frak o} is the only real fixed point and
there are two fixed points in the complex plane 
\begin{equation}
        {\frak p}_\pm\ \DEF\ \left(\begin{array}{l}
                        \dst p=0
                        \\[1.5ex]
                        q=\imat q_\pm=
                        \pm\imat\sqrt{2\left|\ve/a \right|}
                        \end{array}
                 \right)\ +\ O_{\sqrt{{\ssst|}\smash{\ve}{\ssst|}}}(2)\;.
\end{equation}

The equations of motion and their linearization at~{\frak o} are
given by Eqs.~(\ref{eqmotrans}) and~(\ref{leqmotranso}), respectively.
The origin changes its stability when~$\ve$ crosses zero.
{\frak o}~is stable if~$a >0$ and the angular speed around it
 is~$\omega=\sqrt{2|\ve|/\tau}$. If~$a <0$,
{\frak o}~is unstable and the stable and unstable manifolds have
 tangents~$p=\mp\sqrt{2|\ve|/\tau}\,q$~(see figure~\ref{fig_2enegext}).
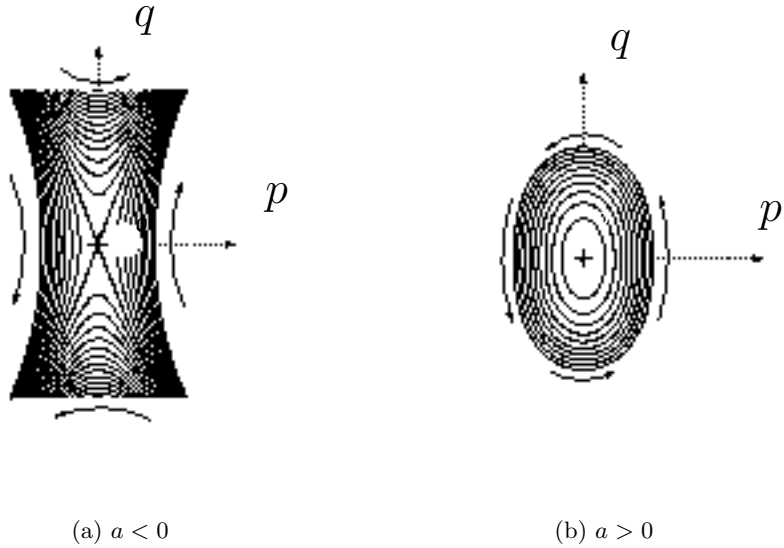
\begin{figure}[!ht]
\begin{center}
\vspace{-.5cm}         
\mbox{\subfigure[$a<0$]{\input{fig_2enegext.pstex_t}}\quad
      \subfigure[$a>0$]{\input{fig_2enegbisext.pstex_t}}
     }
 \vspace{-.5cm}            
\caption{\sl\baselineskip=0.25in\label{fig_2enegext}
        Flow near the bifurcation of a transitional fixed point
        when~$\ve/a <0$.
       }
 \vspace{-.5cm}         
\end{center}
\end{figure}

\end{list}

As we have said (see table~\ref{tab:quadham}), the~$-2\tau$-periodic
stroboscopic view of these flows
and the doubly iterated Poincar{\'e} map coincide. 
We will see, from the study of the
generating functions of next section, that in order to describe the
true dynamics after one-step iteration of the Poincar{\'e} map we must have a
global symmetry with respect to~{\frak o}. That means 
that~${\frak p}_\pm$ are non-longer fixed points when~$\ve\neq0$ but rather
belong to a periodic orbit of period~$2\tau$. In one period, they exchange
each other and all the points in their neighborhood jump
across~{\frak o}.

\subsection{Stable point with $\omega_0=2r\pi/(\ell\tau)$ and $\ell\geqslant3$}
\subsubsection{$\boldsymbol{\ell=3}$}

For~$\ell=3$, the hamiltonian~(\ref{def:Hell})  takes the form
\begin{subequations}\label{def:H3}
\begin{equation}\label{def:H3cart}
	H_3(p,q;t;\ve)\ =\ \frac{1}{2}\,\ve\,(p^2+q^2) 
                 	 \ +\ \frac{1}{3}\,b(p^3-3pq^2)\ 
			 +\ O_{p,\,q,\,\ve}(4)    
\end{equation}
and
\begin{equation}\label{def:H3aa}
	H_3(I,\theta;t;\ve)\ =\ \ve I\ 
			      +\ \frac{1}{3}\,b\,(2I)^{3/2}\cos(3\,\theta)
			      \ +\ O_{\sqrt{I},\ve}(4)
\end{equation}
\end{subequations}

\noindent
for some generic real number~$b$.
The fixed points of~(\ref{def:H3}) are real for all~$\ve$. Aside from the
origin we have 
\begin{equation} 
         {\frak p}_0\ \DEF\ \left(\begin{array}{l}
                         p=-\ve/b
                        \\[1.5ex]
                         q=0
                        \end{array}
                 \right)\ +\ O_\ve(2)
\qquad\mathrm{and}\qquad
        {\frak p}_\pm\ \DEF\ \left(\begin{array}{l}
                       p=\ve/(2b)
                        \\[1.5ex]
                       q_\pm\DEF\pm\sqrt{3}\,\ve/(2b)
                        \end{array}
                 \right)\ +\ O_\ve(2) \ .
\end{equation}
These points lie near the circle centered at~{\frak o} with 
radius~$|\ve/b|$. From expression~(\ref{def:H3aa}) we see that the
flow is -- to first order -- invariant by rotation of~$\frac{2\pi}{3}\ZZ$
around~{\frak o}. The equations of motion are

\noindent
\begin{subequations}\label{eqmo3}
        \hfill
         \raisebox{-0.08cm}[0cm][0cm]{
        \makebox[0cm][l]{\rule{5cm}{0cm}$\left\{\rule{0cm}{0.74cm}\right.$}}
        \parbox{16.4cm}{
         \begin{align}
                \hspace{0.2cm}\dot p&=-\ve\,q\ +\ 2b\,pq\ +\ O_{p,\,q,\,\ve}(3) \;;
                \\[2ex]
                 \dot q&=\ve\,p\ +\ b(p^2-q^2)\ +\ O_{p,\,q,\,\ve}(3)\;.
         \end{align}}        
\end{subequations} 

\begin{figure}[!ht]
\begin{center}
\mbox{\subfigure[$\ve\neq0$]{\input{fig_3eposext.pstex_t}}\hspace{3cm}
      \subfigure[$\ve=0$]{\input{fig_3enulext.pstex_t}}
     }
 \vspace{-.5cm}            
\caption{\sl\baselineskip=0.25in\label{fig_3}
        Flow near the bifurcation of a stable fixed point with~$\ell=3$
       }
 \vspace{-.5cm}         
\end{center}
\end{figure}
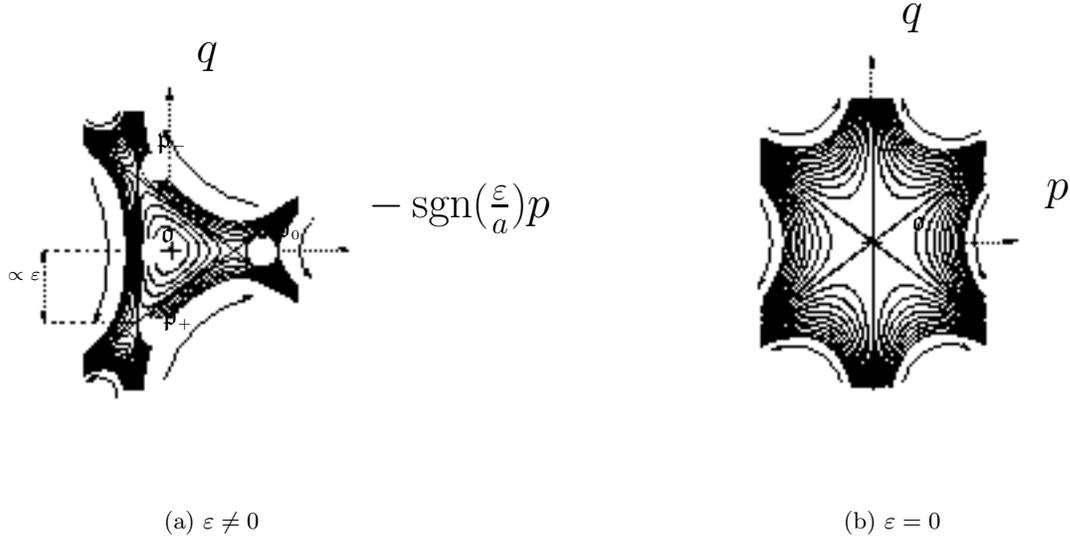

\noindent
By construction~(see 
subsection~\ref{subsec:nofounfoldus}) about~{\frak o} we have a
rotation of angular speed~$\ve$. 
The linearization of~(\ref{eqmo3}) around~${\frak p}_0$ gives 
\begin{equation}\label{leqmotrans}
\biggl(\fatop{\dot u}{\dot v}\biggr)=\biggl(\fatop{
\phantom{-}0\quad-3\ve}{-\ve\quad\phantom{-3}0}\biggr)\biggl(\fatop{u}{v}\biggr)
\end{equation} 
which shows that~${\frak p}_0$ and~${\frak p}_\pm$ are unstable.
The separatrix has the equation 
\begin{equation}
0=b\left[p-\frac{\ve}{2b}\right]\left[q-\frac{1}{\sqrt{3}}\left(p+\frac{\ve}{b}\right)\right]\left[q+\frac{1}{\sqrt{3}}\left(p+\frac{\ve}{b}\right)\right]
+O_{p,q,\ve}(4)\;.
\end{equation}
which, to first order, is made by the three straight lines defining the
equilateral 
triangle~$\ ({\frak p}_0,{\frak p}_+,{\frak p}_-)$~(see 
figure~\ref{fig_3}-a). At~$\ve = 0$, all three points
collapse at~{\frak o}~(see figure~\ref{fig_3}-b).

The flows are seen in a frame rotating at angular
speed~$\omega_0=2\pi r /(3\tau)$ (cf~(\ref{rotframe})).
Thus, in the fixed frame,~$\{{\frak p}_0,{\frak p}_+,{\frak p}_-\}$ are
no longer fixed points but
rotate with angular speed~$\omega_0$.
They belong to a periodic orbit of period~$3\tau$.

\subsubsection{$\boldsymbol{\ell=4}$}
\label{subsubsec:l4}

For~$\ell=4$, the hamiltonian~(\ref{def:Hell}) takes the form
\begin{subequations}\label{def:H4}
\begin{equation}\label{def:H4cart}
        H_4(p,q;t;\ve)\ =\ \frac{1}{2}\,\ve\,(p^2+q^2) 
                 \ +\ \frac{1}{4}\,a(p^4+q^4)\ +\
\frac{1}{2}\,b\,p^2q^2\ +\ O_{p,\,q,\sqrt{{\ssst|}\smash{\ve}{\ssst|}}}(5)
\end{equation}
and
\begin{equation}\label{def:H4aa}
        H_4(I,\theta;t;\ve)\ =\ \ve I\ +\ \frac{1}{4}(3a+b)I^2\ +\
			       \frac{1}{4}(a-b)I^2\cos(4\,\theta)\ 
			       +\ O_{\!\sqrt{I},
			       	       \,\raisebox{0.5mm}[0cm][0cm]
				       {$\sst\sqrt{{\ssst|}\smash{\ve}{\ssst|}
				         \vphantom{p}}
				       $}
				     }(5)\;.
\end{equation}
\end{subequations}

\noindent
for some generic real numbers~$a$ and~$b$. Because we can change the
global sign of~$H$ by inverting the arrow of time and the sign
of~$\ve$, 
we can consider~$a>0$ without loss of generality. Moreover, by
possibly make a global rotation of~$\pi/4$, we can assume~$a-b>0$.
From expression~(\ref{def:H4aa}), we see that the
flow is -- to first order -- invariant by rotation of~$\frac{\pi}{2}\ZZ$
around~{\frak o}. The fixed points are solutions of the algebraic system 

\noindent
\begin{subequations}
        \hfill
         \raisebox{-0.08cm}[0cm][0cm]{
        \makebox[0cm][l]{\rule{4.7cm}{0cm}$\left\{\rule{0cm}{0.74cm}\right.$}}
        \parbox{16.4cm}{
         \begin{align}
                \ve\,p+a\,p^3+b\,pq^2+\ O_{p,\,q,\sqrt{{\ssst|}\smash{\ve}{\ssst|}}}(4)&=0\;;
                \\[2ex]
                 \ve\,q+a\,p^3+b\,p^2q+\
O_{p,\,q,\sqrt{{\ssst|}\smash{\ve}{\ssst|}}}(4)&=0
         \end{align}}        
\end{subequations} 

\noindent
or, in action-angle variables

\noindent
\begin{subequations}
        \hfill
         \raisebox{-0.08cm}[0cm][0cm]{
        \makebox[0cm][l]{\rule{3.7cm}{0cm}$\left\{\rule{0cm}{0.74cm}\right.$}}
        \parbox{16.4cm}{
         \begin{align}&
                \sin(4\theta)+O_{\!\sqrt{I},\,\raisebox{0.5mm}[0cm][0cm]{$\sst\sqrt{{\ssst|}\smash{\ve}{\ssst|}\vphantom{p}}$}}(3)=0\;;
                \\[1ex]
                 &\ve+\frac{1}{2}\bigl[3a+b+(a-b)\cos(4\theta)\bigr]I+O_{\!\sqrt{I},\,\raisebox{0.5mm}[0cm][0cm]{$\sst\sqrt{{\ssst|}\smash{\ve}{\ssst|}\vphantom{p}}$}}(3)=0\;.
         \end{align}}        
\end{subequations} 

Since generically~$a\neq b$, we must distinguish the two cases~$a+b\gtrless 0$.

\paragraph{$\boldsymbol{\alpha)\ a+b>0}$}
\begin{list}{$\rightarrowtriangle$}{}
\item$\boxed{\ve/a <0}$ There are nine fixed points which are all real. Aside
from the center we have

\begin{equation} 
         {\frak p}^\pm\DEF\left(\begin{array}{l}
                         p=\pm\sqrt{-\ve/a}
                        \\[1.5ex]
                         q=0
                        \end{array}
                 \right)\!,\ 
{\frak p}^\pm_\pm\DEF\left(\begin{array}{l}
                         p=\pm\sqrt{-\ve/(a+b)}
                        \\[1.5ex]
                         q=\pm\sqrt{-\ve/(a+b)}
                        \end{array}
                 \right) 
\mathrm{and}\ 
        {\frak p}_\pm\DEF\left(\begin{array}{l}
                       p=0
                        \\[1.5ex]
                       q=\pm\sqrt{-\ve/a}
                        \end{array}
                 \right)\;,
\end{equation}
modulo~$\ O_{p,\,q,\sqrt{{\ssst|}\smash{\ve}{\ssst|}}}(2)$.

\begin{figure}[!ht]
\begin{center}
\vspace{-.5cm}
\input{fig_4eposext.pstex_t}
\vspace{-.5cm}
\caption{\sl\baselineskip=0.25in\label{fig_4eposext}
        Flow near the bifurcation of a stable fixed point of~$\ell=4$
        when~$a+b>0$ and~$\ve/a<0$.
       }
\end{center}
\end{figure}

Linearizing the equations of motion we find that the four 
points~$\{{\frak p}^\pm_\pm\}$ are stable
while~$\{{\frak p}_\pm,{\frak p}^\pm,\}$ are unstable.
The separatrix between them is given by
\begin{equation}
	\left[a(p^2+q^2)-\sqrt{2a(a-b)}\,pq+\ve\right]
	\left[a(p^2+q^2)-\sqrt{2a(a-b)}\,pq+\ve\right]
	+O_{p,\,q,\sqrt{{\ssst|}\smash{\ve}{\ssst|}}}(3)=0\;.
\end{equation}
 To first order the separatrix is made of two ellipses whose larger
axes are oriented along~$\theta=\pm\pi/4$.
Their intersection defines four satellite islands of typical width proportional 
to~$\sqrt{|\ve|}$~(see figure~\ref{fig_4eposext}).

\item$\boxed{\ve=0}$ {\frak o} is the only fixed point and it is
stable with a small positive angular rotation around it. The flow looks like
the stable one in table~\ref{tab:quadham}.
\item$\boxed{\ve/a >0}$ {\frak o} is the only real fixed point and it is
stable with  angular rotation~$\ve$ around it. There are also  eight fixed 
points with  complex coordinates 
\begin{equation} 
         {\frak p}^\pm\DEF\left(\begin{array}{l}
                         p=\pm\imat\sqrt{\ve/a}
                        \\[1.5ex]
                         q=0
                        \end{array}
                 \right)\!,\ 
{\frak p}^\pm_\pm\DEF\left(\begin{array}{l}
                         p=\pm\imat\sqrt{\ve/(a+b)}
                        \\[1.5ex]
                         q=\pm\imat\sqrt{\ve/(a+b)}
                        \end{array}
                 \right) 
\mathrm{and}\ 
        {\frak p}_\pm\DEF\left(\begin{array}{l}
                       p=0
                        \\[1.5ex]
                       q=\pm\imat\sqrt{\ve/a}
                        \end{array}
                 \right)\;,
\end{equation}
modulo~$\ O_{p,\,q,\sqrt{{\ssst|}\smash{\ve}{\ssst|}}}(2)$.
 The flow looks like the stable one in table~\ref{tab:quadham}.
\end{list}


\paragraph{$\boldsymbol{\beta)\ a+b<0}$} 

\begin{list}{$\rightarrowtriangle$}{}
\item$\boxed{\ve/a <0}$

\begin{figure}[!ht]
\begin{center}
\input{fig_4eposbisext.pstex_t}
\caption{\sl\baselineskip=0.25in\label{fig_4eposbisext}
        Flow near the bifurcation of a stable fixed point
        when~$a+b<0$ and~$\ve/a <0$.
       }
\end{center}
\end{figure}


There are five real fixed points 

\begin{equation} 
        {\frak o},\
         {\frak p}^\pm\DEF\left(\begin{array}{l}
                         p=\pm\sqrt{-\ve/a}
                        \\[1.5ex]
                         q=0
                        \end{array}
                 \right)\!,\ 
\mathrm{and}\ 
        {\frak p}_\pm\DEF\left(\begin{array}{l}
                       p=0
                        \\[1.5ex]
                       q=\pm\sqrt{-\ve/a}
                        \end{array}
                 \right)\;,
\end{equation}
modulo~$\ O_{p,\,q,\sqrt{{\ssst|}\smash{\ve}{\ssst|}}}(2)$,
and four fixed points with complex coordinates

\begin{equation} 
        {\frak p}^\pm_\pm\DEF\left(\begin{array}{l}
                         p=\pm\imat\sqrt{-\ve/(a+b)}
                        \\[1.5ex]
                         q=\pm\imat\sqrt{-\ve/(a+b)}
                        \end{array}
                 \right) 
\;,
\end{equation}
modulo~$\ O_{p,\,q,\sqrt{{\ssst|}\smash{\ve}{\ssst|}}}(2)$.

Linearizing the equations of motion we see that the four 
points~$\{{\frak p}_\pm,{\frak p}^\pm,\}$ are all unstable.
The separatrix joining them is given by 
\begin{equation}\label{2hyperbolae}
\left[a(p^2+q^2)-\sqrt{2a(a-b)}\,pq-\ve\right]\left[a(p^2+q^2)+\sqrt{2a(a-b)}\,pq-\ve\right]+O_{p,\,q,\sqrt{{\ssst|}\smash{\ve}{\ssst|}}}(3)=0\;.
\end{equation}
To first order, it is locally made of two hyperbolae of principal
axes~$\theta=\pm\pi/4$~(see figure~\ref{fig_4eposbisext}).

\item$\boxed{\ve=0}$ The nine fixed points have collapsed into a single
point {\frak o} which is
unstable~(see figure~\ref{fig_4enulbisext}).

\begin{figure}[!ht]
\begin{center}
\vspace{-1cm}
\input{fig_4enulbisext.pstex_t}
\vspace{-1cm}
\caption{\sl\baselineskip=0.25in\label{fig_4enulbisext}
        Flow near the bifurcation of a stable fixed point of~$\ell=4$
        when~$a+b<0$ and~$\ve=0$.
       }
\end{center}
\end{figure}

\item$\boxed{\ve/a >0}$ 

In addition to the stable fixed point at the
origin, there are
four real fixed points
\begin{equation}        
        {\frak p}^\pm_\pm\DEF\left(\begin{array}{l}
                       p=\pm\sqrt{\ve/(a+b)}
                        \\[1.5ex]
                       q=\pm\sqrt{\ve/(a+b)}
                        \end{array}
                 \right)\;,
\end{equation}
modulo~$\ O_{p,\,q,\sqrt{{\ssst|}\smash{\ve}{\ssst|}}}(2)$. They are
all unstable. 
To first order, the separatrix between them is given by two hyperbolae
like in Eq.~(\ref{2hyperbolae}) but now the
 principal axes are oriented along~$\theta=0$ and~$\theta=\pi/2$.
Besides them, there are four fixed complex points

\begin{equation} 
        {\frak o},\
         {\frak p}^\pm\DEF\left(\begin{array}{l}
                         p=\pm\imat\sqrt{-\ve/a}
                        \\[1.5ex]
                         q=0
                        \end{array}
                 \right)\ 
\mathrm{and}\ 
        {\frak p}_\pm\DEF\left(\begin{array}{l}
                       p=0
                        \\[1.5ex]
                       q=\pm\imat\sqrt{-\ve/a}
                        \end{array}
                 \right)\;.
\end{equation}
See figure~\ref{fig_4enegbisext}.

\begin{figure}[!hb]
\begin{center}
\vspace{-1cm}
\input{fig_4enegbisext.pstex_t}
\vspace{-.5cm}
\caption{\sl\baselineskip=0.25in\label{fig_4enegbisext}
        Flow near the bifurcation of a stable fixed point
        when~$-b>a$ and~$\ve/a >0$.
       }
\end{center}
\end{figure}

\end{list}

In all cases described in~$\boldsymbol{\alpha)}$
and~$\boldsymbol{\beta)}$ the reference frame rotates at angular
speed~$\omega_0=\pi r/(2\tau)$ (cf Eq.~\ref{rotframe}).
In the fixed frame, the points~${\frak p}$ rotate uniformly
around~{\frak o} with speed~$\omega_0$. $\{{\frak p}^\pm_\pm\}$~belong to
the same periodic orbit of period~$4\tau$.

\subsubsection{$\boldsymbol{\ell}\pmb{\geqslant}\boldsymbol{5}$}
 
\begin{figure}[!hb]
\begin{center}
\input{fig_5eposext.pstex_t}
\caption{\sl\baselineskip=0.25in\label{fig_5eposext}
        Flow near the bifurcation of a stable fixed point for~$\ell=5$
        when~$\ve/a_2<0$.
       }
\end{center}
\end{figure}

We will work in action-angle variables with the normal 
form~(\ref{def:Hellaa}). The origin remains stable across the bifurcation
but the sense of rotation of the flow around it is inverted. The other
fixed points are solutions of

\noindent
\begin{subequations}
        \hfill
         \raisebox{0.2cm}[0cm][0cm]{
        \makebox[0cm][l]{\rule{.7cm}{0cm}$\left\{\rule{0cm}{1cm}\right.$}}
        \parbox{16.4cm}{
         \begin{align}
                &-\ell\eb_lI^{\ell/2}\sin(\ell\theta)\ +\
O_{\!\sqrt{I},\,\raisebox{0.5mm}[0cm][0cm]{$\sst\sqrt{{\ssst|}\smash{\ve}{\ssst|}\vphantom{p}}$}}(\ell+1)\
=\ 0\;;
                \\[2ex]
                &\ve\ +
                      \!\!\sum_{\fatops{k\,\in\,\NN}
                                       {2\,\leqslant\,k\,\leqslant\,\ell/2}
                               }
                            \!\!k\ea _k(\ve)\,I^{k-1}
                       \ +\ \frac{1}{2}\,\ell\ \eb _\ell(\ve)\,I^{\ell/2-1}
                            \cos(\ell\theta)\ +\
O_{\!\sqrt{I},\,\raisebox{0.5mm}[0cm][0cm]{$\sst\sqrt{{\ssst|}\smash{\ve}{\ssst|}\vphantom{p}}$}}(\ell-1)\
=\ 0\;.
         \end{align}}        
\end{subequations}

\noindent
To first order they correspond
to:~$\dst\theta_k\DEF k\frac{\pi}{\ell}$ for~$k\in\finiteset{0}{2\,\ell-1}$
and~$\dst I=-\frac{1}{2a_2}\,\ve+O_{\!\sqrt{|\ve|}}(3)$ 
where~$a_2\DEFt\ea_2(\ve\mathop{=}0)$.

\begin{list}{$\rightarrowtriangle$}{}
\item$\boxed{\ve/a_2 <0}$ In addition to the origin, there are~$2\ell$
real fixed points 
\begin{equation} 
         {\frak p}_k\ \DEF\ \left(\begin{array}{l}
                         I=-\ve/(2a_2)
                        \\[1.5ex]
                         \theta_k\DEF k\pi/\ell
                        \end{array}
                 \right)+O_{\!\sqrt{|\ve|}}(3)\qquad\mathrm{with}\qquad k\in\finiteset{0}{2\,\ell-1}\;.
\end{equation}
To first order, they lie on the circle with center~{\frak o} and 
radius~$\sqrt{-\ve/a_2}$. The linearization of the equations of motion
about~${\frak p}_k$ is  \begin{equation}
\biggl(\fatop{\dot I}{\dot \theta}\biggr)=\biggl(\fatop{
\phantom{-}0\qquad(-1)^k\ell^2b|\ve/2a_2|^{\ell/2}}{2a_2\hfill0\phantom{|\ve/2a_2|^{\ell/2}}}\biggr)\biggl(\fatop{I}{\theta}\biggr)
\end{equation} 
which shows that the points~$\{{\frak p}_k\}_{k\in\finiteset{0}{2\,\ell-1}}$ are
alternately stable and unstable. Without loss of generality, we can
assume that~${\frak p}_0$ is stable.
The separatrix has the following equation 
\begin{equation}\label{sepell}
\ve I+
                      \!\!\sum_{\fatops{k\,\in\,\NN}
                                       {2\,\leqslant\,k\,\leqslant\,\ell/2}
                               }
                            \!\!\ea _k(\ve)\,I^k
                       \ +\ \eb _\ell(\ve)\,I^{\ell/2}
                            \cos(\ell\theta)
                       =H_\ell({\frak p}_1;t;\ve)+\ O_{\!\sqrt{I},\,\ve^\star
}
(\ell+1)\;,   
\end{equation} 
where
\begin{equation}\label{sqrtstar}
	{\ve^\star}\ \DEF\ \begin{cases}
					   	\phantom{\sqrt{}}|\ve|&
					   	\qquad \mathrm{if}\ 
						\ell=3\;;
					   \\[1ex]
						\sqrt{|\ve|}&
						\qquad \mathrm{if}\ 
						\ell\neq3\;.
					   \end{cases}
\end{equation}   
It is a bounded set~$\frac{2\pi}{\ell}$-periodic in~$\theta$. To evaluate
the typical transversal size~$\delta r$ of the island around the stable
satellites we should  solve to first order equation~(\ref{sepell}) for~$\theta=0$
with $I=I({\frak p}_1)+\delta I$. We obtain~$\delta
I\propto|\ve|^{\ell/4}$ and deduce that
\begin{equation}\label{deltar}
\delta r\ \propto\ |\ve|^{\frac{\ell}{4}-\frac{1}{2}}\;.
\end{equation}
(see figure~\ref{fig_5eposext}).

\item$\boxed{\ve=0}$ All the fixed points have collapsed at the origin
around which the motion is stable with a positive rotating speed.
\item$\boxed{\ve/a_2 >0}$ {\frak o} is the only real fixed point. 
There are~$2\ell$ fixed points in the complex plane
\begin{equation} 
         {\frak p}_k\ \DEF\ \left(\begin{array}{l}
                         I=-\ve/a_2<0
                        \\[1.5ex]
                         \theta_k\DEF k\pi/\ell
                        \end{array}
                 \right)+O_{\!\sqrt{|\ve|}}(3)
                 \ =\ \left(\begin{array}{l}
                         p_k\DEF\imat\sqrt{\ve/a_2}\cos(k\pi/\ell)
                        \\[1.5ex]
                         q_k\DEF\imat\sqrt{\ve/a_2}\sin(k\pi/\ell)
                        \end{array}
                 \right)+O_{\!\sqrt{|\ve|}}(2)
\end{equation}
for~$k\in\finiteset{0}{2\,\ell-1}$.
\end{list}

If we go back from the rotating frame to the fixed one, we add a
global rotation of angular speed~$\omega_0=2\pi r/(\ell\tau)$. This
will make the points~{\frak p} belong to two different periodic orbits
of period~$\ell\tau$.

\subsection{Summary}\label{subsec:sumup}
\enlargethispage*{3\baselineskip}
Except the case $\ell=3$ where there are none, in all other cases there are
complex periodic orbits in the vicinity of the bifurcation. Each of them is
connected continuously through the bifurcation with a real one.
When~$\ell=4$ and~$a+b<0$, on each side of the bifurcation there is one real
and one complex satellite orbit around the central point. As the external
parameter goes through the bifurcation these two orbits interchange, the
real one becoming complex and vice versa. In all other cases, when they
exist, the complex orbits are present on one side of the bifurcation only.

The bifurcating (real and complex) orbits are typically separated in phase
space by a scale~$\ve^\star$ given by Eq.~(\ref{sqrtstar}). This quantity
will play a crucial role in the semiclassical considerations of the
following sections. In the particular case of stable bifurcating orbits, in
addition to $\ve^\star$ there exists another typical scale~\pvek\ which
gives the radial width of the stable satellite islands. For the extremal
(resp. transitional) bifurcation the width of the stable island(s) is given
by~$|\ve|^{3/4}$ (resp. $|\ve|$). In all cases, the classical flow looks
roughly the same in both sides of the bifurcation if one looks at it with a
resolution of order $\ve^\star$ or worst.

\section{Normal form for one-dimensional unfolding of generating functions}
\label{sec:nofogenfun}
So far, we have described the~\hbox{$\tau$-periodic} dynamics around an
 equilibrium  point
by simplifying the hamiltonian in its neighborhood. We have revisited the
classification of the generic fixed points by describing, in each case, the
normal form of a hamiltonian whose dynamics, viewed every~$-2\tau$, is the same
as the original one. The
one-freedom dynamics thus represents the local Poincar{\'e} reduction around
a~\hbox{$\tau$-periodic} orbit.
The associated local Poincar{\'e}
map~$\Phi(w;\ve)$ is the stroboscopic view of the reduced dynamics
taken every~$\tau$.
In order to  complete our study, in this section we deduce
the normal forms of the generating function of~$\Phi(w;\ve)$. In some cases, like in semiclassical
theory, the Poincar{\'e} map plays actually a more central role than the
reduced hamiltonian itself.

The starting point for calculating the Poincar{\'e} map from a
time-independent hamiltonian is the formula 
 \begin{eqnarray}
        \Phi(w;\ve)&=&\EXP{\tau\spoisson{\,\cdot\,,H(w;\ve)}}\,w\\
        &=& w+\tau\poisson{w,H(w;\ve)}
          +\sum_{m=2}^\infty\,\frac{\tau^m}{m!}\,
           \{\kern-1.15mm|\ldots\{\kern-1.15mm|w,
                  \underbrace{
                              H(w;\ve)|\kern-1.18mm\},\dots,H(w;\ve)
                              |\kern-1.18mm\}
                             }_{m\ \mathrm{times.}}\;,
                  \label{poissonserie1} 
 \end{eqnarray}
where we define the Poisson bracket of two smooth functions on phase 
space~$\clgv{P}$ by 
\begin{eqnarray}
        \poisson{A,B}\ &\DEF&\transpose{(\nabla_{\!w}A)}
                              \,\JJ\,
                              \nabla_{\!w}B\;;
                        \\[1em]
                       &\DEF&\ \frac{\partial A}{\partial q}
                         \frac{\partial B}{\partial p}
                         -
                         \frac{\partial B}{\partial q}
                         \frac{\partial A}{\partial p}\;.
\end{eqnarray}

Let~$w_0\in\clgv{P}$ and construct the backward and forward
sequence~$\forall n\in\ZZ;\ w_n\DEF\Phi^{(n)}(w_0,\ve)$.
 Some relevant functions containing the same information
as~$\Phi^{(n)}(\kern.2ex\cdot\,;\ve)$ are
\begin{itemize}
\item The $qq$-generating 
      function       
      \begin{equation}\label{def:N}
               F^{(n)}(w_n,w_0)\DEF
               \int_0^{n\tau}\Bigl(
                                   p(t)\dot{q}(t)-H\bigl[p(t),q(t);t;\ve\bigr]
                             \Bigr)\,dt\;.
      \end{equation}
      $w(t)\DEF\textvector{p(t)}{q(t)}$ is the unique solution
      of~(\ref{eqmo}) such that~$w(0)=w_0$ and~$w(n\tau)=w_n$.
      Locally $F^{(n)}$ depends only on two independent variables among
      ~$(p_n,q_n,p_0,q_0)$. These are often taken to be~$q_n$ and~$q_0$. 
      The {\em values} of this function
      are canonically invariant,
      they do not depend on the choice of the symplectic coordinate
      chart -- possibly time dependent -- in which the integral in calculated.
      When~$H$ is constant we get 
      \begin{equation}\label{Nu1}
       F^{(n)}(w_n,w_0)=
                -n\tau\underbrace{H(w_0;\ve)}_{=H(w_n;\ve)}
                +\int_0^{n\tau} p\,\poisson{q,H}\,dt\;.
      \end{equation}
\item The mixed generating 
       function
      \begin{equation}\label{def:S}
           S^{(n)}(p_n,q_0;\ve)\DEF
           p_nq_n(p_n,q_0;\ve)-F^{(n)}(p_n,q_0;\ve\bigr) \ .
      \end{equation}
\end{itemize}

In all cases but the unstable one, we will not be able to complete 
the summation~(\ref{poissonserie1}) exactly. Nevertheless following
our general philosophy we will work at a finite precision. From
section~\ref{sec:geomprop} we have determined for each bifurcation
the typical scale in~$\ve$ entering the dynamics.
This scale is fixed by the order of truncation of the normal forms
of the hamiltonian: the more terms are kept, the finer details can be
described. Accordingly, we will now show that all the dynamical structures up
to a fixed
scale are actually contained in the first terms of the
expansion~(\ref{poissonserie1}). This allows to compute the generating functions
for each type of bifurcation. Let us start with the unstable case which is
special, as we said, in the sense that the sum can be computed explicitly to all
orders. The physical reason for this is that in a generic unfolding of an
unstable point
 there is no bifurcation and hence no local typical scale.

\subsection{Unstable case}

Since no structural change is induced when~$\ve$ varies, we will leave the
\mbox{$\ve$-dependence} implicit up to the end of this subsection.
The one-parameter unfolding is given by~(\ref{def:Hu}) 
\begin{equation}\label{Hu}
    H_u(p,q;t)= \underbrace{\lambda\,pq 
               +\sum_{k=2}^n\,\eh_k\,(pq)^k}_{\DEF K_n(pq)}
               \ +\ O_{p,q}(2n+1)
\end{equation}
We recall (see subsection~\ref{subsec:nofoquadham}) that the true dynamics is
recovered by taking a stroboscopic
view every~$|\varsigma\tau|$, where~$\varsigma\mathop{=}1$ for the
hyperbolic case
and~$\varsigma\mathop{=}\pm2$ for the inverse hyperbolic case. (In the
following discussion we take, for the inverse hyperbolic case, the plus sign).
Since~$d(pq)/dt=\poisson{pq,H_u(p,q;t)}\in O_{p,q}(2n+1)$

 $pq$~is  a constant of motion 
modulo~$O_{p,q}(2n+1)$ and the solutions of the equations of motion are simply

\begin{subequations}\label{eqmosolunstable}
        \hfill
        \raisebox{-0.1cm}[0cm][0cm]{\makebox[0cm][l]{\rule{3.7cm}{0cm}$\left\{\rule{0cm}{.8cm}\right.$}}\parbox{14cm}{
         \begin{align}
                p(t)&=p_0\,\EXP{-tK_n'(p_0q_0)}+O_{p_0,q_0}(2n)\;;\\[1ex]
                 q(t)&=q_0\,\EXP{+tK_n'(p_0q_0)}+O_{p_0,q_0}(2n)\;.  
         \end{align}}        
\end{subequations}

\noindent
From~(\ref{Nu1}) we get
\begin{equation}
          F_u^{(\varsigma)}(w_1,w_0)
           =\varsigma\tau\bigl[p_0q_0\,K_n'(p_0q_0)-K_n(p_0q_0)\bigr]
           +O_{p_0,q_0}(2n+1)\;.
\end{equation}
The mixed generating function is obtained from the definition~(\ref{def:S}) after
inverting equations~(\ref{eqmosolunstable}) up to first order 
 \begin{align}
         S_u^{(\varsigma)}(p_\varsigma,q_0)
                           &=p_0q_0
                                  -\tau\bigl[K_n(p_0q_0)-p_0q_0\,K_n'(p_0q_0)
                                       \bigr]
                           \\[1em]
                           &=\EXP{\varsigma\tau\lambda\,}p_\varsigma q_0
                             +\varsigma\tau\eh_2\;
                              \EXP{2\varsigma\tau\lambda}\,(p_\varsigma q_0)^2
                            +\bigl[
                                   \varsigma\tau\eh_3+2(\varsigma\tau\eh_2)^2
                             \bigr]
                             \EXP{3\varsigma\tau\lambda}\,(p_\varsigma q_0)^3
                            +O_{p_\varsigma,q_0}(8).
\end{align}
In the inverse hyperbolic case the hamiltonian~(\ref{Hu}) 
cannot alone give all the information about the dynamics at
time~$\tau$.
In particular, $w(\tau)$ given by~(\ref{eqmosolunstable}) does not contain one
transformation whose linear part is a symmetry with respect
to the origin. If we instead propose as a guess the following one-step generating
function  
\begin{equation}
   S_{u}^{(1)}(p_1,q_0)=\sigma_1p_1q_0+\sigma_2(p_1q_0)^2+\sigma_3(p_1q_0)^3
                        +O_{p_1,q_0}(8)
\end{equation}
then the real coefficients~$\sigma_1$, $\sigma_2$~and~$\sigma_3$ can
be determined in the following way 
\begin{enumerate}\stepcounter{enumi}
       \item[\arabic{enumi}-] Invert equations~(\ref{sympmap0})
                                 modulo~$O_{p_1,q_1}(7)$ to find 
				 out~$w_1(w_0)$;\stepcounter{enumi}
       \item[\arabic{enumi}-] The monodromy matrix 
             being~$\Mbbld=\Bigl(\begin{smallmatrix}
                                   -\EXP{-\tau\lambda}  & 0 \\
                                   \!0    & -\EXP{\tau\lambda}   
                                 \end{smallmatrix}
                           \biggr)$,
             deduce that $\sigma_1=-\EXP{\tau\lambda}\;$;\stepcounter{enumi}
       \item[\arabic{enumi}-] Iterate the mapping in order to get~$w_2$ 
               modulo~$O_{p_1,q_1}(7)$
             and identify the expression with~$w(2\tau)$ obtained via 
             equations~(\ref{eqmosolunstable}); \stepcounter{enumi}
       \item[\arabic{enumi}-] Eventually find out~$\sigma_2$~and~$\sigma_3$. 
\end{enumerate}
For the inverse hyperbolic case, we thus obtain 
\begin{equation}
        S_u^{(1)}(p_1,q_0)=-
           \EXP{\tau\lambda\,}p_1 q_0
                             -\tau\eh_2\,
                              \EXP{2\tau\lambda}\,(p_1 q_0)^2
                            +\bigl[
                                   \tau\eh_3+\frac{1}{2}(\tau\eh_2)^2
                             \bigr]\,
                             \EXP{3\tau\lambda}\,(p_1 q_0)^3
                            +O_{p_1,q_0}(8)\;.
\end{equation}

Note that, in any case, the mixed generating functions have the same
functional form as the hamiltonian but the relations between their
respective coefficients are non-trivial.

\subsection{Extremal case}

From hamiltonian~(\ref{def:H1}) the typical scales
are~$p\propto \ve^{\frac{3}{4}}$ and~$q\propto\sqrt{|\ve|}$.
We deduce the following Poisson brackets
\begin{align}
       &\poisson{q,H_1}\ =\ \frac{1}{\tau}\,p\ +\ O_{p,\,q,\,\sqrt{|\ve|}}(3)\;;
       \\[1ex]
       &\poisson{p,H_1}\ =\ -\ve- a\,q^2\ +\ O_{p,\,q,\,\sqrt{|\ve|}}(3)\;;
       \\[1ex]
       &\poisson{\poisson{p,H_1},H_1}\ =\ -\frac{2 a}{\tau}\,pq
                                   \ +\ O_{p,\,q,\,\sqrt{|\ve|}}(3)\;;
       \\[1ex]
       &\,\{\kern-1.15mm|\ldots\{\kern-1.15mm|p,
                  \underbrace{
                              H_1|\kern-1.15mm\},\dots,H_1
                              |\kern-1.15mm\}
                             }_{n\ \mathrm{times,}\;n\geqslant3}
                         \ \in\ O_{p,\,q,\,\sqrt{|\ve|}}(3)\;.
\end{align}
From~(\ref{poissonserie1}) we get the Poincar{\'e} map

\begin{subequations}\label{poinmap1}
        \hfill\makebox[0cm][l]{\rule{0cm}{0cm}$\Phi_1^{(1)}$}
         \raisebox{-0.1cm}[0cm][0cm]{
        \makebox[0cm][l]{\rule{0.5cm}{0cm}$\left\{\rule{0cm}{0.85cm}\right.$}}
        \parbox{14cm}{
         \begin{align}
                p_1&=p_0-\tau\ve-\tau a\,q_0^2-\frac{\tau}{3}\, a\,p_0^2
                -\tau a\,p_0q_0+\ O_{p_0,\,q_0,\,\sqrt{|\ve|}}(3) \;;
                \\[1ex]
                 q_1&=q_0+p_0-\frac{\tau}{2}\,\ve-\frac{\tau}{2}\, a\,q_0^2
                 -\frac{\tau}{12}\, a\,p_0^2
                -\frac{\tau}{3}\, a\,p_0q_0+\ O_{p_0,\,q_0,\,\sqrt{|\ve|}}(3)
                \;.      
         \end{align}}        
\end{subequations}

To get the mixed generating function we have two methods 
 \begin{itemize}
     \item The first
is to write the general form of~$S^{(1)}(p_1,q_0)$ 
modulo~$O_{p_0,\,q_0,\,\sqrt{|\ve|}}(4)$ 
and deduce the corresponding map. We get~$S$ by identifying
the latter with~(\ref{poinmap1});
      \item The second is to calculate~$S^{(1)}$ via the
                 formula~(\ref{def:S}).
\end{itemize}
We obtain
\begin{equation}\label{S11}
   S_1^{(1)}(p_1,q_0)=\frac{1}{2}\,p_1^2+\frac{\tau}{2}\,\ve\,p_1+\tau\ve\,q_0+p_1q_0
   +\frac{\tau}{12}\, a\,p_1^3+\frac{\tau}{3}\, a\,p_1^2q_0
   +\frac{\tau}{2}\, a\,p_1q_0^2+\frac{\tau}{3}\,a\,q_0^3 
   +O_{p_1,\,q_0,\,\sqrt{|\ve|}}(4)\;,
\end{equation}
and for the $qq$-generating function~(\ref{def:N}) 
\begin{equation}
   F_1^{(1)}=-\frac{1}{2}\,p_0^2+\tau\ve\,q_0+\tau\ve\,p_0+
             \frac{\tau}{3}\,a\,q_0^3+\frac{2\tau}{3}\, a\,p_0^2q_0
   +\tau\, a\,p_0q_0^2+\frac{\tau}{6}\,a\,p_0^3 
   +O_{p_0,\,q_0,\,\sqrt{|\ve|}}(4)\;.
\end{equation}
Both functions have a functional form quite different from the one of~$H_1$.

\subsection{Transitional case}
\label{subsec:genfuntrans}

From hamiltonian~(\ref{def:H2}) the typical scales
are~$p\propto|\ve|$ and~$q\propto\sqrt{|\ve|}$. Remember that this hamiltonian
describes the true dynamics when we look at the dynamics {\em backwards}
every~$2\tau$. Moreover~$\varsigma=-2$ in  equation~(\ref{monoL}) and the 
monodromy matrix for one period is 
\begin{equation}\label{monodromy2}
     \Mbbld=\begin{pmatrix} \,-1  & \phantom{-}0\ \ \\[1ex]
                                 \,\phantom{-}1    & -1\ \  
                     \end{pmatrix}\;.
\end{equation}
From the hamiltonian~$H_2$, we deduce the following Poisson brackets 
\begin{align}
       &\poisson{q,H_2}=\frac{1}{\tau}\,p+\ O_{\!\sqrt{\vphantom{\ssst|}|p|},\,q,\,\sqrt{{\ssst|}\smash{\ve}{\ssst|}\vphantom{p}}}(5)\;;
       \\[1ex]
       &\poisson{p,H_2}=-2\ve\,q-a\,q^3-\ve b\,q^2+c\,q^4+\
       O_{\!\sqrt{\vphantom{\ssst|}|p|},\,q,\,\sqrt{{\ssst|}\smash{\ve}{\ssst|}\vphantom{p}}}(5)\;;
       \\[1ex]
       &\,\poisson{\poisson{p,H_2},H_2}=-\frac{1}{\tau}\,p(2\ve+3a\,q^2)+\
       O_{\!\sqrt{\vphantom{\ssst|}|p|},\,q,\,\sqrt{{\ssst|}\smash{\ve}{\ssst|}\vphantom{p}}}(5)\;;
       \\[1ex]
       &\{\kern-1.15mm|\ldots\{\kern-1.15mm|p,
                  \underbrace{
                              H_2|\kern-1.15mm\},\dots,H_2
                              |\kern-1.15mm\}
                             }_{n\ \mathrm{times,}\;n\geqslant3}
                         \in\ O_{\!\sqrt{\vphantom{\ssst|}|p|},\,q,\,\sqrt{{\ssst|}\smash{\ve}{\ssst|}\vphantom{p}}}(5)\;.
\end{align}
From~(\ref{poissonserie1}) calculated with~$\tau\mapsto-2\tau$, we get the
Poincar{\'e} map 

\noindent
\begin{subequations}\label{poinmaptrans}
        \hfill\makebox[0.1cm][l]{\rule{-.14cm}{0cm}$\Phi_2^{(2)}$}
         \raisebox{0.15cm}[0cm][0cm]{
        \makebox[0cm][l]{\rule{0.2cm}{0cm}$\left\{\rule{0cm}{0.85cm}\right.$}}
        \parbox{16.4cm}{
         \begin{align}
                \hspace{0.2cm}p_2&=p_0+4\tau\ve\,q_0+2\tau
a\,q_0^3-4\tau\ve\,p_0+2\tau\ve b\, q_0^2
                +2\tau c\,q_0^4-6\tau c\,p_0q_0^2
+O_{\!\sqrt{\vphantom{\ssst|}|p_0|},\,q_0,\,\sqrt{{\ssst|}\smash{\ve}{\ssst|}\vphantom{p}}}(5);
                \\[1ex]
                 \hspace{0.2cm}q_2&=q_0-2p_0-4\tau\ve\,q_0-2\tau
a\,q_0^3+\frac{8}{3}\tau\ve\,p_0-2\tau\ve b\, q_0^2
                -2\tau c\,q_0^4+4\tau
		c\,p_0q_0^2+O_{\!\sqrt{\vphantom{\ssst|}|p_0|},\,q_0,\,\sqrt{{\ssst|}\smash{\ve}{\ssst|}\vphantom{p}}}(5).\hspace{-0.2cm}
         \end{align}}        
\end{subequations}

We now compute the one-step mixed generating function. Since
it is smooth with respect to~$\ve$, its general form is
\begin{equation}\begin{split}
 S_2^{(1)}(p_1,q_0)=\frac{1}{2}\,\eta\,q_0^2+\gamma\,p_1q_0+\frac{1}{3}\,\xi\,q_0^3
+\frac{1}{2}\,\delta\,p_1^2+\kappa\,
p_1q_0^2+\frac{1}{4}\,\theta_1\,q_0^4
+\zeta_3\,p_1^2q_0+\theta_2\,p_1q_0^3+\frac{1}{5}\,\iota\,q_0^5\\[1ex]
+\frac{1}{3}\,\lambda_1\,p_1^3+\frac{1}{2}\,\lambda_2\,p_1^2q_0^2+\lambda_3\,p_1q_0^3+O_{\!\sqrt{\vphantom{\ssst|}|p_0|},\,q_0,\,\sqrt{{\ssst|}\smash{\ve}{\ssst|}\vphantom{p}}}(5)\;.
\end{split}\end{equation} 
where~$(\theta_1,\zeta_3,\theta_2,\iota,\lambda_1,\lambda_2,\lambda_3)\in\RR^6$ 
and~$(\eta,\gamma,\xi,\delta,\kappa)$ are real functions
of~$\ve$                       
\begin{eqnarray}
&&\eta=\eta_0+\ve\eta_1\;;\qquad\gamma=\gamma_0+\ve\gamma_1\;;\qquad
\xi=\xi_0+\ve\xi_1\;;\nonumber\\
&&\delta=\delta_0+\ve\delta_1\;;\qquad\kappa=\kappa_0+\ve\kappa_1\;;\\
&&\mathrm{with}\quad(\eta_0,\eta_1,
\gamma_0,\gamma_1,\xi_0,\xi_1,\delta_0,\delta_1,\kappa_0,\kappa_1)\in\RR^6
\quad\mathrm{and}\quad\gamma_0\neq0\;.\nonumber
\end{eqnarray}

Although they are of order~$O_{\!\sqrt{\vphantom{\ssst|}|p_1|},\,q_0,\,\sqrt{{\ssst|}
\smash{\ve}{\ssst|}\vphantom{p}}}(6)$, the last three terms in~$S_2^{(1)}$ must
be taken into account. This is because when approximating the Poincar{\'e} map up
to order  four~(see equations~(\ref{poinmap22}) and~(\ref{sympmap0}))
differentiation with respect to~$p_1$ makes
the order fall by 
two. From~$p_0(p_1,q_0)$ and~$q_1(p_1,q_0)$ 
modulo~$O_{\!\sqrt{\vphantom{\ssst|}|p_1|},\,q_0,\,\sqrt{{\ssst|}\smash{\ve}
{\ssst|}\vphantom{p}}}(5)$,
it is straightforward to invert them to obtain~$p_1(p_0,q_0)$
and~$q_1(p_0,q_0)$ 
modulo~$O_{\!\sqrt{\vphantom{\ssst|}|p_0|},\,q_0,\,\sqrt{{\ssst|}\smash{\ve}{\ssst|}\vphantom{p}}}(5)$.
Condition~(\ref{monodromy2}) imposes  \begin{equation} \gamma_0=\delta_0=-1\;.
\end{equation}
From this we can calculate~$p_2(p_0,q_0)$ and~$q_2(p_0,q_0)$ 
modulo~$O_{\!\sqrt{\vphantom{\ssst|}|p_0|},\,q_0,\,\sqrt{{\ssst|}\smash{\ve}{\ssst|}\vphantom{p}}}(5)$.   
Identifying these two latter equations with the Poincar{\'e}
map~$\Phi_2^{(2)}$
given by~(\ref{poinmap22}) we get a set of equations with a unique
solution if and only if  
\begin{equation}\label{ab0}
	b=3\,\frac{d\eh_3}{d\ve}\eval{\ve=0}=0
	\qquad\mathrm{and}\qquad 
	c=5\,\eh_5(0)=0\;.
\end{equation}
(cf the discussion of subsection~\ref{subsec:nofotrans}).
We finally find 
\begin{equation}
	S_2^{(1)}(p_1,q_0)=-p_1q_0
			   -\tau\ve\,q_0^2-\frac{1}{2}\,p_1^2
			   -\frac{1}{4}\,\tau a\,q_0^4-\tau\ve\,p_1q_0
			   -\frac{1}{2}\,\tau a\,p_1q_0^3
			   +O_{\!\sqrt{\vphantom{\ssst|}|p_1|},\,q_0,
			       \,\sqrt{{\ssst|}\smash{\ve}{\ssst|}
			               \vphantom{p}}
			      }(5) \ ,
\end{equation}
as well as the Poincar{\'e} map for one period 

\noindent
\begin{subequations}\label{poinmap21}
        \hfill 
	\raisebox{0.3cm}[0cm][0cm]
	{\makebox[0.01cm][l]{\rule{1cm}{0cm}$\Phi_2^{(1)}$
	}        
        \makebox[0.01cm][l]{\rule{1.5cm}{0cm}$\left\{\rule{0cm}{1.1cm}
					      \right.$}
			   }
        \parbox{16cm}{
         		\begin{align}
                		     p_1&
				     =-p_0-2\tau\ve\,q_0-\tau a\,q_0^3
				     +\tau\ve\,p_0
				     +\frac{3}{2}\,\tau\ve a\, p_0q_0^2
				     +O_{\!\sqrt{\vphantom{\ssst|}|p_0|},
				     	 \,q_0,
				     	 \,\sqrt{{\ssst|}\smash{\ve}{\ssst|}
				         \vphantom{p}}
				      }(5);
                		\\[1ex]\begin{split}
                 		     	    \smash{q_1}&
				     	    =-q_0+p_0+\tau\ve\,q_0
				     	    +\frac{1}{2}\,\tau a\,q_0^3
				     	    -\frac{1}{3}\,\tau\ve\,p_0
				     	    +\ve\kappa_1\,  q_0^2
				       \\[-.5ex]
                		     	    &\hspace{1.1cm}
					    +\lambda_1\,p_0^2
					    -\frac{1}{2}\,a\tau\, p_0q_0^2
					    +(\lambda_3-2\zeta_3\tau a)\,q_0^4
					    +O_{\!\sqrt{\vphantom{\ssst|}|p_0|},
					        \,q_0,
					        \,\sqrt{
							{\ssst|}
							\smash{\ve}
							{\ssst|}
					    		\vphantom{p}
						       }
					       }(5)
					       .\hspace{-0.2cm}
                        		\end{split} 
\end{align}}        
\end{subequations}

We should have worked to a higher order to 
determine~$(\kappa_1,\lambda_1,\lambda_3)$.
To complete the study, let us give the expression of the two-step mixed
generating function 
   
\begin{equation}
 S_2^{(2)}(p_2,q_0)\DEF p_2q_2-2\tau H_2-\int_0^{-2\tau}p(t)\dot q(t)\,dt \ ,
\end{equation}
where $p(t)$~and~$q(t)$ are calculated via~(\ref{poissonserie1}). We obtain
\begin{equation}\label{S22}
 S_2^{(2)}(p_2,q_0)=p_2q_0-p_2^2-2\tau\ve\,q^2-\frac{1}{2}\,\tau\, a q^4
   +O_{\!\sqrt{\vphantom{\ssst|}|p_2|},\,q_0,\,\sqrt{{\ssst|}\smash{\ve}{\ssst|}
\vphantom{p}}}(5) \ ,
\end{equation}
as well as the~$qq$-generating function 
\begin{eqnarray}
F_2^{(2)}=S_2^{(2)}(p_2,q_0)-p_2q_2&=&2\tau
H_2(p_2,q_2;\tau;\ve)-2p_2^2+O_{\!\sqrt{\vphantom{\ssst|}|p_2|},\,q_2,\,\sqrt{{\ssst|}\smash{\ve}{\ssst|}\vphantom{p}}}(5);\\
&=& -p_2^2+
2\tau\ve\,q^2+\frac{1}{2}\,a\,q^4+O_{\!\sqrt{\vphantom{\ssst|}|p_2|},\,q_2,\,
\sqrt{{\ssst|}\smash{\ve}{\ssst|}\vphantom{p}}}(5) \ .
\end{eqnarray}

\subsection{Stable cases}

We will treat all stable cases for~$\ell\geqslant3$ simultaneously by 
regrouping hamiltonians~(\ref{def:H3aa}),~(\ref{def:H4aa})
and~(\ref{def:Hellaa}) in the unified form 
\begin{equation}
        H_l(I,\theta;t;\ve)=\ve I+
                      \!\!\sum_{\fatops{k\,\in\,\NN}
                                       {2\,\leqslant\,k\,\leqslant\,\ell/2}
                               }
                            \!\!a _k(\ve)\,I^k
                       \ +\ b _\ell(\ve)\,I^{\ell/2}
                            \cos(\ell\theta)
                       \ +\ O_{\!\sqrt{I},\,\ve^\star
}(\ell+1)\;.
\end{equation}       
We have seen that both~$p$ and~$q$ scale as~$\ve^\star$ given by (\ref{sqrtstar}).
Recall (cf Eq.~(\ref{rotframe})) that this normal
form is obtained looking at the dynamics in a rotating frame of angular
speed~$\omega_0$ given  by Eq.~(\ref{omega0}). As in the two previous
subsections, let us start computing  the Poisson brackets
in~(\ref{poissonserie1}) \begin{align}
       &\poisson{I,H_\ell}\ =\ \ell b_\ell(\ve)\,I^{\ell/2}\sin(\ell\theta)\ +\ O_{\!\sqrt{I},\,\ve^\star
}(\ell+1)\;;
       \\[1ex]
       &\poisson{\theta,H_\ell}\ =\ \ve\ +\!\!
                      \sum_{\fatops{k\,\in\,\NN}
                                       {2\,\leqslant\,k\,\leqslant\,\ell/2}
                               }
                            \!\!ka _k(\ve)\,I^{k-1}\ +\ \frac{1}{2}\,\ell b_\ell(\ve)\,I^{\ell/2-1}\cos(\ell\theta)
\ +\                        
O_{\!\sqrt{I},\,\ve^\star  }(\ell-1)\;;
       \\[1ex]
       &\,\{\kern-1.15mm|\ldots\{\kern-1.15mm|I,
                  \underbrace{
                              H_\ell|\kern-1.15mm\},\dots,H_\ell
                              |\kern-1.15mm\}
                             }_{n\ \mathrm{times,}\;n\geqslant2}
                         \ \in\ O_{\!\sqrt{I},\,\ve^\star  }(\ell+1)\;;
       \\[1ex]
       &\,\{\kern-1.15mm|\ldots\{\kern-1.15mm|\theta,
                  \underbrace{
                              H_\ell|\kern-1.15mm\},\dots,H_\ell
                              |\kern-1.15mm\}
                             }_{n\ \mathrm{times,}\;n\geqslant2}
                         \ \in\ O_{\!\sqrt{I},\,\ve^\star}(\ell+1)\;.
\end{align}
Then, in the rotating frame we obtain~$I(t)$~and~$\theta(t)$ 
modulo~$O_{\!\sqrt{I},\,\ve^\star}(\ell+1)$
and~$O_{\!\sqrt{I},\,\ve^\star}(\ell-1)$
respectively. Going back to the non-rotating frame by
adding~$\omega_0\,t$ to~$\theta(t)$ we have 

\noindent
\begin{subequations}\label{eqmosolustable}
         \begin{align}
                &\ I(t)\ =\ I_0\ +\ \ell t\,
b_\ell(\ve)\,I_0^{\ell/2}\sin\bigl[\ell(\theta_0+\omega_0t)\bigr]\ +\
O_{\!\sqrt{I_0},\,\raisebox{0.5mm}[0cm][0cm]{$\sst\sqrt[\star]{{\ssst|}\smash{\ve}{\ssst|}\vphantom{p}}$}}(\ell+1)\;;\\[1ex]
                \begin{split}
                 &\ \theta(t)\ =\ \theta_0\ +\ (\omega_0+\ve)t\ +\ t\hspace{-.4cm}\sum_{\fatops{k\,\in\,\NN}
                                       {2\,\leqslant\,k\,\leqslant\,\ell/2}
                               }
                            \!\!ka _k(\ve)\,I_0^{k-1}\ +\ \frac{1}{2}\,t\ell\,
b_\ell(\ve)\,I_0^{\ell/2-1}\cos\bigl[\ell(\theta_0+\omega_0t)\bigr]\\[-4ex]
&\hspace{10.4cm}+\
O_{\!\sqrt{I_0},\,\raisebox{0.5mm}[0cm][0cm]{$\sst\sqrt[\star]{{\ssst|}\smash{\ve}{\ssst|}\vphantom{p}}$}}(\ell-1)\;.
         \end{split}
         \end{align}
\end{subequations}

For every integer~$n$, it is then straightforward to express the first
orders of~$\theta_n\DEFt\theta(n\tau)$ 
and~$I_0$ in terms of~$I_n\DEFt I(n\tau)$and~$\theta_0$. 
The mixed generating function for time~$n$ follows immediately 
\begin{align}\label{Sln}\begin{split}
         S_\ell^{(n)}(I_n,\theta_0)
                           &\ =\ I_n\theta_0\ +\ n(\omega_0+\ve)\tau\,I_n\ +\ n\tau\hspace{-.4cm}
                             \sum_{\fatops{k\,\in\,\NN}
                                       {2\,\leqslant\,k\,\leqslant\,\ell/2}
                               }
                            \!\!a _k(\ve)\,I_n^{k}\ +\ n\tau
			    b_\ell(\ve)\,I_n^{\ell/2}\cos(\ell\theta_0)\\[-3ex]&
\hspace{9cm}+O_{\!\sqrt{I_n},\,\raisebox{0.5mm}[0cm][0cm]{$\sst\sqrt[\star]
{{\ssst|}\smash{\ve}{\ssst|}\vphantom{p}}$}}(\ell+1)   
              \end{split}             \\[3ex]
&\ =\ I_n\theta_0\ +\ n\omega_0\tau I_n\ +\ n\tau H_\ell(I_n,\theta_0) \ .      
\end{align}

We finally calculate the~$qq$-generating function from, for instance,
its definition~(\ref{def:N}) 
\begin{equation}\label{Felln}
F_\ell^{(n)}\ =\ n\tau\hspace{-.4cm}\sum_{\fatops{k\,\in\,\NN}
                                       {2\,\leqslant\,k\,\leqslant\,\ell/2}
                               }
                            \!\!(k-1)a _k(\ve)\,I_0^{k}\ +\ n\tau\left(\frac{\ell}{2}-1\right) b_\ell(\ve)\,I_0^{\ell/2}\cos(\ell\theta_0)
\ +\ O_{\!\sqrt{I_0},\,\raisebox{0.5mm}[0cm][0cm]{$\sst\sqrt[\star]{{\ssst|}\smash{\ve}{\ssst|}\vphantom{p}}$}}(\ell+1)\;.   
\end{equation}

\section{Asymptotic expansion of oscillating integrals}
\label{sec:quant}

In the previous sections, we have exclusively dealt
 with classical mechanics. The semiclassical analogue of
 these geometrical studies
 is highly non-trivial for essentially two reasons. The first is connected to the
 choice of the quantum
 hamiltonian one must start with. Infinitely many quantum hamiltonians
 differing from each other by~$\hbar$-dependent terms correspond to the same
classical dynamics. The second is that the semiclassical analogue of the
Poincar{\'e} reduction~(section~\ref{sec:reduction}) is very difficult to
extend beyond the leading order (see \cite{Bogomolny92a,Prosen95a}).
Nevertheless, we will assume in the following that one has already solved
these problems for a given system. 

Let us then begin with a given evolution operator
$\hat{U}_\gamma(\opp,\opq)$ depending on
one external parameter~$\gamma$. The Hilbert space on which the canonical operators 
$\opp$ and~$\opq$ are defined
can be finite or infinite. 
The associated classical dynamics corresponds to a one-step iteration 
Poincar{\'e} map $\Phi_\gamma^{(1)}(p,q)$.

Our aim now is to use Meyer's classification to write down semiclassical
expansions of physical quantities related to the exact quantum evolution
operator~$\hat{U}_\gamma$. For instance, we may compute semiclassically the trace
of correlation operators of the form
\begin{equation}\label{correlation}
        \hat{C}(\gamma)\ \DEF\ \prod_{i=0}^{M} 
        a_i(\opp,\opq)\,
        \hat{U}_\gamma^{n_i}
\end{equation}
where~$M$ is a strictly positive integer,~$\{a_i\}_{i\,\in\,\finiteset{0}{M}}$
are some smooth functions which do not depend on~$\hbar$ 
and~$\{n_i\}_{i\,\in\,\finiteset{0}{M}}$ are some integers whose sum
is~$\ell\neq0$. Without loss of generality we can
assume~$\ell>0$ by possibly changing the arrow of time. One particular
case of such operators are of course the trace of the ~$\ell$th power 
of~$\hat{U}_\gamma$, which play a central role for calculating the spectral
function of the quantum system~\cite{Tabor83a}.


Formally we have
\begin{equation}\label{traceformulae2}
        Q_\ell(\hbar;\gamma)\ \DEF\ 
        \tr\bigl[\hat{C}(\gamma)\bigr]\ 
        =\ \sum_{k\geqslant0}\ \hbar^k
	\iint_{\!\!{\frak s}}
            F_{\ell,k}(p',q;\gamma)\;\EXP{\frac{\imat}{\hbar}N^{(\ell)}(p',q;\gamma)}
        \,dp'\,dq \ ,
\end{equation}
where the~$F$'s are some smooth functions. The integration domain 
is the whole bidimensional phase space and the argument of the
exponential is
\begin{equation}\label{def:phase}
        N^{(\ell)}(p',q;\gamma)= N^{(\ell)}(0,0;\gamma)
                                 + p'q-S^{(\ell)}(p',q;\gamma)
\end{equation}
where~$S^{(\ell)}$ is the mixed generating function of the $\ell$th
iterated Poincar{\'e} classical
map~$\Phi_\gamma^{(\ell)}\DEFt\big[\Phi_\gamma^{(1)}\big]^\ell$.
Without loss of generality we can take~$N^{(\ell)}(0,0;\gamma)=0$.

The basic steps leading to Eq.~(\ref{traceformulae2}) are the following. First,
the fact that~$\Phi_\gamma^{(1)}(p,q)$ is the classical dynamics associated 
to~$\hat{U}_\gamma(\opp,\opq)$ means that we have 

\begin{equation}
\frac{\brae{p'}\hat{U}_\gamma\kete{q}}{\langle p'\,\kete{q}}\ \scl\ 
\EXP{\frac{\imat}{\hbar}N^{(1)}(p',q;\gamma)}\ 
\sum_{k\geqslant0}u^{(k)}(p',q)\,\hbar^k
\end{equation}
where the $u$'s are some smooth functions except, perhaps, for a
zero-measure set. For~$\hat{U}_\gamma$ to be unitary, $u_0$ has to be the
square root of~\intext{$\partial^2N^{(1)}/\partial p'\partial
q$}~\cite[Appendix~A]{Miller70a}.

Second, remark that for every smooth~$a(p,q)$ and for
every integer~$n$, we have the formal asymptotic series
\begin{equation}
\frac{\brae{p}\hat{U}_\gamma^{-n}a(\opp,\opq)\hat{U}_\gamma^{n}\kete{q}}
			           {\langle p\,\kete{q}}
	\ \scl\ a\big[\Phi_\gamma^{(n)}(p,q)\big]\ 
	+\ \sum_{k\geqslant1}a^{(k)}(p,q)\,\hbar^k
\end{equation}
for some functions~$a^{(k)}$'s which are smooth everywhere except for a zero-measure set.
This last formula can be understood if we write the formal expansion of~$a(p,q)$ in powers of
$(p,q)$, then use the commutation relation between~$\opp$ and~$\opq$ and eventually make
a stationary phase expansion at  all orders in~$\hbar$ after having inserted  the appropriate
resolution of identity made
with the projectors~$\kete{p}\brae{p}$ or~$\kete{q}\brae{q}$. Except for a zero-measure set
of~$(p,q)$'s the stationary points of the oscillating integral are non-degenerated.

Third, write~$Q_\ell(\hbar;\gamma)$ in the following manner  
\begin{equation}
 Q_\ell(\hbar;\gamma)\ =\ 
        \tr\bigl[\hat{U}_\gamma^{\ell}\,a_1
	 	 \hat{U}_\gamma^{n_1}a_2\hat{U}_\gamma^{-n_1}\,
		 \hat{U}_\gamma^{n_1+n_2}a_3\hat{U}_\gamma^{-n_1-n_2}
		 \cdots
		 \hat{U}_\gamma^{\ell-n_M}a_M\hat{U}_\gamma^{-\ell+n_M}
	   \bigr] \ ,
\end{equation}
and insert alternatively the resolution of identity made
with the projectors~$\kete{p}\brae{p}$ or~$\kete{q}\brae{q}$.
We are therefore led to a many-dimensional oscillating integral.
A stationary phase expansion can be made formally to all orders
in~$\hbar$ in the Morse (quadratic) directions of the critical points.
Eq.~(\ref{traceformulae2}) is thus recovered, where the integral in the
quasi-degenerate directions remains to be done\endnote{\ Since we are working
with a
	one-dimensional unfolding there is only one direction in the 
	integration
	domain where a critical point of a smooth function can be 
	non-quadratic and
	this result is independent of the number of arguments of the function 
	~\cite[chap~4, \S~5: the splitting lemma]{Poston/Stewart78a}.
	In~(\ref{traceformulae2}) we nevertheless keep two directions of 
	integration
	for stressing the physical interpretation of the critical points 
	in term
	of the periodic orbits of the dynamics but it can be checked in the
	following that in all cases (extremal, transitional and stable)
	there is only one direction in which the integration is not a 
	gaussian one. 
}.


If we make a stationary phase approximation of the integrals
appearing in~(\ref{traceformulae2}), we see that the domains of phase
space which contribute are the neighborhoods of the fixed points
of~$\Phi^{(\ell)}$, which are the periodic orbits of~$\Phi^{(1)}$ whose
period divides~$\ell$. When we want to calculate the contribution
to~$Q_\ell$ of the~$k$ repetitions of an \hbox{$m\tau$-periodic} orbit
($mk=\ell$) we should have specified the normal form of the generating
functions corresponding to the
mapping~$\Phi^{(\ell)}_m$~(in section~\ref{sec:nofogenfun} it has been done
only when~$\ell=m$ for extremal and transitional cases).
But it is clear, from the series~(\ref{poissonserie1}),
that changing~$m\tau$ to~$mk\tau$ will only modify the coefficients of the
polynomials in~$(p,q,\ve)$ but not their form~(in the stable case we get a
simple relation of proportionality: for~$F_m^{(\ell)}$ the $\tau$-dependence
is given by~$\ell\tau$, see Eq.~(\ref{Felln})).
Since we are mainly interested in the~$\hbar$-dependence of the terms in the
semiclassical expansion we are allowed to consider only primitive periodic
orbits. Therefore we will take for~$N^{(\ell)}$ the generating functions
calculated in the previous section.

 At this stage, we must recall the hierarchy of structures which
are present in a generic mixed classical dynamics.
In a compact domain of phase
space there exists  a finite number of
periodic orbits of a given period~$\ell$.
Let us consider one of these orbits. If it is
far from its bifurcation point, then complex  periodic orbits
are neglected and the real ones --- which are isolated ---
contribute in the usual way \cite{Gutzwiller90a}.
On the contrary, near its bifurcation one can always identify a
typical distance~$\ve^\star$ which was introduced and calculated
for each type of bifurcation in~\ref{sec:geomprop}.
Moreover there exists a typical phase-space quantum
scale~$\lambda\propto\sqrt{\hbar}$.
If~$\lambda\gtrsim\ve^\star$, then a uniform approximation is needed
in order to compute the semiclassical contribution of the orbits
to~$Q_\ell$
\cite{Ozorio/Hannay87a,Tomsovic+95a,
Sieber96a,Sieber97a,Schomerus/Sieber97a}.
Finally, if~$\lambda<\ve^\star$, then one can still consider the satellites
orbits are isolated but now the complex ones are no longer negligible.
This is the typical situation we will deal with  in the remaining
of this section. More precisely, we will evaluate the semiclassical
contribution
of the real as well as the complex periodic  orbits for each type of
bifurcation described previously.

After having constructed a suitable symplectic chart,
we are therefore led to oscillating integrals of
 the form 
\begin{equation}\label{def:oscint}
        {\cal I}_\ell\ \DEF\ 
        \iint F(p',q)\,\EXP{\frac{\imat}{\hbar}N_\ell(p',q;\ve)}\,dp'\,dq \ ,
\end{equation}
where~$F$ is a smooth generic function. Without loss of generality we
can suppose~$F(0,0)=1$. $N_\ell$,~defined by~(\ref{def:phase}), is obtained
from the normal forms of the mixed generating function~$S_\ell^{(\ell)}$ of
section~\ref{sec:geomprop}.
 The integration domain is a neighborhood of the fixed island whose
 central point is at the origin~{\frak o}. It is assumed  not to
overlap with other islands. $\ve$ must now be considered to be a
fixed quantity small compared to all other classical quantities. 
We will control semiclassical errors by varying~$\hbar$ only.



\subsection{Extremal case}
 
From~(\ref{S11}) we have 
\begin{equation}
   -N_1(p,q;\ve)=\frac{\tau}{2}\,\ve\,p+\tau\ve\,q
   +\frac{1}{2}\,p^2+\frac{\tau}{12}\, a\,p^3+\frac{\tau}{3}\, a\,p^2q
   +\frac{\tau}{2}\, a\,pq^2+\frac{\tau}{3}\,a\,q^3 
   +O_{p,\,q,\,\sqrt{|\ve|}}(4)\;.
\end{equation}

\begin{figure}[!ht]
\begin{center}
\vspace{-.5cm}
\mbox{\subfigure[$\ve/a<0$]{\input{saddlesextrepos.pstex_t}}\hspace{-3cm}
      \subfigure[$\ve/a>0$]{\input{saddlesextreneg.pstex_t}}
     }
\vspace{-.5cm}
\caption{\sl\baselineskip=0.25in\label{fig:saddlesextr}
        Deformations of the contour through the saddles points for the
	oscillating integrals~${\cal I}_1$.
       }
\end{center}
\end{figure}
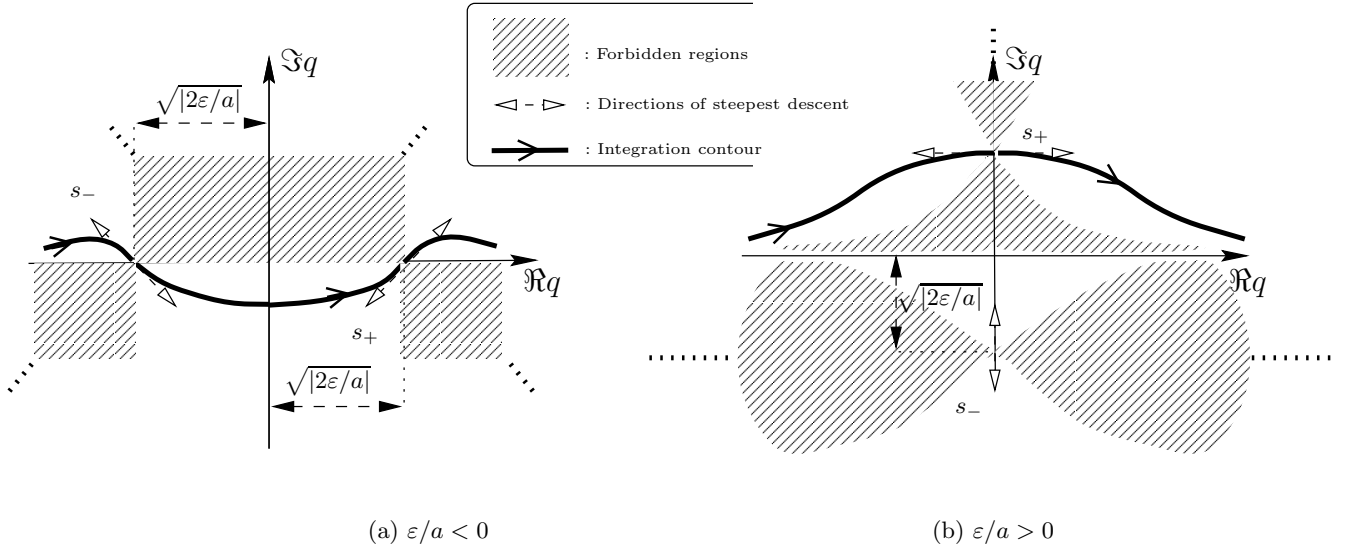

Near the fixed points we have~$F(p,q)=1+O_{p,q}(1)$ and
therefore the uniform approximation of~(\ref{def:oscint}) is an Airy function.
 The two fixed points are isolated if
\begin{equation}
        \hbar\ll\tau\ve^2\;.
\end{equation}
They both give an oscillating contribution if they are real ($\ve/a<0$) 
\begin{equation}
        {\cal I}_1=\frac{2\pi\imat\hbar}{\sqrt{2\tau|\ve a|^{1/2}}}
            \left(
                 \EXP{-\imat\frac{\pi}{2}}\,
                 \EXP{-2\imat\tau|\ve|^{3/2}/(3\hbar|a|^{1/2})}
                 \bigl[1+O_{\sqrt\hbar}(3)\bigr]
                +\EXP{2\imat\tau|\ve|^{3/2}/(3\hbar|a|^{1/2})}
                 \bigl[1+O_{\sqrt\hbar}(3)\bigr]
            \right) \ .
\end{equation}
This is the usual Gutzwiller contributions where we have made
explicit the Maslov index for each 
orbit~(see figure~\ref{fig:saddlesextr}~a)).

If~$\ve/a>0$, the integration contour can be deformed to pass through
only one complex saddle point, giving an exponentially small
contribution~(see figure~\ref{fig:saddlesextr}~b)) 
\begin{equation}\label{tunnel}
 {\cal I}_1=\frac{2\pi\imat\hbar}{\sqrt{2\tau|\ve a|^{1/2}}}\,
\EXP{-\imat\frac{\pi}{4}}\,\EXP{-2\tau|\ve|^{3/2}/(3\hbar|a|^{1/2})}\bigl(1+O_\hbar(2)\bigr)\;.
\end{equation}

\subsection{Transitional case}
From~(\ref{S22}) we have 
\begin{equation}
        N_2(p,q)=p^2+2\tau\ve\,q^2+\frac{1}{2}\,\tau\, a q^4
                +O_{\!\sqrt{\vphantom{\ssst|}  |p|},\,q,
                    \,\sqrt{{\ssst|}\smash{\ve}{\ssst|}\vphantom{p}}
                   }(5)\;.         
\end{equation}
The satellite points can be considered to be isolated from the origin
if
\begin{equation}
        \hbar\ll\tau\frac{2\ve^2}{|a|}\;.
\end{equation}
The contribution of the central fixed point is
\begin{equation}
        \frac{\pi\imat\hbar}{\sqrt{2\tau|\ve|}}\;
        \EXP{-\imat\frac{\pi}{2}\nu[\ve]} \ ,
\end{equation}
where the Maslov index is given by
\begin{equation}\label{maslov}
\nu[x]\DEF\left\{
                        \begin{array}{ll}
                                0&\mbox{if $x>0$;}\\[1ex]
                                1&\mbox{if $x<0$.}
                        \end{array}
          \right.
\end{equation}
If~$\ve/a<0$ the satellite points are real and contribute  by 
\begin{equation}
        2\sqrt{\frac{\pi\imat\hbar}{8\tau|\ve|}}\,
        \EXP{-\imat\frac{\pi}{2}\nu[-\ve]}\,
        \EXP{-2\imat\tau\ve^2/(a\hbar)}\;.
\end{equation}
If~$\ve/a>0$ we must study the relative orientation of the saddle
points. Indeed after having done one stationary integration in
the~$p$ direction, we have 
\begin{equation}
        {\cal I}_2=
        \sqrt{\pi\imat\hbar}
        \int_{-\infty}^{+\infty}\,
                \EXP{\frac{\imat\tau}{\hbar}(2\ve q^2+aq^4)}
                \,dq\big[1+O_{\sqrt{\hbar}}(2)\big]\;.
\end{equation}
The saddle points~$(s_0=0,s_{\pm1}=\imat q_\pm)$ are represented in
figure~\ref{fig:saddlestraneneg}.
\begin{figure}[!ht]
\begin{center}
\input{saddlestraneneg.pstex_t}
\caption{\sl\baselineskip=0.25in\label{fig:saddlestraneneg}
        Saddle points for ${\cal I}_2$ when~$\ve/a>0$.
       }
\end{center}
\end{figure}

 The hatched
regions represent the forbidden regions  
where~$\Im\big[N(z)-N(s)\bigr]<0$.
It is clear from the figure that it is impossible to deform the 
contour~{\clgv C} in order to cross~$s_1$ or~$s_{-1}$ and to come back to
the real axis. The complex saddles will therefore 
contribute to semiclassical expansions only via power laws in~$\hbar$
when one tries to go beyond the dominant order of the semiclassical
expansion at~$s_0$. This power law expansion will contain some information about
the presence of~$s_{\pm1}$. Let us specify the corresponding 
power of~$\hbar$ 

\begin{eqnarray}
        {\cal I}_2&=&
        \iint_{\!\!\!\RR^2}\!\!\!
          \left(1+\frac{1}{2}p^2\partial_{p,p}F\eval{(0,0)}
              +\frac{1}{2}q^2\partial_{q,q}F\eval{(0,0)}
              +\frac{\imat\tau}{2\hbar}\,aq^4
          \right)
          \EXP{\frac{\imat}{\hbar}(2\ve\tau q^2+p^2)
              }
        \,dp\,dq
        \big[1+O_{\sqrt\hbar}(3)\big]
    \nonumber\\[2ex]
        &=&
        \frac{\pi\imat\hbar}{\sqrt{2\tau|\ve|}}
        \,\EXP{-\imat\frac{\pi}{2}\nu[\ve]}
        \left[1+\frac{1}{4}\,\imat\hbar\left(\partial_{p,p}F\eval{(0,0)}
              +\frac{1}{2\tau\ve}\partial_{q,q}F\eval{(0,0)}
              -\frac{3a}{8\tau\ve^2}\right)
              +O_{\sqrt\hbar}(3)
        \right]\;.
\end{eqnarray}
 It is the third term of the parenthesis which is the explicit leading
contribution of the complex satellite points. It corresponds to a
term~$\hbar$ smaller than the leading Gutzwiller contribution of the real one.
Even if~$a$ is not known its sign can be determined by looking at the stability
of the central orbit (see figure~\ref{fig_2enegext}).

The fact that the complex saddle points are indeed encoded
in the asymptotic series of~${\cal I}_2$ is well known:
When~$F\equiv1$, one can re-sum the complete asymptotic expansion
of~${\cal I}_2$ and give to it a sense in the whole complex
~\hbox{$\hbar$-plane} except over one half line. The discontinuity
of~${\cal I}_2$ between the two sides of this cut can be interpreted
in terms of an integral whose contour goes actually trough the
saddles~$s_1$ or~$s_{-1}$. See for
example~\cite[\S2.1 a]{Bogomolny+80a} or~\cite[\S9.4 a]{Itzykson/Zuber80a}.

\subsection{Stable cases}
From~(\ref{Sln}) we have in action-angle variables 
\begin{equation}\label{Nellaa}
-N_\ell(I,\theta;\ve)=\ \ell\ve\tau\,I\ +\ \ell\tau\hspace{-.4cm}
                             \sum_{\fatops{k\,\in\,\NN}
                                       {2\,\leqslant\,k\,\leqslant\,\ell/2}
                               }
                            \!\!a _k(\ve)\,I^{k}\ 
			    +\ \ell\tau b_\ell(\ve)\,I^{\ell/2}\cos(\ell\theta)+O_{\!\sqrt{I},\,\ve^\star}(\ell+1)\;.  
\end{equation}

We are here systematically ignoring problems related to the quantization
procedure. In expressions like~(\ref{def:oscint}) the passage to action-angle
variables or to any other symplectic coordinate system of jacobian unity 
can be done. If one
{\em starts} from the classical problem expressed in action-angle coordinates,
we may face however some difficulties to quantize
it~\citeaffixed{Carruthers/Nieto68a}{see} because the phase space has a
cylindrical topology and therefore cannot be described by a unique symplectic
chart.  In Feynman path integrals or their discretized analogues, this implies
to take care of the homotopy of the paths. The interfering contribution
 of each of them is weighted by the character of its homotopy 
 group~\cite{Schulman81a}. For the
cylinder, it corresponds to a infinite sum over the winding numbers.
Semiclassically this will select exactly the central orbit whose winding number
is~$r=\ell\tau\omega/(2\pi)$, the others being at a distance much larger
than~$\ve$ and a fortiori than~$\hbar$; we are therefore led 
to~(\ref{Nellaa}). For a concrete illustration see~\cite[\S2]{Sieber96a}.

The contribution to the central fixed point~$I=0$ is given by
\begin{equation}\label{Il}\begin{split}
{\cal I}_\ell^{{\frak
o}}=\int_{\theta}^{2\pi}\int_{I=0}^{\infty}F(I,\theta)
&\Biggl[1-\frac{\imat\ell\tau}{\hbar}\biggl(\hspace{-2ex}\sum_{\fatops{k\,\in\,\NN}
                                       {2\,\leqslant\,k\,\leqslant\,\ell/2}
                               }
                            \!\!a _k(\ve)\,I^{k}+\ \ell\tau
b_\ell(\ve)\,I^{\ell/2}\cos(\ell\theta)\biggr)
\\[2ex]
&+\frac{1}{2}\left(-\frac{\imat\ell\tau}{\hbar}\right)^2\biggl(\hspace{-2ex}\sum_{\fatops{k\,\in\,\NN}
                                       {2\,\leqslant\,k\,\leqslant\,\ell/2}
                               }
                            \!\!a _k(\ve)\,I^{k}+\ \ell\tau
b_\ell(\ve)\,I^{\ell/2}\cos(\ell\theta)\biggr)^2+\cdots
\Biggr]
\EXP{-\imat\ell\tau\ve I/\hbar}\,dI\,d\theta\;.
\end{split}
\end{equation}
The dominant term is~$-\frac{\imat\hbar}{\ell\tau\ve}$. Many powers
of~$\hbar$ follow among which  we want to determine the first which depends on~$b_\ell$.
If we expand~$F$ in Fourier series we obtain
\begin{equation}
F(I,\theta)\ =\ \sum_{k=0}^{\infty}\;\bigl[F_k(I)\cos(k\theta)
+\tilde{F}_k(I)\sin(k\theta)\bigr]\;.
\end{equation}
Moreover, we know from the smooth behavior of~$F$ in a neighborhood of 
the origin that we
have $F(I,\theta)=F(\sqrt{2I}\cos{\theta},\sqrt{2I}\sin{\theta})$ and
thus that, $\forall\,k\in\NN,\ F_k(I)=A_kI^{k/2}$ for some complex 
constant~$A_k$.
In~(\ref{Il}) we get a term of the form 
\begin{equation}
-\frac{\imat\ell\tau}{\hbar}\,\pi b_\ell\left(\frac{\hbar}{\ell\tau\ve}\right)^{\ell/2+1}\int_0^\infty
x^{\ell/2}F_\ell\left(\frac{\hbar x}{\ell\tau\ve}\right)\EXP{-x}\,dx
=-\imat\pi A_\ell b_\ell\left(\frac{\hbar
}{\ell\tau}\right)^\ell\frac{\ell\,!}{\ve^{\ell+1}}+O_{\sqrt{\hbar}}(2\ell +1)\;.
\end{equation}
But this is not the largest term depending on~$b_\ell$
since there is a term proportional to~$\hbar^{\ell-1}$ 
\begin{equation}\label{Ineg}
-\left(\frac{\ell\tau}{\hbar}\right)^2\,b_\ell^2
\int_{\theta=0}^{2\pi}\int_{I=0}^{\infty}I^\ell\cos^2(\ell\theta)\EXP{-\imat\ell\tau\ve
I/\hbar}\,dI\,d\theta
=-\pi  b_\ell^2\left(\frac{\hbar
}{\ell\tau}\right)^{\ell-1}\frac{\ell\,!}{\ve^{\ell+1}}\;.
\end{equation}

When the satellite points are real, then their
contribution to semiclassical expansion corresponds to the standard
Gutzwiller oscillating terms. To leading order one finds

---\ For~$\ell=3$ 

\begin{equation}
{\cal I}_3=-\frac{\imat\hbar}{3\tau\ve}\bigl[1+O_{\sqrt\hbar}(3)\bigr]+
\frac{2\pi\imat\hbar}{\sqrt{3}\tau|\ve|}\,
\EXP{-\imat(\nu[b_3]+\nu[\ve])\pi/2}\,
\EXP{-4\imat\tau\ve^3/(9\hbar b_3^2)}\,[1+O_{\sqrt\hbar}(3)\bigr] \ .
\end{equation}

---\ For~$\ell=4$ and with the notations of subsection~\ref{subsubsec:l4},
we can have one or two real families of fixed points corresponding
to~$\cos(4\theta)=\pm1$. When they are real their contribution is
\begin{equation}
4\frac{2\pi\imat\hbar}{\left|\frac{a-b}{a+c}\right|^{1/2}\tau|\ve|}\,
\EXP{-\imat\big(\nu[-a-c]+\nu[a-b]\big)\pi/2}\,
\EXP{-2\imat\tau\ve^2(\frac{1}{a}+\frac{1}{a+c})/\hbar}\,\bigl[1+O_{\sqrt\hbar}
(3)\bigr] \ .
\end{equation}

---\ For~$\ell\geqslant5$ both stable and unstable  satellite orbits
give the same contribution when~$\ve/a_2<0$,
\begin{equation}
\frac{2\pi\imat\hbar}{|b_\ell|^{1/2}\tau}\,\left|\frac{2a_2}{\ve}\right|^{\ell/4}
\EXP{-\imat\big(\nu[\ve]+\nu[\pm b_\ell]\big)\pi/2}\,
\EXP{\imat\ell\tau\ve^2/(4a_2\hbar)}\,\bigl[1+O_{\sqrt\hbar}(3)\bigr]\;.
\end{equation}

Martin Sieber showed~\citeyear{Sieber96a} that if one wants to compute the
difference between the contribution of the stable and unstable satellite orbits
as a function of~$\ve$ one needs to work with normal forms up
to $O_{\!\sqrt{I},\,\raisebox{0.5mm}[0cm][0cm]{$\sst\sqrt[\star]{{\ssst|}
\smash{\ve}{\ssst|}\vphantom{p}}$}}(2\ell-4)$. The expression of this normal
form is given by equation~(14)  of Ref.\cite{Sieber96a}

When the satellite orbits are complex, they actually correspond to
real negative action~$I$. The integration contour in~$I$ goes from
zero to~$+\infty$ and therefore cannot cross the saddle points on the
negative real axis. The contribution of the complex orbits comes
entirely from the saddle expansion at the origin. It thus contains information
about the complex solution through a term of order~$\hbar^{\ell-1}$ given by
Eq.~(\ref{Ineg}). 

\begin{figure}[!ht]
\begin{center}
\input{saddlesIneg.pstex_t}
\caption{\sl\baselineskip=0.25in\label{fig:saddlesIneg}
        Saddle points for ${\cal I}_\ell$ when~$\ve/a_2>0$ and~$\ell\geqslant4$.
       }
\end{center}
\end{figure}
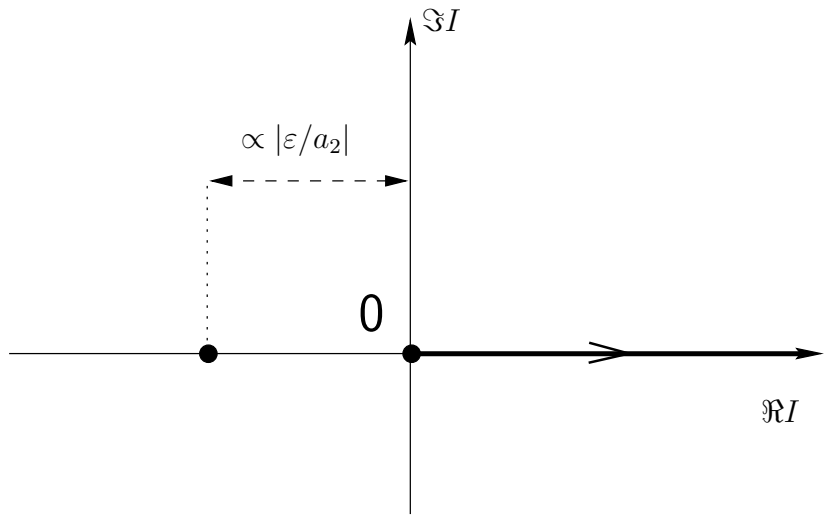

\section{Conclusion}

In the first part of this paper we have revisited Meyer's classification of
bifurcations in two-freedom systems by systematically including and locating
the complex periodic orbits which are present on each side of the
bifurcation (subsection~\ref{subsec:sumup} presents a summary of the main
features concerning this point). Thereafter we have explicitly studied the
contribution of complex periodic orbits to semiclassical
expansions~(\ref{traceformulae}) when the classical regime is a mixture of
regular and irregular motion. Our starting hypothesis was that the ghost
orbits which may generically contribute are those which are not too deeply
located in the complex phase space and therefore can be associated to a
bifurcation. We have shown that the only case where they do not give a power
law contribution in~$\hbar$ is the extremal bifurcation where the monodromy
matrix has a determinant equal to one. Their contribution is exponentially
small in~$\hbar$ (see~(\ref{tunnel})), whereas the argument of the
exponential is proportional to the distance from the bifurcation to the
power~$3/2$. Because of this large power and because no real orbits come to
shadow them, these ghost orbits can be easily observed when one compares
exact and semiclassical quantities~\cite{Kus+93a,Leboeuf/Mouchet94a}. In the
latter reference it has moreover been shown that these orbits may also play
an important role in the semiclassical description of tunneling effects. In
all the other types of bifurcations the complex orbits do not create new
peaks in the Fourier transform of the density of states but rather modify
the amplitude of the peak associated to the central real orbit.

We have systematically ignored in our discussion the consequences a symmetry
may have on bifurcations. Because discrete as well as continuous symmetries
are, from a technical point of view, equivalent to the introduction of another
continuous parameter in the unfolding in addition to~$\ve$, their presence
increases drastically the number of the possible types of bifurcations. One
should for example consider cases where the linearized motion corresponds
to~$\Lbbld=\Zero$ (or~$\Mbbld=\Id$), and should therefore classify first all
the possible cubic hamiltonians. To our knowledge there is no systematic
classification of the bifurcations of hamiltonians of codimension two or
higher.  Nevertheless some normal forms for certain types of bifurcations
can be found in the literature~\citeaffixed{Meyer86a}{see references in}.
The simplest cases are of course the normal forms given in
section~\ref{subsec:tables} with some of the higher order coefficients put
to zero. Then we get some additional phase portraits like the one obtained
in Ref.~\cite[table~1 case~c]{deAguiar+87a}. One can easily go in these
cases through a semiclassical analysis similar to the one presented here.

\bigskip     
A.M. would like to express his gratitude to Eugene Bogomolny, Stephen Creagh,
Martin Sieber and Daniel Rouben for
many fruitful discussions. He also thanks le Service de Prestations
Informatiques du  D\'epartement de Math\'ematiques et d'Informatique de
l'\'Ecole Normale Sup\'erieure (Paris) for assistance.

\newpage
\newpage
    \begingroup
     \parindent 0pt
     \parskip 2ex
     \def\enotesize{\normalsize}
     \theendnotes
     \endgroup
\appendix
\newpage

\begin{center}\large
 {\bfseries Appendix:} {Summary tables}\label{subsec:tables}
\end{center}

\vspace{-\baselineskip}

\enlargethispage*{4\baselineskip}\noindent
\begin{table}[!hb]\hspace{-4pt}
\begin{tabular}{||r||c|c|c||}
	\hline\hline
	$\Mbbld$: monodromy matrix&$\ve/a<0$&$\ve=0$&$\ve/a>0$\\
	\hline\hline
	&\multicolumn{3}{c||}{\rule[-.5em]{0pt}{2em}{\em\Large Extremal case}}
	\\
	\cline{2-4}
	\raisebox{1.3cm}[0ex][0ex]
	 	{\parbox[b]{3cm}{\begin{center}\hspace{-2em}  
					$\spectre\Mbbld\eval{\ve=0}=\{1,1\}$
                        	 \end{center}
				}
		}
	&\input{fig_1epos.pstex_t}
	&\input{fig_1enul.pstex_t}
	&\input{fig_1eneg.pstex_t}\\[1ex]
	&\multicolumn{3}{c||}{Hamiltonian normal form:}\\
	&\multicolumn{3}{c||}{\rule[-1.5em]{0pt}{3em}
			     $H_1(p,q;t;\ve)=
			     \frac{1}{2\tau}\,p^2+\ve\,q+\frac{1}{3}\,a\,q^3
                     	     +\frac{1}{4}\,b\,q^4
			     +O_{
			    	 p,\,q,\,
				 \sqrt{{\ssst|}\smash{\ve}{\ssst|}\vphantom{p}}
			        }(5)
			   $}\\
	\hline\hline
	&\multicolumn{3}{c||}{\rule[-.5em]{0pt}{2em}{\em\Large
							 Transitional case}}
	\\
	\cline{2-4}
	&&\rule{0ex}{3.5ex}\fbox{$a<0$}&\\
	\raisebox{-1cm}[0ex][0ex]
	 	{\parbox[b]{3cm}{\begin{center}\hspace{-2em} 
					\smash{$\spectre\Mbbld\eval{\ve=0}=\{-1,-1\}$}
                        	 \end{center}
				}
		}
	&\raisebox{-4mm}{\input{fig_2epos.pstex_t}}
	&\raisebox{-1mm}{\input{fig_2enul.pstex_t}}
	&\raisebox{-1mm}{\input{fig_2eneg.pstex_t}}\\
		&&\fbox{$a>0$}&\\
	&\raisebox{-3mm}{\input{fig_2eposbis.pstex_t}}
	&\raisebox{-1mm}{\input{fig_2enulbis.pstex_t}}
	&\raisebox{-3mm}{\input{fig_2enegbis.pstex_t}}\\[1ex]
	&\multicolumn{3}{c||}{Hamiltonian normal form:}\\
	&\multicolumn{3}{c||}{\rule[-1.5em]{0pt}{3em}
			      $H_2(p,q;t;\ve)=
			      \frac{1}{2\tau}\,p^2+\ve\,q^2+\frac{1}{4}\,a\,q^4
			      +b\ve\,q^4+cq^6
        		      +O_{
			      	    \!\sqrt{\vphantom{\ssst|}  p},\,q,
			            \,\sqrt{{\ssst|}\smash{\ve}{\ssst|}
			      	    \vphantom{p}}
			     	   }(7)
			    $}\\
       \hline\hline
\end{tabular} 
\caption{\em Generic bifurcations of extremal and transitional points
        \label{table:extran}}
\end{table}
\noindent
\begin{table}[!htb]
\begin{center}
\begin{tabular}{||l||c|c|c||}
	\hline\hline
	&\multicolumn{3}{c||}{  \rule[-1.5em]{0pt}{4em}
				        {\Large{\em Stable case:}
				         $\spectre\Mbbld\eval{\ve=0}
				         =\{\EXP{\pm\imat\frac{2r\pi}{\ell}}\}$
				        }
				     }
	\\
	\hline\hline
	&$\ve/a<0$&$\ve=0$&$\ve/a>0$\\
	\hline\hline
	&\multicolumn{3}{c||}{\rule{0ex}{3.5ex}\fbox{$\ell=3$}}\\	
	\cline{2-4}
	&\input{fig_3epos.pstex_t}
	&\input{fig_3enul.pstex_t}
	&\input{fig_3eneg.pstex_t}\\[1ex]
	&\multicolumn{3}{c||}{Hamiltonian normal forms:}\\
	&\multicolumn{3}{c||}{\rule[-1.5em]{0pt}{3em}
			       $H_3(p,q;t;\ve)\ =\ \frac{1}{2}\,\ve\,(p^2+q^2) 
                 		+\frac{1}{3}\,b(p^3-3pq^2)+O_{p,\,q,\,\ve}(4) 
			       $}\\
	&\multicolumn{3}{c||}{\rule[-1.5em]{0pt}{3em}
			      $H_3(I,\theta;t;\ve)\ =\ \ve I\ 
			      +\frac{1}{3}\,b\,(2I)^{3/2}\cos(3\,\theta)
			      +O_{\sqrt{I},\ve}(4)   
			      $}\\		   
	\hline\hline	
	&\multicolumn{3}{c||}{\rule{0ex}{3.5ex}\fbox{$\ell=4$}}\\
	\raisebox{2cm}[0ex][0ex]{$\quad\boldsymbol{\alpha)\ a+b>0}$}
	&\input{fig_4epos.pstex_t}
	&\input{fig_4enul.pstex_t}
	&\input{fig_4eneg.pstex_t}\\
	\raisebox{2cm}[0ex][0ex]{$\quad\boldsymbol{\beta)\ a+b<0}$}
	&\input{fig_4eposbis.pstex_t}
	&\input{fig_4enulbis.pstex_t}
	&\input{fig_4enegbis.pstex_t}\\[1ex]
	&\multicolumn{3}{c||}{Hamiltonian normal forms:}\\
	&\multicolumn{3}{c||}{\rule[-1.5em]{0pt}{3em}
			      $H_4(p,q;t;\ve)=\frac{1}{2}\,\ve\,(p^2+q^2) 
                 	       +\frac{1}{4}\,a(p^4+q^4)
		 	       +\frac{1}{2}\,b\,p^2q^2
		 	       +O_{p,\,q,\sqrt{{\ssst|}\smash{\ve}{\ssst|}}}(5)
			       $}\\
        &\multicolumn{3}{c||}{\rule[-1.5em]{0pt}{3em}
			      $H_4(I,\theta;t;\ve)=\ve I
			       +\frac{1}{4}(3a+b)I^2+
			       \frac{1}{4}(a-b)I^2\cos(4\,\theta)\ 
			       +O_{\!\sqrt{I},
			       	       \,\raisebox{0.5mm}[0cm][0cm]
				       {$\sst\sqrt{{\ssst|}\smash{\ve}{\ssst|}
				         \vphantom{p}}
				       $}
				    }(5)\;.
			    $}\\		    
	\hline\hline
\end{tabular}
\caption{\em Generic bifurcations in stable cases for~$\ell\in\{3,4\}$.
        \label{table:l34}}
\end{center}
\end{table}

\noindent
\begin{table}[!htb]
\begin{center}
\begin{tabular}{||c|c|c||}
	\hline\hline
	\multicolumn{3}{||c||}{  \rule[-1.5em]{0pt}{4em}
				        {\Large{\em Stable case:}
				         $\spectre\Mbbld\eval{\ve=0}
				         =\{\EXP{\pm\imat\frac{2r\pi}{\ell}}\}$
				        }
				     }
	\\
	\hline\hline
	$\ve/\ea_2<0$&$\ve=0$&$\ve/\ea_2>0$\\
	\hline\hline
        \multicolumn{3}{||c||}{\rule{0ex}{3.5ex}\fbox{$\ell\geqslant5$}}\\
	\input{fig_5epos.pstex_t}
	&\raisebox{1cm}{\input{fig_5enul.pstex_t}}
	&\raisebox{1cm}{\makebox[2cm]{\input{fig_5eneg.pstex_t}}}\\
	\multicolumn{3}{||c||}{\rule[-1.5em]{0pt}{3em}Hamiltonian normal forms:}\\[2ex]
	\multicolumn{3}{||c||}{\rule[-1.5em]{0pt}{3em}
			     $\dst H_\ell(p,q;t;\ve)=\frac{1}{2}\,\ve\,(p^2+q^2)+
                 \hspace{-3ex}\sum_{\fatops{k\,\in\,\NN}
                                  {2\,\leqslant\,k\,\leqslant\,\ell/2}
                          }
                            \hspace{-3ex}\tilde\ea _k(\ve)\,(p^2+q^2)^k
                       +\big|\eh_{{\ssst\kern.12ex\ell0\ell}}
                                              (\ve)\big|
                                        \;\Re\big[(p+\imat q)^\ell
                                \big]  $}\\
        \multicolumn{3}{||r||}{
	\raisebox{5ex}{$+O_{p,q}(\ell+1) 
			       $}}\\
	\multicolumn{3}{||c||}{\rule[-1.5em]{0pt}{3em}
			     $\dst H_\ell(I,\theta;t;\ve)=\ve I+
                      \hspace{-3ex}\sum_{\fatops{k\,\in\,\NN}
                                       {2\,\leqslant\,k\,\leqslant\,\ell/2}
                               }
                            \hspace{-3ex}\ea _k(\ve)\,I^k
                       +\eb _\ell(\ve)\,I^{\ell/2}
                            \cos(\ell\theta)
                       +O_{\!\sqrt{I}}(\ell+1)
			       $}\\
	\hline\hline		   
\end{tabular}
\caption{\em Generic bifurcations in stable cases for~$\ell\geqslant5$.
        \label{table:l5}}
\end{center}
\end{table}

\clearpage

\bibliography{/usr/users/nefer/mouchet/tex/biblio/mrabbrev,/usr/users/nefer/mouchet/tex/biblio/qchaos}
{\bibliographystyle{/usr/users/nefer/mouchet/tex/biblio/annphys}}

\end{document}

%% file: reduction.pstex_t
\begin{picture}(0,0)%
\epsfig{file=reduction.pstex}%
\end{picture}%
\setlength{\unitlength}{0.00087500in}%
\begingroup\makeatletter\ifx\SetFigFont\undefined
\def\x#1#2#3#4#5#6#7\relax{\def\x{#1#2#3#4#5#6}}%
\expandafter\x\fmtname xxxxxx\relax \def\y{splain}%
\ifx\x\y   
\gdef\SetFigFont#1#2#3{%
  \ifnum #1<17\tiny\else \ifnum #1<20\small\else
  \ifnum #1<24\normalsize\else \ifnum #1<29\large\else
  \ifnum #1<34\Large\else \ifnum #1<41\LARGE\else
     \huge\fi\fi\fi\fi\fi\fi
  \csname #3\endcsname}%
\else
\gdef\SetFigFont#1#2#3{\begingroup
  \count@#1\relax \ifnum 25<\count@\count@25\fi
  \def\x{\endgroup\@setsize\SetFigFont{#2pt}}%
  \expandafter\x
    \csname \romannumeral\the\count@ pt\expandafter\endcsname
    \csname @\romannumeral\the\count@ pt\endcsname
  \csname #3\endcsname}%
\fi
\fi\endgroup
\begin{picture}(5655,3315)(467,-2980)
\put(4951,-2941){\makebox(0,0)[lb]{\smash{\SetFigFont{12}{14.4}{rm}\special{ps: gsave 0 0 0 setrgbcolor}${\frak s}(E)$\special{ps: grestore}}}}
\put(4201,179){\makebox(0,0)[lb]{\smash{\SetFigFont{12}{14.4}{rm}\special{ps: gsave 0 0 0 setrgbcolor}${\frak p}$\special{ps: grestore}}}}
\put(1321,-2266){\makebox(0,0)[lb]{\smash{\SetFigFont{12}{14.4}{rm}\special{ps: gsave 0 0 0 setrgbcolor}$\Sigma(E)$\special{ps: grestore}}}}
\put(3466,-2701){\makebox(0,0)[lb]{\smash{\SetFigFont{12}{14.4}{rm}\special{ps: gsave 0 0 0 setrgbcolor}${\frak S}$\special{ps: grestore}}}}
\put(4516,-1036){\makebox(0,0)[lb]{\smash{\SetFigFont{12}{14.4}{rm}\special{ps: gsave 0 0 0 setrgbcolor}\origin\special{ps: grestore}}}}
\end{picture}

%% file: RU1.pstex_t
\begin{picture}(0,0)%
\epsfig{file=RU1.pstex}%
\end{picture}%
\setlength{\unitlength}{0.00087500in}%
\begingroup\makeatletter\ifx\SetFigFont\undefined
\def\x#1#2#3#4#5#6#7\relax{\def\x{#1#2#3#4#5#6}}%
\expandafter\x\fmtname xxxxxx\relax \def\y{splain}%
\ifx\x\y   
\gdef\SetFigFont#1#2#3{%
  \ifnum #1<17\tiny\else \ifnum #1<20\small\else
  \ifnum #1<24\normalsize\else \ifnum #1<29\large\else
  \ifnum #1<34\Large\else \ifnum #1<41\LARGE\else
     \huge\fi\fi\fi\fi\fi\fi
  \csname #3\endcsname}%
\else
\gdef\SetFigFont#1#2#3{\begingroup
  \count@#1\relax \ifnum 25<\count@\count@25\fi
  \def\x{\endgroup\@setsize\SetFigFont{#2pt}}%
  \expandafter\x
    \csname \romannumeral\the\count@ pt\expandafter\endcsname
    \csname @\romannumeral\the\count@ pt\endcsname
  \csname #3\endcsname}%
\fi
\fi\endgroup
\begin{picture}(3187,2158)(901,-2114)
\put(2506,-572){\makebox(0,0)[lb]{\smash{\SetFigFont{10}{12.0}{rm}0}}}
\put(901,-2086){\makebox(0,0)[lb]{\smash{\SetFigFont{10}{12.0}{rm}$\CC$}}}
\put(3691,-106){\makebox(0,0)[lb]{\smash{\SetFigFont{10}{12.0}{rm}\ref{marginal}'}}}
\put(1036,-106){\makebox(0,0)[lb]{\smash{\SetFigFont{10}{12.0}{rm}\ref{marginal}''}}}
\put(2476,-2086){\makebox(0,0)[lb]{\smash{\SetFigFont{10}{12.0}{rm}\ref{stable}}}}
\put(3556,-1636){\makebox(0,0)[lb]{\smash{\SetFigFont{10}{12.0}{rm}\ref{unstable}'}}}
\put(1306,-1636){\makebox(0,0)[lb]{\smash{\SetFigFont{10}{12.0}{rm}\ref{unstable}''}}}
\put(3286,-826){\makebox(0,0)[lb]{\smash{\SetFigFont{10}{12.0}{rm}1}}}
\put(1576,-826){\makebox(0,0)[lb]{\smash{\SetFigFont{10}{12.0}{rm}-1}}}
\end{picture}

%% file: fig_hyperbolic.pstex_t
\begin{picture}(0,0)%
\epsfig{file=fig_hyperbolic.pstex}%
\end{picture}%
\setlength{\unitlength}{0.00083300in}%
\begingroup\makeatletter\ifx\SetFigFont\undefined
\def\x#1#2#3#4#5#6#7\relax{\def\x{#1#2#3#4#5#6}}%
\expandafter\x\fmtname xxxxxx\relax \def\y{splain}%
\ifx\x\y   
\gdef\SetFigFont#1#2#3{%
  \ifnum #1<17\tiny\else \ifnum #1<20\small\else
  \ifnum #1<24\normalsize\else \ifnum #1<29\large\else
  \ifnum #1<34\Large\else \ifnum #1<41\LARGE\else
     \huge\fi\fi\fi\fi\fi\fi
  \csname #3\endcsname}%
\else
\gdef\SetFigFont#1#2#3{\begingroup
  \count@#1\relax \ifnum 25<\count@\count@25\fi
  \def\x{\endgroup\@setsize\SetFigFont{#2pt}}%
  \expandafter\x
    \csname \romannumeral\the\count@ pt\expandafter\endcsname
    \csname @\romannumeral\the\count@ pt\endcsname
  \csname #3\endcsname}%
\fi
\fi\endgroup
\begin{picture}(1824,1824)(-11,-973)
\put(1726, 14){\makebox(0,0)[lb]{\smash{\SetFigFont{12}{14.4}{rm}$p$}}}
\put(760,-157){\makebox(0,0)[lb]{\smash{\SetFigFont{12}{14.4}{rm}\origin}}}
\put(976,689){\makebox(0,0)[lb]{\smash{\SetFigFont{12}{14.4}{rm}$q$}}}
\end{picture}

%% file: fig_parabolic.pstex_t
\begin{picture}(0,0)%
\epsfig{file=fig_parabolic.pstex}%
\end{picture}%
\setlength{\unitlength}{0.00083300in}%
\begingroup\makeatletter\ifx\SetFigFont\undefined
\def\x#1#2#3#4#5#6#7\relax{\def\x{#1#2#3#4#5#6}}%
\expandafter\x\fmtname xxxxxx\relax \def\y{splain}%
\ifx\x\y   
\gdef\SetFigFont#1#2#3{%
  \ifnum #1<17\tiny\else \ifnum #1<20\small\else
  \ifnum #1<24\normalsize\else \ifnum #1<29\large\else
  \ifnum #1<34\Large\else \ifnum #1<41\LARGE\else
     \huge\fi\fi\fi\fi\fi\fi
  \csname #3\endcsname}%
\else
\gdef\SetFigFont#1#2#3{\begingroup
  \count@#1\relax \ifnum 25<\count@\count@25\fi
  \def\x{\endgroup\@setsize\SetFigFont{#2pt}}%
  \expandafter\x
    \csname \romannumeral\the\count@ pt\expandafter\endcsname
    \csname @\romannumeral\the\count@ pt\endcsname
  \csname #3\endcsname}%
\fi
\fi\endgroup
\begin{picture}(1824,1892)(20,-1123)
\put(830,-197){\makebox(0,0)[lb]{\smash{\SetFigFont{12}{14.4}{rm}\origin}}}
\put(1006,613){\makebox(0,0)[lb]{\smash{\SetFigFont{12}{14.4}{rm}$q$}}}
\put(1681,-212){\makebox(0,0)[lb]{\smash{\SetFigFont{12}{14.4}{rm}$p$}}}
\end{picture}

%% file: fig_elliptic.pstex_t
\begin{picture}(0,0)%
\epsfig{file=fig_elliptic.pstex}%
\end{picture}%
\setlength{\unitlength}{0.00083300in}%
\begingroup\makeatletter\ifx\SetFigFont\undefined
\def\x#1#2#3#4#5#6#7\relax{\def\x{#1#2#3#4#5#6}}%
\expandafter\x\fmtname xxxxxx\relax \def\y{splain}%
\ifx\x\y   
\gdef\SetFigFont#1#2#3{%
  \ifnum #1<17\tiny\else \ifnum #1<20\small\else
  \ifnum #1<24\normalsize\else \ifnum #1<29\large\else
  \ifnum #1<34\Large\else \ifnum #1<41\LARGE\else
     \huge\fi\fi\fi\fi\fi\fi
  \csname #3\endcsname}%
\else
\gdef\SetFigFont#1#2#3{\begingroup
  \count@#1\relax \ifnum 25<\count@\count@25\fi
  \def\x{\endgroup\@setsize\SetFigFont{#2pt}}%
  \expandafter\x
    \csname \romannumeral\the\count@ pt\expandafter\endcsname
    \csname @\romannumeral\the\count@ pt\endcsname
  \csname #3\endcsname}%
\fi
\fi\endgroup
\begin{picture}(1824,1824)(-11,-973)
\put(820,-166){\makebox(0,0)[lb]{\smash{\SetFigFont{12}{14.4}{rm}\origin}}}
\put(1726, 14){\makebox(0,0)[lb]{\smash{\SetFigFont{12}{14.4}{rm}$p$}}}
\put(976,689){\makebox(0,0)[lb]{\smash{\SetFigFont{12}{14.4}{rm}$q$}}}
\put(325,564){\makebox(0,0)[lb]{\smash{\SetFigFont{12}{14.4}{rm}$\omega$}}}
\end{picture}

%% file: RRR.pstex_t
\begin{picture}(0,0)%
\epsfig{file=RRR.pstex}%
\end{picture}%
\setlength{\unitlength}{0.00087500in}%
\begingroup\makeatletter\ifx\SetFigFont\undefined
\def\x#1#2#3#4#5#6#7\relax{\def\x{#1#2#3#4#5#6}}%
\expandafter\x\fmtname xxxxxx\relax \def\y{splain}%
\ifx\x\y   
\gdef\SetFigFont#1#2#3{%
  \ifnum #1<17\tiny\else \ifnum #1<20\small\else
  \ifnum #1<24\normalsize\else \ifnum #1<29\large\else
  \ifnum #1<34\Large\else \ifnum #1<41\LARGE\else
     \huge\fi\fi\fi\fi\fi\fi
  \csname #3\endcsname}%
\else
\gdef\SetFigFont#1#2#3{\begingroup
  \count@#1\relax \ifnum 25<\count@\count@25\fi
  \def\x{\endgroup\@setsize\SetFigFont{#2pt}}%
  \expandafter\x
    \csname \romannumeral\the\count@ pt\expandafter\endcsname
    \csname @\romannumeral\the\count@ pt\endcsname
  \csname #3\endcsname}%
\fi
\fi\endgroup
\begin{picture}(5115,5955)(46,-5905)
\put(4682,-3212){\makebox(0,0)[lb]{\smash{\SetFigFont{12}{14.4}{rm}: resonant terms}}}
\put(4637,-2312){\makebox(0,0)[lb]{\smash{\SetFigFont{12}{14.4}{rm}: integer coordinates}}}
\put( 46,-4121){\makebox(0,0)[lb]{\smash{\SetFigFont{12}{14.4}{rm}$\ell$}}}
\put(4456,-4741){\makebox(0,0)[lb]{\smash{\SetFigFont{12}{14.4}{rm}$\alpha$}}}
\put(3421,-466){\makebox(0,0)[lb]{\smash{\SetFigFont{12}{14.4}{rm}$R_-$}}}
\put(2836,-106){\makebox(0,0)[lb]{\smash{\SetFigFont{12}{14.4}{rm}$R_0$}}}
\put(1396,-421){\makebox(0,0)[lb]{\smash{\SetFigFont{12}{14.4}{rm}$R_+$}}}
\put(451,-331){\makebox(0,0)[lb]{\smash{\SetFigFont{12}{14.4}{rm}$s$}}}
\put(4681,-466){\makebox(0,0)[lb]{\smash{\SetFigFont{12}{14.4}{rm}$s=\alpha$}}}
\put(451,-4426){\makebox(0,0)[lb]{\smash{\SetFigFont{12}{14.4}{rm}$O$}}}
\put(367,-4104){\makebox(0,0)[lb]{\smash{\SetFigFont{6}{7.2}{rm}3}}}
\end{picture}

%% file: matrix.pstex_t
\begin{picture}(0,0)%
\epsfig{file=matrix.pstex}%
\end{picture}%
\setlength{\unitlength}{0.00087500in}%
\begingroup\makeatletter\ifx\SetFigFont\undefined
\def\x#1#2#3#4#5#6#7\relax{\def\x{#1#2#3#4#5#6}}%
\expandafter\x\fmtname xxxxxx\relax \def\y{splain}%
\ifx\x\y   
\gdef\SetFigFont#1#2#3{%
  \ifnum #1<17\tiny\else \ifnum #1<20\small\else
  \ifnum #1<24\normalsize\else \ifnum #1<29\large\else
  \ifnum #1<34\Large\else \ifnum #1<41\LARGE\else
     \huge\fi\fi\fi\fi\fi\fi
  \csname #3\endcsname}%
\else
\gdef\SetFigFont#1#2#3{\begingroup
  \count@#1\relax \ifnum 25<\count@\count@25\fi
  \def\x{\endgroup\@setsize\SetFigFont{#2pt}}%
  \expandafter\x
    \csname \romannumeral\the\count@ pt\expandafter\endcsname
    \csname @\romannumeral\the\count@ pt\endcsname
  \csname #3\endcsname}%
\fi
\fi\endgroup
\begin{picture}(7290,3646)(114,-3235)
\put(1127,163){\makebox(0,0)[lb]{\smash{\SetFigFont{12}{14.4}{rm}-2}}}
\put(227,-287){\makebox(0,0)[lb]{\smash{\SetFigFont{12}{14.4}{rm}0}}}
\put(227,163){\makebox(0,0)[lb]{\smash{\SetFigFont{12}{14.4}{rm}\impitau}}}
\put(226,-2986){\makebox(0,0)[lb]{\smash{\SetFigFont{12}{14.4}{rm}0}}}
\put(4591,-2536){\makebox(0,0)[lb]{\smash{\SetFigFont{12}{14.4}{rm}$-2k$}}}
\put(1261,-1186){\makebox(0,0)[lb]{\smash{\SetFigFont{12}{14.4}{rm}0}}}
\put(1936,-286){\makebox(0,0)[lb]{\smash{\SetFigFont{12}{14.4}{rm}-4}}}
\put(2026,164){\makebox(0,0)[lb]{\smash{\SetFigFont{12}{14.4}{rm}0}}}
\put(2656,-286){\makebox(0,0)[lb]{\smash{\SetFigFont{12}{14.4}{rm}0}}}
\put(3151,-2536){\makebox(0,0)[lb]{\smash{\SetFigFont{12}{14.4}{rm}0}}}
\put(3736,-2536){\makebox(0,0)[lb]{\smash{\SetFigFont{12}{14.4}{rm}\impitau}}}
\put(4006,-2986){\makebox(0,0)[lb]{\smash{\SetFigFont{12}{14.4}{rm}0}}}
\put(4591,-2986){\makebox(0,0)[lb]{\smash{\SetFigFont{12}{14.4}{rm}\impitau}}}
\put(4771,-2086){\makebox(0,0)[lb]{\smash{\SetFigFont{12}{14.4}{rm}0}}}
\put(4771,164){\makebox(0,0)[lb]{\smash{\SetFigFont{12}{14.4}{rm}0}}}
\put(6346,-1456){\makebox(0,0)[lb]{\smash{\SetFigFont{12}{14.4}{rm}$=\!-\!$}}}
\put(5671,164){\makebox(0,0)[lb]{\smash{\SetFigFont{12}{14.4}{rm}\sigmaz}}}
\put(5581,-286){\makebox(0,0)[lb]{\smash{\SetFigFont{12}{14.4}{rm}\sigmau}}}
\put(5671,-2986){\makebox(0,0)[lb]{\smash{\SetFigFont{12}{14.4}{rm}\sigmak}}}
\put(5581,-1141){\makebox(0,0)[lb]{\smash{\SetFigFont{12}{14.4}{rm}\sigmaa}}}
\put(3736,-1186){\makebox(0,0)[lb]{\smash{\SetFigFont{12}{14.4}{rm}0}}}
\put(2566,-1186){\makebox(0,0)[lb]{\smash{\SetFigFont{12}{14.4}{rm}$-2(\alpha+1)$}}}
\put(1891,-1186){\makebox(0,0)[lb]{\smash{\SetFigFont{12}{14.4}{rm}\impitau}}}
\put(856,-286){\makebox(0,0)[lb]{\smash{\SetFigFont{12}{14.4}{rm}\impitau}}}
\put(6752,-1142){\makebox(0,0)[lb]{\smash{\SetFigFont{12}{14.4}{rm}\ha}}}
\put(6752,-242){\makebox(0,0)[lb]{\smash{\SetFigFont{12}{14.4}{rm}\hu}}}
\put(6887,-2987){\makebox(0,0)[lb]{\smash{\SetFigFont{12}{14.4}{rm}\hk}}}
\put(6887,208){\makebox(0,0)[lb]{\smash{\SetFigFont{12}{14.4}{rm}\hz}}}
\end{picture}

%% file: vextr.pstex_t
\begin{picture}(0,0)%
\epsfig{file=vextr.pstex}%
\end{picture}%
\setlength{\unitlength}{0.00087500in}%
\begingroup\makeatletter\ifx\SetFigFont\undefined
\def\x#1#2#3#4#5#6#7\relax{\def\x{#1#2#3#4#5#6}}%
\expandafter\x\fmtname xxxxxx\relax \def\y{splain}%
\ifx\x\y   
\gdef\SetFigFont#1#2#3{%
  \ifnum #1<17\tiny\else \ifnum #1<20\small\else
  \ifnum #1<24\normalsize\else \ifnum #1<29\large\else
  \ifnum #1<34\Large\else \ifnum #1<41\LARGE\else
     \huge\fi\fi\fi\fi\fi\fi
  \csname #3\endcsname}%
\else
\gdef\SetFigFont#1#2#3{\begingroup
  \count@#1\relax \ifnum 25<\count@\count@25\fi
  \def\x{\endgroup\@setsize\SetFigFont{#2pt}}%
  \expandafter\x
    \csname \romannumeral\the\count@ pt\expandafter\endcsname
    \csname @\romannumeral\the\count@ pt\endcsname
  \csname #3\endcsname}%
\fi
\fi\endgroup
\begin{picture}(7360,3076)(439,-2399)
\put(4478,-1861){\makebox(0,0)[lb]{\smash{\SetFigFont{12}{14.4}{rm}$-(\sgn{a})q$}}}
\put(3197,568){\makebox(0,0)[lb]{\smash{\SetFigFont{12}{14.4}{rm}\vextr}}}
\put(1951,-1861){\makebox(0,0)[lb]{\smash{\SetFigFont{12}{14.4}{rm}$-(\sgn{a})q$}}}
\put(6961,-1861){\makebox(0,0)[lb]{\smash{\SetFigFont{12}{14.4}{rm}$-(\sgn{a})q$}}}
\put(676,569){\makebox(0,0)[lb]{\smash{\SetFigFont{12}{14.4}{rm}\vextr}}}
\put(5717,568){\makebox(0,0)[lb]{\smash{\SetFigFont{12}{14.4}{rm}\vextr}}}
\put(1261,-2363){\makebox(0,0)[lb]{\smash{\SetFigFont{12}{14.4}{rm}\psve}}}
\put(4096,-151){\makebox(0,0)[lb]{\smash{\SetFigFont{12}{14.4}{rm}$\boxed{\ve=0}$}}}
\put(6526,-151){\makebox(0,0)[lb]{\smash{\SetFigFont{12}{14.4}{rm}$\boxed{\frac{\ve}{a}>0}$}}}
\put(1576,-151){\makebox(0,0)[lb]{\smash{\SetFigFont{12}{14.4}{rm}$\boxed{\frac{\ve}{a}<0}$}}}
\put(6601,-1883){\makebox(0,0)[lb]{\smash{\SetFigFont{12}{14.4}{rm}0}}}
\put(4073,-1875){\makebox(0,0)[lb]{\smash{\SetFigFont{12}{14.4}{rm}0}}}
\put(1508,-1890){\makebox(0,0)[lb]{\smash{\SetFigFont{12}{14.4}{rm}0}}}
\end{picture}

%% file: fig_1eposext.pstex_t
\begin{picture}(0,0)%
\epsfig{file=fig_1eposext.pstex}%
\end{picture}%
\setlength{\unitlength}{0.00087500in}%
\begingroup\makeatletter\ifx\SetFigFont\undefined
\def\x#1#2#3#4#5#6#7\relax{\def\x{#1#2#3#4#5#6}}%
\expandafter\x\fmtname xxxxxx\relax \def\y{splain}%
\ifx\x\y   
\gdef\SetFigFont#1#2#3{%
  \ifnum #1<17\tiny\else \ifnum #1<20\small\else
  \ifnum #1<24\normalsize\else \ifnum #1<29\large\else
  \ifnum #1<34\Large\else \ifnum #1<41\LARGE\else
     \huge\fi\fi\fi\fi\fi\fi
  \csname #3\endcsname}%
\else
\gdef\SetFigFont#1#2#3{\begingroup
  \count@#1\relax \ifnum 25<\count@\count@25\fi
  \def\x{\endgroup\@setsize\SetFigFont{#2pt}}%
  \expandafter\x
    \csname \romannumeral\the\count@ pt\expandafter\endcsname
    \csname @\romannumeral\the\count@ pt\endcsname
  \csname #3\endcsname}%
\fi
\fi\endgroup
\begin{picture}(2737,2956)(-11,-2121)
\put(1557,-176){\makebox(0,0)[lb]{\smash{\SetFigFont{9}{10.8}{rm}\fqp}}}
\put(1470,574){\makebox(0,0)[lb]{\smash{\SetFigFont{17}{20.4}{rm}$q$}}}
\put(1587,-573){\makebox(0,0)[lb]{\smash{\SetFigFont{9}{10.8}{rm}\origin}}}
\put(348,-586){\makebox(0,0)[lb]{\smash{\SetFigFont{9}{10.8}{rm}\psve}}}
\put(1126,-2097){\makebox(0,0)[lb]{\smash{\SetFigFont{9}{10.8}{rm}\pvetq}}}
\put(2158,-337){\makebox(0,0)[lb]{\smash{\SetFigFont{17}{20.4}{rm}$p$}}}
\put(1587,-1000){\makebox(0,0)[lb]{\smash{\SetFigFont{9}{10.8}{rm}\fqm}}}
\end{picture}

%% file: fig_1enulext.pstex_t
\begin{picture}(0,0)%
\epsfig{file=fig_1enulext.pstex}%
\end{picture}%
\setlength{\unitlength}{0.00087500in}%
\begingroup\makeatletter\ifx\SetFigFont\undefined
\def\x#1#2#3#4#5#6#7\relax{\def\x{#1#2#3#4#5#6}}%
\expandafter\x\fmtname xxxxxx\relax \def\y{splain}%
\ifx\x\y   
\gdef\SetFigFont#1#2#3{%
  \ifnum #1<17\tiny\else \ifnum #1<20\small\else
  \ifnum #1<24\normalsize\else \ifnum #1<29\large\else
  \ifnum #1<34\Large\else \ifnum #1<41\LARGE\else
     \huge\fi\fi\fi\fi\fi\fi
  \csname #3\endcsname}%
\else
\gdef\SetFigFont#1#2#3{\begingroup
  \count@#1\relax \ifnum 25<\count@\count@25\fi
  \def\x{\endgroup\@setsize\SetFigFont{#2pt}}%
  \expandafter\x
    \csname \romannumeral\the\count@ pt\expandafter\endcsname
    \csname @\romannumeral\the\count@ pt\endcsname
  \csname #3\endcsname}%
\fi
\fi\endgroup
\begin{picture}(3121,3122)(45,-2316)
\put(1760,-759){\makebox(0,0)[lb]{\smash{\SetFigFont{9}{10.8}{rm}\special{ps: gsave 0 0 0 setrgbcolor}\origin\special{ps: grestore}}}}
\put(2698,-597){\makebox(0,0)[lb]{\smash{\SetFigFont{14}{16.8}{rm}$p$}}}
\put(1698,423){\makebox(0,0)[lb]{\smash{\SetFigFont{14}{16.8}{rm}$q$}}}
\end{picture}

%% file: fig_1enegext.pstex_t
\begin{picture}(0,0)%
\epsfig{file=fig_1enegext.pstex}%
\end{picture}%
\setlength{\unitlength}{0.00087500in}%
\begingroup\makeatletter\ifx\SetFigFont\undefined
\def\x#1#2#3#4#5#6#7\relax{\def\x{#1#2#3#4#5#6}}%
\expandafter\x\fmtname xxxxxx\relax \def\y{splain}%
\ifx\x\y   
\gdef\SetFigFont#1#2#3{%
  \ifnum #1<17\tiny\else \ifnum #1<20\small\else
  \ifnum #1<24\normalsize\else \ifnum #1<29\large\else
  \ifnum #1<34\Large\else \ifnum #1<41\LARGE\else
     \huge\fi\fi\fi\fi\fi\fi
  \csname #3\endcsname}%
\else
\gdef\SetFigFont#1#2#3{\begingroup
  \count@#1\relax \ifnum 25<\count@\count@25\fi
  \def\x{\endgroup\@setsize\SetFigFont{#2pt}}%
  \expandafter\x
    \csname \romannumeral\the\count@ pt\expandafter\endcsname
    \csname @\romannumeral\the\count@ pt\endcsname
  \csname #3\endcsname}%
\fi
\fi\endgroup
\begin{picture}(2536,2535)(34,-1755)
\put(1459,-475){\makebox(0,0)[lb]{\smash{\SetFigFont{8}{9.6}{rm}\special{ps: gsave 0 0 0 setrgbcolor}\origin\special{ps: grestore}}}}
\put(2216,-362){\makebox(0,0)[lb]{\smash{\SetFigFont{17}{20.4}{rm}$p$}}}
\put(1407,351){\makebox(0,0)[lb]{\smash{\SetFigFont{17}{20.4}{rm}$q$}}}
\end{picture}

%% file: vtran.pstex_t
\begin{picture}(0,0)%
\epsfig{file=vtran.pstex}%
\end{picture}%
\setlength{\unitlength}{0.00087500in}%
\begingroup\makeatletter\ifx\SetFigFont\undefined
\def\x#1#2#3#4#5#6#7\relax{\def\x{#1#2#3#4#5#6}}%
\expandafter\x\fmtname xxxxxx\relax \def\y{splain}%
\ifx\x\y   
\gdef\SetFigFont#1#2#3{%
  \ifnum #1<17\tiny\else \ifnum #1<20\small\else
  \ifnum #1<24\normalsize\else \ifnum #1<29\large\else
  \ifnum #1<34\Large\else \ifnum #1<41\LARGE\else
     \huge\fi\fi\fi\fi\fi\fi
  \csname #3\endcsname}%
\else
\gdef\SetFigFont#1#2#3{\begingroup
  \count@#1\relax \ifnum 25<\count@\count@25\fi
  \def\x{\endgroup\@setsize\SetFigFont{#2pt}}%
  \expandafter\x
    \csname \romannumeral\the\count@ pt\expandafter\endcsname
    \csname @\romannumeral\the\count@ pt\endcsname
  \csname #3\endcsname}%
\fi
\fi\endgroup
\begin{picture}(7360,2866)(439,-2189)
\put(3197,568){\makebox(0,0)[lb]{\smash{\SetFigFont{12}{14.4}{rm}\vtran}}}
\put(7637,-1886){\makebox(0,0)[lb]{\smash{\SetFigFont{12}{14.4}{rm}$q$}}}
\put(5717,568){\makebox(0,0)[lb]{\smash{\SetFigFont{12}{14.4}{rm}\vtran}}}
\put(676,569){\makebox(0,0)[lb]{\smash{\SetFigFont{12}{14.4}{rm}\vtran}}}
\put(2596,-1886){\makebox(0,0)[lb]{\smash{\SetFigFont{12}{14.4}{rm}$q$}}}
\put(5117,-1886){\makebox(0,0)[lb]{\smash{\SetFigFont{12}{14.4}{rm}$q$}}}
\put(1298,-2153){\makebox(0,0)[lb]{\smash{\SetFigFont{12}{14.4}{rm}\psve}}}
\put(6654,-1838){\makebox(0,0)[lb]{\smash{\SetFigFont{12}{14.4}{rm}0}}}
\put(1576,-1838){\makebox(0,0)[lb]{\smash{\SetFigFont{12}{14.4}{rm}0}}}
\put(4104,-1861){\makebox(0,0)[lb]{\smash{\SetFigFont{12}{14.4}{rm}0}}}
\put(1261,-1141){\makebox(0,0)[lb]{\smash{\SetFigFont{12}{14.4}{rm}$\boxed{\frac{\ve}{a}<0}$}}}
\put(3443,-1141){\makebox(0,0)[lb]{\smash{\SetFigFont{12}{14.4}{rm}$\boxed{\ve=0}$}}}
\put(5978,-1186){\makebox(0,0)[lb]{\smash{\SetFigFont{12}{14.4}{rm}$\boxed{\frac{\ve}{a}>0}$}}}
\end{picture}

%% file: fig_2eposext.pstex_t
\begin{picture}(0,0)%
\epsfig{file=fig_2eposext.pstex}%
\end{picture}%
\setlength{\unitlength}{0.00087500in}%
\begingroup\makeatletter\ifx\SetFigFont\undefined
\def\x#1#2#3#4#5#6#7\relax{\def\x{#1#2#3#4#5#6}}%
\expandafter\x\fmtname xxxxxx\relax \def\y{splain}%
\ifx\x\y   
\gdef\SetFigFont#1#2#3{%
  \ifnum #1<17\tiny\else \ifnum #1<20\small\else
  \ifnum #1<24\normalsize\else \ifnum #1<29\large\else
  \ifnum #1<34\Large\else \ifnum #1<41\LARGE\else
     \huge\fi\fi\fi\fi\fi\fi
  \csname #3\endcsname}%
\else
\gdef\SetFigFont#1#2#3{\begingroup
  \count@#1\relax \ifnum 25<\count@\count@25\fi
  \def\x{\endgroup\@setsize\SetFigFont{#2pt}}%
  \expandafter\x
    \csname \romannumeral\the\count@ pt\expandafter\endcsname
    \csname @\romannumeral\the\count@ pt\endcsname
  \csname #3\endcsname}%
\fi
\fi\endgroup
\begin{picture}(2704,2898)(35,-2123)
\put(1289,-2099){\makebox(0,0)[lb]{\smash{\SetFigFont{8}{9.6}{rm}\pve}}}
\put(427,-669){\makebox(0,0)[lb]{\smash{\SetFigFont{8}{9.6}{rm}\psve}}}
\put(1453,-1300){\makebox(0,0)[lb]{\smash{\SetFigFont{8}{9.6}{rm}\special{ps: gsave 0 0 0 setrgbcolor}\fqm\special{ps: grestore}}}}
\put(1464,-64){\makebox(0,0)[lb]{\smash{\SetFigFont{8}{9.6}{rm}\special{ps: gsave 0 0 0 setrgbcolor}\fqp\special{ps: grestore}}}}
\put(1530,-688){\makebox(0,0)[lb]{\smash{\SetFigFont{8}{9.6}{rm}\special{ps: gsave 0 0 0 setrgbcolor}\origin\special{ps: grestore}}}}
\put(1491,559){\makebox(0,0)[lb]{\smash{\SetFigFont{17}{20.4}{rm}$q$}}}
\put(2045,-487){\makebox(0,0)[lb]{\smash{\SetFigFont{17}{20.4}{rm}$p$}}}
\end{picture}

%% file: fig_2eposbisext.pstex_t
\begin{picture}(0,0)%
\epsfig{file=fig_2eposbisext.pstex}%
\end{picture}%
\setlength{\unitlength}{0.00087500in}%
\begingroup\makeatletter\ifx\SetFigFont\undefined
\def\x#1#2#3#4#5#6#7\relax{\def\x{#1#2#3#4#5#6}}%
\expandafter\x\fmtname xxxxxx\relax \def\y{splain}%
\ifx\x\y   
\gdef\SetFigFont#1#2#3{%
  \ifnum #1<17\tiny\else \ifnum #1<20\small\else
  \ifnum #1<24\normalsize\else \ifnum #1<29\large\else
  \ifnum #1<34\Large\else \ifnum #1<41\LARGE\else
     \huge\fi\fi\fi\fi\fi\fi
  \csname #3\endcsname}%
\else
\gdef\SetFigFont#1#2#3{\begingroup
  \count@#1\relax \ifnum 25<\count@\count@25\fi
  \def\x{\endgroup\@setsize\SetFigFont{#2pt}}%
  \expandafter\x
    \csname \romannumeral\the\count@ pt\expandafter\endcsname
    \csname @\romannumeral\the\count@ pt\endcsname
  \csname #3\endcsname}%
\fi
\fi\endgroup
\begin{picture}(2705,2846)(34,-2295)
\put(1321,-2271){\makebox(0,0)[lb]{\smash{\SetFigFont{8}{9.6}{rm}\pve}}}
\put(449,-888){\makebox(0,0)[lb]{\smash{\SetFigFont{8}{9.6}{rm}\psve}}}
\put(1481,-1496){\makebox(0,0)[lb]{\smash{\SetFigFont{8}{9.6}{rm}\special{ps: gsave 0 0 0 setrgbcolor}\fqm\special{ps: grestore}}}}
\put(1454,335){\makebox(0,0)[lb]{\smash{\SetFigFont{17}{20.4}{rm}$q$}}}
\put(2043,-736){\makebox(0,0)[lb]{\smash{\SetFigFont{17}{20.4}{rm}$p$}}}
\put(1481,-877){\makebox(0,0)[lb]{\smash{\SetFigFont{8}{9.6}{rm}\special{ps: gsave 0 0 0 setrgbcolor}\origin\special{ps: grestore}}}}
\put(1481,-286){\makebox(0,0)[lb]{\smash{\SetFigFont{8}{9.6}{rm}\special{ps: gsave 0 0 0 setrgbcolor}\fqp\special{ps: grestore}}}}
\end{picture}

%% file: fig_2enulext.pstex_t
\begin{picture}(0,0)%
\epsfig{file=fig_2enulext.pstex}%
\end{picture}%
\setlength{\unitlength}{0.00087500in}%
\begingroup\makeatletter\ifx\SetFigFont\undefined
\def\x#1#2#3#4#5#6#7\relax{\def\x{#1#2#3#4#5#6}}%
\expandafter\x\fmtname xxxxxx\relax \def\y{splain}%
\ifx\x\y   
\gdef\SetFigFont#1#2#3{%
  \ifnum #1<17\tiny\else \ifnum #1<20\small\else
  \ifnum #1<24\normalsize\else \ifnum #1<29\large\else
  \ifnum #1<34\Large\else \ifnum #1<41\LARGE\else
     \huge\fi\fi\fi\fi\fi\fi
  \csname #3\endcsname}%
\else
\gdef\SetFigFont#1#2#3{\begingroup
  \count@#1\relax \ifnum 25<\count@\count@25\fi
  \def\x{\endgroup\@setsize\SetFigFont{#2pt}}%
  \expandafter\x
    \csname \romannumeral\the\count@ pt\expandafter\endcsname
    \csname @\romannumeral\the\count@ pt\endcsname
  \csname #3\endcsname}%
\fi
\fi\endgroup
\begin{picture}(2596,2697)(-11,-2494)
\put(1429,-1211){\makebox(0,0)[lb]{\smash{\SetFigFont{10}{12.0}{rm}\special{ps: gsave 0 0 0 setrgbcolor}\origin\special{ps: grestore}}}}
\put(2251,-1067){\makebox(0,0)[lb]{\smash{\SetFigFont{17}{20.4}{rm}$p$}}}
\put(1394,-38){\makebox(0,0)[lb]{\smash{\SetFigFont{17}{20.4}{rm}$q$}}}
\end{picture}

%% file: fig_2enegext.pstex_t
\begin{picture}(0,0)%
\epsfig{file=fig_2enegext.pstex}%
\end{picture}%
\setlength{\unitlength}{0.00087500in}%
\begingroup\makeatletter\ifx\SetFigFont\undefined
\def\x#1#2#3#4#5#6#7\relax{\def\x{#1#2#3#4#5#6}}%
\expandafter\x\fmtname xxxxxx\relax \def\y{splain}%
\ifx\x\y   
\gdef\SetFigFont#1#2#3{%
  \ifnum #1<17\tiny\else \ifnum #1<20\small\else
  \ifnum #1<24\normalsize\else \ifnum #1<29\large\else
  \ifnum #1<34\Large\else \ifnum #1<41\LARGE\else
     \huge\fi\fi\fi\fi\fi\fi
  \csname #3\endcsname}%
\else
\gdef\SetFigFont#1#2#3{\begingroup
  \count@#1\relax \ifnum 25<\count@\count@25\fi
  \def\x{\endgroup\@setsize\SetFigFont{#2pt}}%
  \expandafter\x
    \csname \romannumeral\the\count@ pt\expandafter\endcsname
    \csname @\romannumeral\the\count@ pt\endcsname
  \csname #3\endcsname}%
\fi
\fi\endgroup
\begin{picture}(2762,2840)(-11,-2039)
\put(1488,-671){\makebox(0,0)[lb]{\smash{\SetFigFont{7}{8.4}{rm}\special{ps: gsave 0 0 0 setrgbcolor}\origin\special{ps: grestore}}}}
\put(2274,-502){\makebox(0,0)[lb]{\smash{\SetFigFont{17}{20.4}{rm}$p$}}}
\put(1484,561){\makebox(0,0)[lb]{\smash{\SetFigFont{17}{20.4}{rm}$q$}}}
\end{picture}

%% file: fig_2enegbisext.pstex_t
\begin{picture}(0,0)%
\epsfig{file=fig_2enegbisext.pstex}%
\end{picture}%
\setlength{\unitlength}{0.00087500in}%
\begingroup\makeatletter\ifx\SetFigFont\undefined
\def\x#1#2#3#4#5#6#7\relax{\def\x{#1#2#3#4#5#6}}%
\expandafter\x\fmtname xxxxxx\relax \def\y{splain}%
\ifx\x\y   
\gdef\SetFigFont#1#2#3{%
  \ifnum #1<17\tiny\else \ifnum #1<20\small\else
  \ifnum #1<24\normalsize\else \ifnum #1<29\large\else
  \ifnum #1<34\Large\else \ifnum #1<41\LARGE\else
     \huge\fi\fi\fi\fi\fi\fi
  \csname #3\endcsname}%
\else
\gdef\SetFigFont#1#2#3{\begingroup
  \count@#1\relax \ifnum 25<\count@\count@25\fi
  \def\x{\endgroup\@setsize\SetFigFont{#2pt}}%
  \expandafter\x
    \csname \romannumeral\the\count@ pt\expandafter\endcsname
    \csname @\romannumeral\the\count@ pt\endcsname
  \csname #3\endcsname}%
\fi
\fi\endgroup
\begin{picture}(2597,2688)(-11,-1858)
\put(1270,-457){\makebox(0,0)[lb]{\smash{\SetFigFont{7}{8.4}{rm}\special{ps: gsave 1 1 1 setrgbcolor}\origin\special{ps: grestore}}}}
\put(1352,599){\makebox(0,0)[lb]{\smash{\SetFigFont{17}{20.4}{rm}$q$}}}
\put(2253,-431){\makebox(0,0)[lb]{\smash{\SetFigFont{17}{20.4}{rm}$p$}}}
\end{picture}

%% file: fig_3eposext.pstex_t
\begin{picture}(0,0)%
\epsfig{file=fig_3eposext.pstex}%
\end{picture}%
\setlength{\unitlength}{0.00087500in}%
\begingroup\makeatletter\ifx\SetFigFont\undefined
\def\x#1#2#3#4#5#6#7\relax{\def\x{#1#2#3#4#5#6}}%
\expandafter\x\fmtname xxxxxx\relax \def\y{splain}%
\ifx\x\y   
\gdef\SetFigFont#1#2#3{%
  \ifnum #1<17\tiny\else \ifnum #1<20\small\else
  \ifnum #1<24\normalsize\else \ifnum #1<29\large\else
  \ifnum #1<34\Large\else \ifnum #1<41\LARGE\else
     \huge\fi\fi\fi\fi\fi\fi
  \csname #3\endcsname}%
\else
\gdef\SetFigFont#1#2#3{\begingroup
  \count@#1\relax \ifnum 25<\count@\count@25\fi
  \def\x{\endgroup\@setsize\SetFigFont{#2pt}}%
  \expandafter\x
    \csname \romannumeral\the\count@ pt\expandafter\endcsname
    \csname @\romannumeral\the\count@ pt\endcsname
  \csname #3\endcsname}%
\fi
\fi\endgroup
\begin{picture}(2570,2709)(98,-2046)
\put(186,-1030){\makebox(0,0)[lb]{\smash{\special{ps:gsave currentpoint currentpoint translate
-0.0 rotate neg exch neg exch translate}\SetFigFont{7}{8.4}{rm}\pve\special{ps:currentpoint grestore moveto}}}}
\put(1108,-802){\makebox(0,0)[lb]{\smash{\special{ps:gsave currentpoint currentpoint translate
-0.0 rotate neg exch neg exch translate}\SetFigFont{7}{8.4}{rm}\origin\special{ps:currentpoint grestore moveto}}}}
\put(1084,-234){\makebox(0,0)[lb]{\smash{\special{ps:gsave currentpoint currentpoint translate
-0.0 rotate neg exch neg exch translate}\SetFigFont{7}{8.4}{rm}\fqm\special{ps:currentpoint grestore moveto}}}}
\put(1121,-1302){\makebox(0,0)[lb]{\smash{\special{ps:gsave currentpoint currentpoint translate
-0.0 rotate neg exch neg exch translate}\SetFigFont{7}{8.4}{rm}\fqp\special{ps:currentpoint grestore moveto}}}}
\put(1801,-772){\makebox(0,0)[lb]{\smash{\special{ps:gsave currentpoint currentpoint translate
-0.0 rotate neg exch neg exch translate}\SetFigFont{7}{8.4}{rm}\fqz\special{ps:currentpoint grestore moveto}}}}
\put(1321,279){\makebox(0,0)[lb]{\smash{\SetFigFont{17}{20.4}{rm}$q$}}}
\put(2358,-654){\makebox(0,0)[lb]{\smash{\SetFigFont{17}{20.4}{rm}$-\sgn(\frac{\ve}{a})p$}}}
\end{picture}

%% file: fig_3enulext.pstex_t
\begin{picture}(0,0)%
\epsfig{file=fig_3enulext.pstex}%
\end{picture}%
\setlength{\unitlength}{0.00087500in}%
\begingroup\makeatletter\ifx\SetFigFont\undefined
\def\x#1#2#3#4#5#6#7\relax{\def\x{#1#2#3#4#5#6}}%
\expandafter\x\fmtname xxxxxx\relax \def\y{splain}%
\ifx\x\y   
\gdef\SetFigFont#1#2#3{%
  \ifnum #1<17\tiny\else \ifnum #1<20\small\else
  \ifnum #1<24\normalsize\else \ifnum #1<29\large\else
  \ifnum #1<34\Large\else \ifnum #1<41\LARGE\else
     \huge\fi\fi\fi\fi\fi\fi
  \csname #3\endcsname}%
\else
\gdef\SetFigFont#1#2#3{\begingroup
  \count@#1\relax \ifnum 25<\count@\count@25\fi
  \def\x{\endgroup\@setsize\SetFigFont{#2pt}}%
  \expandafter\x
    \csname \romannumeral\the\count@ pt\expandafter\endcsname
    \csname @\romannumeral\the\count@ pt\endcsname
  \csname #3\endcsname}%
\fi
\fi\endgroup
\begin{picture}(2674,2775)(48,-2281)
\put(2341,-781){\makebox(0,0)[lb]{\smash{\SetFigFont{17}{20.4}{rm}$p$}}}
\put(1470,278){\makebox(0,0)[lb]{\smash{\special{ps:gsave currentpoint currentpoint translate
-0.0 rotate neg exch neg exch translate}\SetFigFont{17}{20.4}{rm}$q$\special{ps:currentpoint grestore moveto}}}}
\put(1532,-964){\makebox(0,0)[lb]{\smash{\special{ps:gsave currentpoint currentpoint translate
-0.0 rotate neg exch neg exch translate}\SetFigFont{12}{14.4}{rm}\origin\special{ps:currentpoint grestore moveto}}}}
\end{picture}

%% file: fig_4eposext.pstex_t
\begin{picture}(0,0)%
\epsfig{file=fig_4eposext.pstex}%
\end{picture}%
\setlength{\unitlength}{0.00087500in}%
\begingroup\makeatletter\ifx\SetFigFont\undefined
\def\x#1#2#3#4#5#6#7\relax{\def\x{#1#2#3#4#5#6}}%
\expandafter\x\fmtname xxxxxx\relax \def\y{splain}%
\ifx\x\y   
\gdef\SetFigFont#1#2#3{%
  \ifnum #1<17\tiny\else \ifnum #1<20\small\else
  \ifnum #1<24\normalsize\else \ifnum #1<29\large\else
  \ifnum #1<34\Large\else \ifnum #1<41\LARGE\else
     \huge\fi\fi\fi\fi\fi\fi
  \csname #3\endcsname}%
\else
\gdef\SetFigFont#1#2#3{\begingroup
  \count@#1\relax \ifnum 25<\count@\count@25\fi
  \def\x{\endgroup\@setsize\SetFigFont{#2pt}}%
  \expandafter\x
    \csname \romannumeral\the\count@ pt\expandafter\endcsname
    \csname @\romannumeral\the\count@ pt\endcsname
  \csname #3\endcsname}%
\fi
\fi\endgroup
\begin{picture}(4128,4128)(34,-3343)
\put(482,-1624){\makebox(0,0)[lb]{\smash{\special{ps:gsave currentpoint currentpoint translate
-40.0 rotate neg exch neg exch translate}\SetFigFont{12}{14.4}{rm}\psve\special{ps:currentpoint grestore moveto}}}}
\put(2197,-2927){\makebox(0,0)[lb]{\smash{\SetFigFont{12}{14.4}{rm}\psve}}}
\put(2811,-672){\makebox(0,0)[lb]{\smash{\SetFigFont{12}{14.4}{rm}\special{ps: gsave 0 0 0 setrgbcolor}\fppqp\special{ps: grestore}}}}
\put(2806,-1897){\makebox(0,0)[lb]{\smash{\SetFigFont{12}{14.4}{rm}\special{ps: gsave 0 0 0 setrgbcolor}\fppqm\special{ps: grestore}}}}
\put(1265,-1900){\makebox(0,0)[lb]{\smash{\SetFigFont{12}{14.4}{rm}\special{ps: gsave 0 0 0 setrgbcolor}\fpmqm\special{ps: grestore}}}}
\put(1372,-1508){\makebox(0,0)[lb]{\smash{\SetFigFont{12}{14.4}{rm}\special{ps: gsave 0 0 0 setrgbcolor}\fpm\special{ps: grestore}}}}
\put(1804,-1825){\makebox(0,0)[lb]{\smash{\SetFigFont{12}{14.4}{rm}\special{ps: gsave 0 0 0 setrgbcolor}\fqm\special{ps: grestore}}}}
\put(1782,-812){\makebox(0,0)[lb]{\smash{\SetFigFont{12}{14.4}{rm}\special{ps: gsave 0 0 0 setrgbcolor}\fqp\special{ps: grestore}}}}
\put(2713,-1482){\makebox(0,0)[lb]{\smash{\SetFigFont{12}{14.4}{rm}\special{ps: gsave 0 0 0 setrgbcolor}\fpp\special{ps: grestore}}}}
\put(3846,-1167){\makebox(0,0)[lb]{\smash{\SetFigFont{29}{34.8}{rm}$p$}}}
\put(1278,-675){\makebox(0,0)[lb]{\smash{\SetFigFont{12}{14.4}{rm}\special{ps: gsave 0 0 0 setrgbcolor}\fpmqp\special{ps: grestore}}}}
\put(2257,102){\makebox(0,0)[lb]{\smash{\SetFigFont{29}{34.8}{rm}$q$}}}
\put(2191,-1337){\makebox(0,0)[lb]{\smash{\SetFigFont{12}{14.4}{rm}\special{ps: gsave 0 0 0 setrgbcolor}\origin\special{ps: grestore}}}}
\end{picture}

%% file: fig_4eposbisext.pstex_t
\begin{picture}(0,0)%
\epsfig{file=fig_4eposbisext.pstex}%
\end{picture}%
\setlength{\unitlength}{0.00087500in}%
\begingroup\makeatletter\ifx\SetFigFont\undefined
\def\x#1#2#3#4#5#6#7\relax{\def\x{#1#2#3#4#5#6}}%
\expandafter\x\fmtname xxxxxx\relax \def\y{splain}%
\ifx\x\y   
\gdef\SetFigFont#1#2#3{%
  \ifnum #1<17\tiny\else \ifnum #1<20\small\else
  \ifnum #1<24\normalsize\else \ifnum #1<29\large\else
  \ifnum #1<34\Large\else \ifnum #1<41\LARGE\else
     \huge\fi\fi\fi\fi\fi\fi
  \csname #3\endcsname}%
\else
\gdef\SetFigFont#1#2#3{\begingroup
  \count@#1\relax \ifnum 25<\count@\count@25\fi
  \def\x{\endgroup\@setsize\SetFigFont{#2pt}}%
  \expandafter\x
    \csname \romannumeral\the\count@ pt\expandafter\endcsname
    \csname @\romannumeral\the\count@ pt\endcsname
  \csname #3\endcsname}%
\fi
\fi\endgroup
\begin{picture}(4069,4070)(-11,-3430)
\put(990,-2585){\makebox(0,0)[lb]{\smash{\special{ps:gsave currentpoint currentpoint translate
-37.0 rotate neg exch neg exch translate}\SetFigFont{12}{14.4}{rm}\psve\special{ps:currentpoint grestore moveto}}}}
\put(1955,-878){\makebox(0,0)[lb]{\smash{\SetFigFont{12}{14.4}{rm}\special{ps: gsave 0 0 0 setrgbcolor}\fqp\special{ps: grestore}}}}
\put(1946,-1862){\makebox(0,0)[lb]{\smash{\SetFigFont{12}{14.4}{rm}\special{ps: gsave 0 0 0 setrgbcolor}\fqm\special{ps: grestore}}}}
\put(2129,-1247){\makebox(0,0)[lb]{\smash{\SetFigFont{12}{14.4}{rm}\special{ps: gsave 0 0 0 setrgbcolor}\origin\special{ps: grestore}}}}
\put(1434,-1437){\makebox(0,0)[lb]{\smash{\SetFigFont{12}{14.4}{rm}\special{ps: gsave 0 0 0 setrgbcolor}\fpm\special{ps: grestore}}}}
\put(2505,-1433){\makebox(0,0)[lb]{\smash{\SetFigFont{12}{14.4}{rm}\special{ps: gsave 0 0 0 setrgbcolor}\fpp\special{ps: grestore}}}}
\put(3781,-1234){\makebox(0,0)[lb]{\smash{\SetFigFont{29}{34.8}{rm}$p$}}}
\put(2210,195){\makebox(0,0)[lb]{\smash{\SetFigFont{29}{34.8}{rm}$q$}}}
\end{picture}

%% file: fig_4enulbisext.pstex_t
\begin{picture}(0,0)%
\epsfig{file=fig_4enulbisext.pstex}%
\end{picture}%
\setlength{\unitlength}{0.00087500in}%
\begingroup\makeatletter\ifx\SetFigFont\undefined
\def\x#1#2#3#4#5#6#7\relax{\def\x{#1#2#3#4#5#6}}%
\expandafter\x\fmtname xxxxxx\relax \def\y{splain}%
\ifx\x\y   
\gdef\SetFigFont#1#2#3{%
  \ifnum #1<17\tiny\else \ifnum #1<20\small\else
  \ifnum #1<24\normalsize\else \ifnum #1<29\large\else
  \ifnum #1<34\Large\else \ifnum #1<41\LARGE\else
     \huge\fi\fi\fi\fi\fi\fi
  \csname #3\endcsname}%
\else
\gdef\SetFigFont#1#2#3{\begingroup
  \count@#1\relax \ifnum 25<\count@\count@25\fi
  \def\x{\endgroup\@setsize\SetFigFont{#2pt}}%
  \expandafter\x
    \csname \romannumeral\the\count@ pt\expandafter\endcsname
    \csname @\romannumeral\the\count@ pt\endcsname
  \csname #3\endcsname}%
\fi
\fi\endgroup
\begin{picture}(2683,2685)(34,-1992)
\put(1490,350){\makebox(0,0)[lb]{\smash{\SetFigFont{17}{20.4}{rm}$q$}}}
\put(1327,-496){\makebox(0,0)[lb]{\smash{\SetFigFont{7}{8.4}{rm}\special{ps: gsave 0 0 0 setrgbcolor}\origin\special{ps: grestore}}}}
\put(2554,-524){\makebox(0,0)[lb]{\smash{\SetFigFont{17}{20.4}{rm}$p$}}}
\end{picture}

%% file: fig_4enegbisext.pstex_t
\begin{picture}(0,0)%
\epsfig{file=fig_4enegbisext.pstex}%
\end{picture}%
\setlength{\unitlength}{0.00087500in}%
\begingroup\makeatletter\ifx\SetFigFont\undefined
\def\x#1#2#3#4#5#6#7\relax{\def\x{#1#2#3#4#5#6}}%
\expandafter\x\fmtname xxxxxx\relax \def\y{splain}%
\ifx\x\y   
\gdef\SetFigFont#1#2#3{%
  \ifnum #1<17\tiny\else \ifnum #1<20\small\else
  \ifnum #1<24\normalsize\else \ifnum #1<29\large\else
  \ifnum #1<34\Large\else \ifnum #1<41\LARGE\else
     \huge\fi\fi\fi\fi\fi\fi
  \csname #3\endcsname}%
\else
\gdef\SetFigFont#1#2#3{\begingroup
  \count@#1\relax \ifnum 25<\count@\count@25\fi
  \def\x{\endgroup\@setsize\SetFigFont{#2pt}}%
  \expandafter\x
    \csname \romannumeral\the\count@ pt\expandafter\endcsname
    \csname @\romannumeral\the\count@ pt\endcsname
  \csname #3\endcsname}%
\fi
\fi\endgroup
\begin{picture}(2763,2765)(34,-2115)
\put(1517,-1935){\makebox(0,0)[lb]{\smash{\SetFigFont{7}{8.4}{rm}\psve}}}
\put(1512,170){\makebox(0,0)[lb]{\smash{\SetFigFont{17}{20.4}{rm}$q$}}}
\put(2575,-578){\makebox(0,0)[lb]{\smash{\SetFigFont{17}{20.4}{rm}$p$}}}
\put(1483,-779){\makebox(0,0)[lb]{\smash{\SetFigFont{7}{8.4}{rm}\special{ps: gsave 0 0 0 setrgbcolor}\origin\special{ps: grestore}}}}
\put(948,-1025){\makebox(0,0)[lb]{\smash{\SetFigFont{7}{8.4}{rm}\special{ps: gsave 0 0 0 setrgbcolor}\fpmqm\special{ps: grestore}}}}
\put(1765,-1017){\makebox(0,0)[lb]{\smash{\SetFigFont{7}{8.4}{rm}\special{ps: gsave 0 0 0 setrgbcolor}\fpmqp\special{ps: grestore}}}}
\put(1775,-461){\makebox(0,0)[lb]{\smash{\SetFigFont{7}{8.4}{rm}\special{ps: gsave 0 0 0 setrgbcolor}\fppqp\special{ps: grestore}}}}
\put(954,-453){\makebox(0,0)[lb]{\smash{\SetFigFont{7}{8.4}{rm}\special{ps: gsave 0 0 0 setrgbcolor}\fppqm\special{ps: grestore}}}}
\end{picture}

%% file: fig_5eposext.pstex_t
\begin{picture}(0,0)%
\epsfig{file=fig_5eposext.pstex}%
\end{picture}%
\setlength{\unitlength}{0.00087500in}%
\begingroup\makeatletter\ifx\SetFigFont\undefined
\def\x#1#2#3#4#5#6#7\relax{\def\x{#1#2#3#4#5#6}}%
\expandafter\x\fmtname xxxxxx\relax \def\y{splain}%
\ifx\x\y   
\gdef\SetFigFont#1#2#3{%
  \ifnum #1<17\tiny\else \ifnum #1<20\small\else
  \ifnum #1<24\normalsize\else \ifnum #1<29\large\else
  \ifnum #1<34\Large\else \ifnum #1<41\LARGE\else
     \huge\fi\fi\fi\fi\fi\fi
  \csname #3\endcsname}%
\else
\gdef\SetFigFont#1#2#3{\begingroup
  \count@#1\relax \ifnum 25<\count@\count@25\fi
  \def\x{\endgroup\@setsize\SetFigFont{#2pt}}%
  \expandafter\x
    \csname \romannumeral\the\count@ pt\expandafter\endcsname
    \csname @\romannumeral\the\count@ pt\endcsname
  \csname #3\endcsname}%
\fi
\fi\endgroup
\begin{picture}(4965,4869)(1560,-4648)
\put(1560,-2111){\makebox(0,0)[lb]{\smash{\SetFigFont{12}{14.4}{rm}\psve}}}
\put(6457,-1759){\makebox(0,0)[lb]{\smash{\SetFigFont{12}{14.4}{rm}\pvek}}}
\put(4489,-439){\makebox(0,0)[lb]{\smash{\SetFigFont{29}{34.8}{rm}$q$}}}
\put(3158,-2818){\makebox(0,0)[lb]{\smash{\SetFigFont{12}{14.4}{rm}\special{ps: gsave 0 0 0 setrgbcolor}\fpt\special{ps: grestore}}}}
\put(3293,-3628){\makebox(0,0)[lb]{\smash{\SetFigFont{12}{14.4}{rm}\special{ps: gsave 0 0 0 setrgbcolor}\fpqu\special{ps: grestore}}}}
\put(4283,-3493){\makebox(0,0)[lb]{\smash{\SetFigFont{12}{14.4}{rm}\special{ps: gsave 0 0 0 setrgbcolor}\fpc\special{ps: grestore}}}}
\put(5273,-3538){\makebox(0,0)[lb]{\smash{\SetFigFont{12}{14.4}{rm}\special{ps: gsave 0 0 0 setrgbcolor}\fps\special{ps: grestore}}}}
\put(5363,-2818){\makebox(0,0)[lb]{\smash{\SetFigFont{12}{14.4}{rm}\special{ps: gsave 0 0 0 setrgbcolor}\fpse\special{ps: grestore}}}}
\put(4418,-2458){\makebox(0,0)[lb]{\smash{\SetFigFont{12}{14.4}{rm}\special{ps: gsave 0 0 0 setrgbcolor}\origin\special{ps: grestore}}}}
\put(5003,-1693){\makebox(0,0)[lb]{\smash{\SetFigFont{12}{14.4}{rm}\special{ps: gsave 0 0 0 setrgbcolor}\fpn\special{ps: grestore}}}}
\put(4298,-981){\makebox(0,0)[lb]{\smash{\SetFigFont{12}{14.4}{rm}\special{ps: gsave 0 0 0 setrgbcolor}\fpz\special{ps: grestore}}}}
\put(2895,-1903){\makebox(0,0)[lb]{\smash{\SetFigFont{12}{14.4}{rm}\special{ps: gsave 0 0 0 setrgbcolor}\fpd\special{ps: grestore}}}}
\put(6211,-2938){\makebox(0,0)[lb]{\smash{\SetFigFont{29}{34.8}{rm}$p$}}}
\put(5783,-2263){\makebox(0,0)[lb]{\smash{\SetFigFont{12}{14.4}{rm}\special{ps: gsave 0 0 0 setrgbcolor}\fph\special{ps: grestore}}}}
\put(3488,-1693){\makebox(0,0)[lb]{\smash{\SetFigFont{12}{14.4}{rm}\special{ps: gsave 0 0 0 setrgbcolor}\fpu\special{ps: grestore}}}}
\end{picture}

%% file: saddlesextrepos.pstex_t
\begin{picture}(0,0)%
\epsfig{file=saddlesextrepos.pstex}%
\end{picture}%
\setlength{\unitlength}{0.00087500in}%
\begingroup\makeatletter\ifx\SetFigFont\undefined
\def\x#1#2#3#4#5#6#7\relax{\def\x{#1#2#3#4#5#6}}%
\expandafter\x\fmtname xxxxxx\relax \def\y{splain}%
\ifx\x\y   
\gdef\SetFigFont#1#2#3{%
  \ifnum #1<17\tiny\else \ifnum #1<20\small\else
  \ifnum #1<24\normalsize\else \ifnum #1<29\large\else
  \ifnum #1<34\Large\else \ifnum #1<41\LARGE\else
     \huge\fi\fi\fi\fi\fi\fi
  \csname #3\endcsname}%
\else
\gdef\SetFigFont#1#2#3{\begingroup
  \count@#1\relax \ifnum 25<\count@\count@25\fi
  \def\x{\endgroup\@setsize\SetFigFont{#2pt}}%
  \expandafter\x
    \csname \romannumeral\the\count@ pt\expandafter\endcsname
    \csname @\romannumeral\the\count@ pt\endcsname
  \csname #3\endcsname}%
\fi
\fi\endgroup
\begin{picture}(5079,2949)(5230,-2773)
\put(8737,-372){\makebox(0,0)[lb]{\smash{\SetFigFont{6}{7.2}{rm}\special{ps: gsave 0 0 0 setrgbcolor}:\  Forbidden regions\special{ps: grestore}}}}
\put(8743,-673){\makebox(0,0)[lb]{\smash{\SetFigFont{6}{7.2}{rm}\special{ps: gsave 0 0 0 setrgbcolor}:\ Directions of steepest descent\special{ps: grestore}}}}
\put(7337,-2061){\makebox(0,0)[lb]{\smash{\SetFigFont{9}{10.8}{rm}\special{ps: gsave 0 0 0 setrgbcolor}$s_+$\special{ps: grestore}}}}
\put(5641,-1202){\makebox(0,0)[lb]{\smash{\SetFigFont{9}{10.8}{rm}\special{ps: gsave 0 0 0 setrgbcolor}$s_-$\special{ps: grestore}}}}
\put(6151,-661){\makebox(0,0)[lb]{\smash{\SetFigFont{9}{10.8}{rm}\special{ps: gsave 0 0 0 setrgbcolor}$\sqrt{|2\ve/a|}$\special{ps: grestore}}}}
\put(6907,-445){\makebox(0,0)[lb]{\smash{\SetFigFont{12}{14.4}{rm}\special{ps: gsave 0 0 0 setrgbcolor}$\Im{q}$\special{ps: grestore}}}}
\put(8371,-1771){\makebox(0,0)[lb]{\smash{\SetFigFont{12}{14.4}{rm}\special{ps: gsave 0 0 0 setrgbcolor}$\Re{q}$\special{ps: grestore}}}}
\put(6931,-2356){\makebox(0,0)[lb]{\smash{\SetFigFont{9}{10.8}{rm}\special{ps: gsave 0 0 0 setrgbcolor}$\sqrt{|2\ve/a|}$\special{ps: grestore}}}}
\put(8743,-943){\makebox(0,0)[lb]{\smash{\SetFigFont{6}{7.2}{rm}\special{ps: gsave 0 0 0 setrgbcolor}:\  Integration contour\special{ps: grestore}}}}
\end{picture}

%% file: saddlesextreneg.pstex_t
\begin{picture}(0,0)%
\epsfig{file=saddlesextreneg.pstex}%
\end{picture}%
\setlength{\unitlength}{0.00087500in}%
\begingroup\makeatletter\ifx\SetFigFont\undefined
\def\x#1#2#3#4#5#6#7\relax{\def\x{#1#2#3#4#5#6}}%
\expandafter\x\fmtname xxxxxx\relax \def\y{splain}%
\ifx\x\y   
\gdef\SetFigFont#1#2#3{%
  \ifnum #1<17\tiny\else \ifnum #1<20\small\else
  \ifnum #1<24\normalsize\else \ifnum #1<29\large\else
  \ifnum #1<34\Large\else \ifnum #1<41\LARGE\else
     \huge\fi\fi\fi\fi\fi\fi
  \csname #3\endcsname}%
\else
\gdef\SetFigFont#1#2#3{\begingroup
  \count@#1\relax \ifnum 25<\count@\count@25\fi
  \def\x{\endgroup\@setsize\SetFigFont{#2pt}}%
  \expandafter\x
    \csname \romannumeral\the\count@ pt\expandafter\endcsname
    \csname @\romannumeral\the\count@ pt\endcsname
  \csname #3\endcsname}%
\fi
\fi\endgroup
\begin{picture}(4204,2846)(3269,-2670)
\put(6819,-1666){\makebox(0,0)[lb]{\smash{\SetFigFont{12}{14.4}{rm}\special{ps: gsave 0 0 0 setrgbcolor}$\Re{q}$\special{ps: grestore}}}}
\put(4814,-1754){\makebox(0,0)[lb]{\smash{\SetFigFont{9}{10.8}{rm}\special{ps: gsave 0 0 0 setrgbcolor}$\sqrt{|2\ve/a|}$\special{ps: grestore}}}}
\put(5573,-749){\makebox(0,0)[lb]{\smash{\SetFigFont{9}{10.8}{rm}\special{ps: gsave 0 0 0 setrgbcolor}$s_+$\special{ps: grestore}}}}
\put(5471,-332){\makebox(0,0)[lb]{\smash{\SetFigFont{12}{14.4}{rm}\special{ps: gsave 0 0 0 setrgbcolor}$\Im{q}$\special{ps: grestore}}}}
\put(5170,-2401){\makebox(0,0)[lb]{\smash{\SetFigFont{9}{10.8}{rm}\special{ps: gsave 0 0 0 setrgbcolor}$s_-$\special{ps: grestore}}}}
\end{picture}

%% file: saddlestraneneg.pstex_t
\begin{picture}(0,0)%
\epsfig{file=saddlestraneneg.pstex}%
\end{picture}%
\setlength{\unitlength}{0.00087500in}%
\begingroup\makeatletter\ifx\SetFigFont\undefined
\def\x#1#2#3#4#5#6#7\relax{\def\x{#1#2#3#4#5#6}}%
\expandafter\x\fmtname xxxxxx\relax \def\y{splain}%
\ifx\x\y   
\gdef\SetFigFont#1#2#3{%
  \ifnum #1<17\tiny\else \ifnum #1<20\small\else
  \ifnum #1<24\normalsize\else \ifnum #1<29\large\else
  \ifnum #1<34\Large\else \ifnum #1<41\LARGE\else
     \huge\fi\fi\fi\fi\fi\fi
  \csname #3\endcsname}%
\else
\gdef\SetFigFont#1#2#3{\begingroup
  \count@#1\relax \ifnum 25<\count@\count@25\fi
  \def\x{\endgroup\@setsize\SetFigFont{#2pt}}%
  \expandafter\x
    \csname \romannumeral\the\count@ pt\expandafter\endcsname
    \csname @\romannumeral\the\count@ pt\endcsname
  \csname #3\endcsname}%
\fi
\fi\endgroup
\begin{picture}(6684,4746)(1910,-7744)
\put(6841,-3841){\makebox(0,0)[lb]{\smash{\SetFigFont{7}{8.4}{rm}\special{ps: gsave 0 0 0 setrgbcolor}:\  forbidden regions\special{ps: grestore}}}}
\put(4997,-4112){\makebox(0,0)[lb]{\smash{\SetFigFont{12}{14.4}{rm}\special{ps: gsave 0 0 0 setrgbcolor}$s_{+1}$\special{ps: grestore}}}}
\put(4997,-5237){\makebox(0,0)[lb]{\smash{\SetFigFont{12}{14.4}{rm}\special{ps: gsave 0 0 0 setrgbcolor}$s_0$\special{ps: grestore}}}}
\put(4997,-6812){\makebox(0,0)[lb]{\smash{\SetFigFont{12}{14.4}{rm}\special{ps: gsave 0 0 0 setrgbcolor}$s_{-1}$\special{ps: grestore}}}}
\put(4502,-3212){\makebox(0,0)[lb]{\smash{\SetFigFont{12}{14.4}{rm}\special{ps: gsave 0 0 0 setrgbcolor}$(\sgn{a})\Im{q}$\special{ps: grestore}}}}
\put(3017,-4787){\makebox(0,0)[lb]{\smash{\SetFigFont{12}{14.4}{rm}\special{ps: gsave 0 0 0 setrgbcolor}$\sqrt{|2\ve/a|}$\special{ps: grestore}}}}
\put(3017,-6137){\makebox(0,0)[lb]{\smash{\SetFigFont{12}{14.4}{rm}\special{ps: gsave 0 0 0 setrgbcolor}$\sqrt{|2\ve/a|}$\special{ps: grestore}}}}
\put(6452,-5687){\makebox(0,0)[lb]{\smash{\SetFigFont{12}{14.4}{rm}\special{ps: gsave 0 0 0 setrgbcolor}$\Re{q}$\special{ps: grestore}}}}
\put(6848,-4670){\makebox(0,0)[lb]{\smash{\SetFigFont{7}{8.4}{rm}\special{ps: gsave 0 0 0 setrgbcolor}:\ {\clgv C}\special{ps: grestore}}}}
\put(6848,-4278){\makebox(0,0)[lb]{\smash{\SetFigFont{7}{8.4}{rm}\special{ps: gsave 0 0 0 setrgbcolor}:\ Directions of steepest descent\special{ps: grestore}}}}
\end{picture}

%% file: saddlesIneg.pstex_t
\begin{picture}(0,0)%
\epsfig{file=saddlesIneg.pstex}%
\end{picture}%
\setlength{\unitlength}{0.00087500in}%
\begingroup\makeatletter\ifx\SetFigFont\undefined
\def\x#1#2#3#4#5#6#7\relax{\def\x{#1#2#3#4#5#6}}%
\expandafter\x\fmtname xxxxxx\relax \def\y{splain}%
\ifx\x\y   
\gdef\SetFigFont#1#2#3{%
  \ifnum #1<17\tiny\else \ifnum #1<20\small\else
  \ifnum #1<24\normalsize\else \ifnum #1<29\large\else
  \ifnum #1<34\Large\else \ifnum #1<41\LARGE\else
     \huge\fi\fi\fi\fi\fi\fi
  \csname #3\endcsname}%
\else
\gdef\SetFigFont#1#2#3{\begingroup
  \count@#1\relax \ifnum 25<\count@\count@25\fi
  \def\x{\endgroup\@setsize\SetFigFont{#2pt}}%
  \expandafter\x
    \csname \romannumeral\the\count@ pt\expandafter\endcsname
    \csname @\romannumeral\the\count@ pt\endcsname
  \csname #3\endcsname}%
\fi
\fi\endgroup
\begin{picture}(4907,3093)(1895,-3194)
\put(3306,-961){\makebox(0,0)[lb]{\smash{\SetFigFont{12}{14.4}{rm}\special{ps: gsave 0 0 0 setrgbcolor}$\propto{|\ve/a_2|}$\special{ps: grestore}}}}
\put(4014,-2064){\makebox(0,0)[lb]{\smash{\SetFigFont{20}{24.0}{sf}\special{ps: gsave 0 0 0 setrgbcolor}0\special{ps: grestore}}}}
\put(4397,-257){\makebox(0,0)[lb]{\smash{\SetFigFont{12}{14.4}{rm}\special{ps: gsave 0 0 0 setrgbcolor}$\Im{I}$\special{ps: grestore}}}}
\put(6437,-2612){\makebox(0,0)[lb]{\smash{\SetFigFont{12}{14.4}{rm}\special{ps: gsave 0 0 0 setrgbcolor}$\Re{I}$\special{ps: grestore}}}}
\end{picture}

%% file: fig_1epos.pstex_t
\begin{picture}(0,0)%
\epsfig{file=fig_1epos.pstex}%
\end{picture}%
\setlength{\unitlength}{0.00087500in}%
\begingroup\makeatletter\ifx\SetFigFont\undefined
\def\x#1#2#3#4#5#6#7\relax{\def\x{#1#2#3#4#5#6}}%
\expandafter\x\fmtname xxxxxx\relax \def\y{splain}%
\ifx\x\y   
\gdef\SetFigFont#1#2#3{%
  \ifnum #1<17\tiny\else \ifnum #1<20\small\else
  \ifnum #1<24\normalsize\else \ifnum #1<29\large\else
  \ifnum #1<34\Large\else \ifnum #1<41\LARGE\else
     \huge\fi\fi\fi\fi\fi\fi
  \csname #3\endcsname}%
\else
\gdef\SetFigFont#1#2#3{\begingroup
  \count@#1\relax \ifnum 25<\count@\count@25\fi
  \def\x{\endgroup\@setsize\SetFigFont{#2pt}}%
  \expandafter\x
    \csname \romannumeral\the\count@ pt\expandafter\endcsname
    \csname @\romannumeral\the\count@ pt\endcsname
  \csname #3\endcsname}%
\fi
\fi\endgroup
\begin{picture}(1620,1651)(417,-821)
\put(417, 11){\makebox(0,0)[lb]{\smash{\SetFigFont{7}{8.4}{rm}\psve}}}
\put(1081,-797){\makebox(0,0)[lb]{\smash{\SetFigFont{7}{8.4}{rm}\pvetq}}}
\end{picture}

%% file: fig_1enul.pstex_t
\begin{picture}(0,0)%
\epsfig{file=fig_1enul.pstex}%
\end{picture}%
\setlength{\unitlength}{0.00087500in}%
\begingroup\makeatletter\ifx\SetFigFont\undefined
\def\x#1#2#3#4#5#6#7\relax{\def\x{#1#2#3#4#5#6}}%
\expandafter\x\fmtname xxxxxx\relax \def\y{splain}%
\ifx\x\y   
\gdef\SetFigFont#1#2#3{%
  \ifnum #1<17\tiny\else \ifnum #1<20\small\else
  \ifnum #1<24\normalsize\else \ifnum #1<29\large\else
  \ifnum #1<34\Large\else \ifnum #1<41\LARGE\else
     \huge\fi\fi\fi\fi\fi\fi
  \csname #3\endcsname}%
\else
\gdef\SetFigFont#1#2#3{\begingroup
  \count@#1\relax \ifnum 25<\count@\count@25\fi
  \def\x{\endgroup\@setsize\SetFigFont{#2pt}}%
  \expandafter\x
    \csname \romannumeral\the\count@ pt\expandafter\endcsname
    \csname @\romannumeral\the\count@ pt\endcsname
  \csname #3\endcsname}%
\fi
\fi\endgroup
\begin{picture}(1572,1572)(-5,-722)
\end{picture}

%% file: fig_1eneg.pstex_t
\begin{picture}(0,0)%
\epsfig{file=fig_1eneg.pstex}%
\end{picture}%
\setlength{\unitlength}{0.00087500in}%
\begingroup\makeatletter\ifx\SetFigFont\undefined
\def\x#1#2#3#4#5#6#7\relax{\def\x{#1#2#3#4#5#6}}%
\expandafter\x\fmtname xxxxxx\relax \def\y{splain}%
\ifx\x\y   
\gdef\SetFigFont#1#2#3{%
  \ifnum #1<17\tiny\else \ifnum #1<20\small\else
  \ifnum #1<24\normalsize\else \ifnum #1<29\large\else
  \ifnum #1<34\Large\else \ifnum #1<41\LARGE\else
     \huge\fi\fi\fi\fi\fi\fi
  \csname #3\endcsname}%
\else
\gdef\SetFigFont#1#2#3{\begingroup
  \count@#1\relax \ifnum 25<\count@\count@25\fi
  \def\x{\endgroup\@setsize\SetFigFont{#2pt}}%
  \expandafter\x
    \csname \romannumeral\the\count@ pt\expandafter\endcsname
    \csname @\romannumeral\the\count@ pt\endcsname
  \csname #3\endcsname}%
\fi
\fi\endgroup
\begin{picture}(1594,1593)(-10,-759)
\end{picture}

%% file: fig_2epos.pstex_t
\begin{picture}(0,0)%
\epsfig{file=fig_2epos.pstex}%
\end{picture}%
\setlength{\unitlength}{0.00087500in}%
\begingroup\makeatletter\ifx\SetFigFont\undefined
\def\x#1#2#3#4#5#6#7\relax{\def\x{#1#2#3#4#5#6}}%
\expandafter\x\fmtname xxxxxx\relax \def\y{splain}%
\ifx\x\y   
\gdef\SetFigFont#1#2#3{%
  \ifnum #1<17\tiny\else \ifnum #1<20\small\else
  \ifnum #1<24\normalsize\else \ifnum #1<29\large\else
  \ifnum #1<34\Large\else \ifnum #1<41\LARGE\else
     \huge\fi\fi\fi\fi\fi\fi
  \csname #3\endcsname}%
\else
\gdef\SetFigFont#1#2#3{\begingroup
  \count@#1\relax \ifnum 25<\count@\count@25\fi
  \def\x{\endgroup\@setsize\SetFigFont{#2pt}}%
  \expandafter\x
    \csname \romannumeral\the\count@ pt\expandafter\endcsname
    \csname @\romannumeral\the\count@ pt\endcsname
  \csname #3\endcsname}%
\fi
\fi\endgroup
\begin{picture}(1596,1730)(417,-896)
\put(417, 45){\makebox(0,0)[lb]{\smash{\SetFigFont{7}{8.4}{rm}\psve}}}
\put(1170,-872){\makebox(0,0)[lb]{\smash{\SetFigFont{7}{8.4}{rm}\pve}}}
\end{picture}

%% file: fig_2enul.pstex_t
\begin{picture}(0,0)%
\epsfig{file=fig_2enul.pstex}%
\end{picture}%
\setlength{\unitlength}{0.00087500in}%
\begingroup\makeatletter\ifx\SetFigFont\undefined
\def\x#1#2#3#4#5#6#7\relax{\def\x{#1#2#3#4#5#6}}%
\expandafter\x\fmtname xxxxxx\relax \def\y{splain}%
\ifx\x\y   
\gdef\SetFigFont#1#2#3{%
  \ifnum #1<17\tiny\else \ifnum #1<20\small\else
  \ifnum #1<24\normalsize\else \ifnum #1<29\large\else
  \ifnum #1<34\Large\else \ifnum #1<41\LARGE\else
     \huge\fi\fi\fi\fi\fi\fi
  \csname #3\endcsname}%
\else
\gdef\SetFigFont#1#2#3{\begingroup
  \count@#1\relax \ifnum 25<\count@\count@25\fi
  \def\x{\endgroup\@setsize\SetFigFont{#2pt}}%
  \expandafter\x
    \csname \romannumeral\the\count@ pt\expandafter\endcsname
    \csname @\romannumeral\the\count@ pt\endcsname
  \csname #3\endcsname}%
\fi
\fi\endgroup
\begin{picture}(1576,1576)(-10,-743)
\end{picture}

%% file: fig_2eneg.pstex_t
\begin{picture}(0,0)%
\epsfig{file=fig_2eneg.pstex}%
\end{picture}%
\setlength{\unitlength}{0.00087500in}%
\begingroup\makeatletter\ifx\SetFigFont\undefined
\def\x#1#2#3#4#5#6#7\relax{\def\x{#1#2#3#4#5#6}}%
\expandafter\x\fmtname xxxxxx\relax \def\y{splain}%
\ifx\x\y   
\gdef\SetFigFont#1#2#3{%
  \ifnum #1<17\tiny\else \ifnum #1<20\small\else
  \ifnum #1<24\normalsize\else \ifnum #1<29\large\else
  \ifnum #1<34\Large\else \ifnum #1<41\LARGE\else
     \huge\fi\fi\fi\fi\fi\fi
  \csname #3\endcsname}%
\else
\gdef\SetFigFont#1#2#3{\begingroup
  \count@#1\relax \ifnum 25<\count@\count@25\fi
  \def\x{\endgroup\@setsize\SetFigFont{#2pt}}%
  \expandafter\x
    \csname \romannumeral\the\count@ pt\expandafter\endcsname
    \csname @\romannumeral\the\count@ pt\endcsname
  \csname #3\endcsname}%
\fi
\fi\endgroup
\begin{picture}(1595,1595)(-10,-766)
\end{picture}

%% file: fig_2eposbis.pstex_t
\begin{picture}(0,0)%
\epsfig{file=fig_2eposbis.pstex}%
\end{picture}%
\setlength{\unitlength}{0.00087500in}%
\begingroup\makeatletter\ifx\SetFigFont\undefined
\def\x#1#2#3#4#5#6#7\relax{\def\x{#1#2#3#4#5#6}}%
\expandafter\x\fmtname xxxxxx\relax \def\y{splain}%
\ifx\x\y   
\gdef\SetFigFont#1#2#3{%
  \ifnum #1<17\tiny\else \ifnum #1<20\small\else
  \ifnum #1<24\normalsize\else \ifnum #1<29\large\else
  \ifnum #1<34\Large\else \ifnum #1<41\LARGE\else
     \huge\fi\fi\fi\fi\fi\fi
  \csname #3\endcsname}%
\else
\gdef\SetFigFont#1#2#3{\begingroup
  \count@#1\relax \ifnum 25<\count@\count@25\fi
  \def\x{\endgroup\@setsize\SetFigFont{#2pt}}%
  \expandafter\x
    \csname \romannumeral\the\count@ pt\expandafter\endcsname
    \csname @\romannumeral\the\count@ pt\endcsname
  \csname #3\endcsname}%
\fi
\fi\endgroup
\begin{picture}(1621,1725)(417,-880)
\put(1201,-856){\makebox(0,0)[lb]{\smash{\SetFigFont{7}{8.4}{rm}\pve}}}
\put(417, 36){\makebox(0,0)[lb]{\smash{\SetFigFont{7}{8.4}{rm}\psve}}}
\end{picture}

%% file: fig_2enulbis.pstex_t
\begin{picture}(0,0)%
\epsfig{file=fig_2enulbis.pstex}%
\end{picture}%
\setlength{\unitlength}{0.00087500in}%
\begingroup\makeatletter\ifx\SetFigFont\undefined
\def\x#1#2#3#4#5#6#7\relax{\def\x{#1#2#3#4#5#6}}%
\expandafter\x\fmtname xxxxxx\relax \def\y{splain}%
\ifx\x\y   
\gdef\SetFigFont#1#2#3{%
  \ifnum #1<17\tiny\else \ifnum #1<20\small\else
  \ifnum #1<24\normalsize\else \ifnum #1<29\large\else
  \ifnum #1<34\Large\else \ifnum #1<41\LARGE\else
     \huge\fi\fi\fi\fi\fi\fi
  \csname #3\endcsname}%
\else
\gdef\SetFigFont#1#2#3{\begingroup
  \count@#1\relax \ifnum 25<\count@\count@25\fi
  \def\x{\endgroup\@setsize\SetFigFont{#2pt}}%
  \expandafter\x
    \csname \romannumeral\the\count@ pt\expandafter\endcsname
    \csname @\romannumeral\the\count@ pt\endcsname
  \csname #3\endcsname}%
\fi
\fi\endgroup
\begin{picture}(1592,1592)(-11,-745)
\end{picture}

%% file: fig_2enegbis.pstex_t
\begin{picture}(0,0)%
\epsfig{file=fig_2enegbis.pstex}%
\end{picture}%
\setlength{\unitlength}{0.00087500in}%
\begingroup\makeatletter\ifx\SetFigFont\undefined
\def\x#1#2#3#4#5#6#7\relax{\def\x{#1#2#3#4#5#6}}%
\expandafter\x\fmtname xxxxxx\relax \def\y{splain}%
\ifx\x\y   
\gdef\SetFigFont#1#2#3{%
  \ifnum #1<17\tiny\else \ifnum #1<20\small\else
  \ifnum #1<24\normalsize\else \ifnum #1<29\large\else
  \ifnum #1<34\Large\else \ifnum #1<41\LARGE\else
     \huge\fi\fi\fi\fi\fi\fi
  \csname #3\endcsname}%
\else
\gdef\SetFigFont#1#2#3{\begingroup
  \count@#1\relax \ifnum 25<\count@\count@25\fi
  \def\x{\endgroup\@setsize\SetFigFont{#2pt}}%
  \expandafter\x
    \csname \romannumeral\the\count@ pt\expandafter\endcsname
    \csname @\romannumeral\the\count@ pt\endcsname
  \csname #3\endcsname}%
\fi
\fi\endgroup
\begin{picture}(1576,1577)(-10,-740)
\end{picture}

%% file: fig_3epos.pstex_t
\begin{picture}(0,0)%
\epsfig{file=fig_3epos.pstex}%
\end{picture}%
\setlength{\unitlength}{0.00087500in}%
\begingroup\makeatletter\ifx\SetFigFont\undefined
\def\x#1#2#3#4#5#6#7\relax{\def\x{#1#2#3#4#5#6}}%
\expandafter\x\fmtname xxxxxx\relax \def\y{splain}%
\ifx\x\y   
\gdef\SetFigFont#1#2#3{%
  \ifnum #1<17\tiny\else \ifnum #1<20\small\else
  \ifnum #1<24\normalsize\else \ifnum #1<29\large\else
  \ifnum #1<34\Large\else \ifnum #1<41\LARGE\else
     \huge\fi\fi\fi\fi\fi\fi
  \csname #3\endcsname}%
\else
\gdef\SetFigFont#1#2#3{\begingroup
  \count@#1\relax \ifnum 25<\count@\count@25\fi
  \def\x{\endgroup\@setsize\SetFigFont{#2pt}}%
  \expandafter\x
    \csname \romannumeral\the\count@ pt\expandafter\endcsname
    \csname @\romannumeral\the\count@ pt\endcsname
  \csname #3\endcsname}%
\fi
\fi\endgroup
\begin{picture}(1576,1543)(417,-744)
\put(417,-134){\makebox(0,0)[lb]{\smash{\SetFigFont{7}{8.4}{rm}\pve}}}
\end{picture}

%% file: fig_3enul.pstex_t
\begin{picture}(0,0)%
\epsfig{file=fig_3enul.pstex}%
\end{picture}%
\setlength{\unitlength}{0.00087500in}%
\begingroup\makeatletter\ifx\SetFigFont\undefined
\def\x#1#2#3#4#5#6#7\relax{\def\x{#1#2#3#4#5#6}}%
\expandafter\x\fmtname xxxxxx\relax \def\y{splain}%
\ifx\x\y   
\gdef\SetFigFont#1#2#3{%
  \ifnum #1<17\tiny\else \ifnum #1<20\small\else
  \ifnum #1<24\normalsize\else \ifnum #1<29\large\else
  \ifnum #1<34\Large\else \ifnum #1<41\LARGE\else
     \huge\fi\fi\fi\fi\fi\fi
  \csname #3\endcsname}%
\else
\gdef\SetFigFont#1#2#3{\begingroup
  \count@#1\relax \ifnum 25<\count@\count@25\fi
  \def\x{\endgroup\@setsize\SetFigFont{#2pt}}%
  \expandafter\x
    \csname \romannumeral\the\count@ pt\expandafter\endcsname
    \csname @\romannumeral\the\count@ pt\endcsname
  \csname #3\endcsname}%
\fi
\fi\endgroup
\begin{picture}(1570,1570)(-2,-724)
\end{picture}

%% file: fig_3eneg.pstex_t
\begin{picture}(0,0)%
\epsfig{file=fig_3eneg.pstex}%
\end{picture}%
\setlength{\unitlength}{0.00087500in}%
\begingroup\makeatletter\ifx\SetFigFont\undefined
\def\x#1#2#3#4#5#6#7\relax{\def\x{#1#2#3#4#5#6}}%
\expandafter\x\fmtname xxxxxx\relax \def\y{splain}%
\ifx\x\y   
\gdef\SetFigFont#1#2#3{%
  \ifnum #1<17\tiny\else \ifnum #1<20\small\else
  \ifnum #1<24\normalsize\else \ifnum #1<29\large\else
  \ifnum #1<34\Large\else \ifnum #1<41\LARGE\else
     \huge\fi\fi\fi\fi\fi\fi
  \csname #3\endcsname}%
\else
\gdef\SetFigFont#1#2#3{\begingroup
  \count@#1\relax \ifnum 25<\count@\count@25\fi
  \def\x{\endgroup\@setsize\SetFigFont{#2pt}}%
  \expandafter\x
    \csname \romannumeral\the\count@ pt\expandafter\endcsname
    \csname @\romannumeral\the\count@ pt\endcsname
  \csname #3\endcsname}%
\fi
\fi\endgroup
\begin{picture}(1542,1543)(439,-744)
\put(1782,-134){\makebox(0,0)[lb]{\smash{\SetFigFont{7}{8.4}{rm}\pve}}}
\end{picture}

%% file: fig_4epos.pstex_t
\begin{picture}(0,0)%
\epsfig{file=fig_4epos.pstex}%
\end{picture}%
\setlength{\unitlength}{0.00087500in}%
\begingroup\makeatletter\ifx\SetFigFont\undefined
\def\x#1#2#3#4#5#6#7\relax{\def\x{#1#2#3#4#5#6}}%
\expandafter\x\fmtname xxxxxx\relax \def\y{splain}%
\ifx\x\y   
\gdef\SetFigFont#1#2#3{%
  \ifnum #1<17\tiny\else \ifnum #1<20\small\else
  \ifnum #1<24\normalsize\else \ifnum #1<29\large\else
  \ifnum #1<34\Large\else \ifnum #1<41\LARGE\else
     \huge\fi\fi\fi\fi\fi\fi
  \csname #3\endcsname}%
\else
\gdef\SetFigFont#1#2#3{\begingroup
  \count@#1\relax \ifnum 25<\count@\count@25\fi
  \def\x{\endgroup\@setsize\SetFigFont{#2pt}}%
  \expandafter\x
    \csname \romannumeral\the\count@ pt\expandafter\endcsname
    \csname @\romannumeral\the\count@ pt\endcsname
  \csname #3\endcsname}%
\fi
\fi\endgroup
\begin{picture}(1665,1542)(319,-743)
\put(469,-211){\makebox(0,0)[lb]{\smash{
\SetFigFont{7}{8.4}{rm}\psve
}}}
\put(1250,-652){\makebox(0,0)[lb]{\smash{\SetFigFont{7}{8.4}{rm}\psve}}}
\end{picture}

%% file: fig_4enul.pstex_t
\begin{picture}(0,0)%
\epsfig{file=fig_4enul.pstex}%
\end{picture}%
\setlength{\unitlength}{0.00087500in}%
\begingroup\makeatletter\ifx\SetFigFont\undefined
\def\x#1#2#3#4#5#6#7\relax{\def\x{#1#2#3#4#5#6}}%
\expandafter\x\fmtname xxxxxx\relax \def\y{splain}%
\ifx\x\y   
\gdef\SetFigFont#1#2#3{%
  \ifnum #1<17\tiny\else \ifnum #1<20\small\else
  \ifnum #1<24\normalsize\else \ifnum #1<29\large\else
  \ifnum #1<34\Large\else \ifnum #1<41\LARGE\else
     \huge\fi\fi\fi\fi\fi\fi
  \csname #3\endcsname}%
\else
\gdef\SetFigFont#1#2#3{\begingroup
  \count@#1\relax \ifnum 25<\count@\count@25\fi
  \def\x{\endgroup\@setsize\SetFigFont{#2pt}}%
  \expandafter\x
    \csname \romannumeral\the\count@ pt\expandafter\endcsname
    \csname @\romannumeral\the\count@ pt\endcsname
  \csname #3\endcsname}%
\fi
\fi\endgroup
\begin{picture}(1592,1592)(-11,-755)
\end{picture}

%% file: fig_4eneg.pstex_t
\begin{picture}(0,0)%
\epsfig{file=fig_4eneg.pstex}%
\end{picture}%
\setlength{\unitlength}{0.00087500in}%
\begingroup\makeatletter\ifx\SetFigFont\undefined
\def\x#1#2#3#4#5#6#7\relax{\def\x{#1#2#3#4#5#6}}%
\expandafter\x\fmtname xxxxxx\relax \def\y{splain}%
\ifx\x\y   
\gdef\SetFigFont#1#2#3{%
  \ifnum #1<17\tiny\else \ifnum #1<20\small\else
  \ifnum #1<24\normalsize\else \ifnum #1<29\large\else
  \ifnum #1<34\Large\else \ifnum #1<41\LARGE\else
     \huge\fi\fi\fi\fi\fi\fi
  \csname #3\endcsname}%
\else
\gdef\SetFigFont#1#2#3{\begingroup
  \count@#1\relax \ifnum 25<\count@\count@25\fi
  \def\x{\endgroup\@setsize\SetFigFont{#2pt}}%
  \expandafter\x
    \csname \romannumeral\the\count@ pt\expandafter\endcsname
    \csname @\romannumeral\the\count@ pt\endcsname
  \csname #3\endcsname}%
\fi
\fi\endgroup
\begin{picture}(1560,1561)(-10,-732)
\end{picture}

%% file: fig_4eposbis.pstex_t
\begin{picture}(0,0)%
\epsfig{file=fig_4eposbis.pstex}%
\end{picture}%
\setlength{\unitlength}{0.00087500in}%
\begingroup\makeatletter\ifx\SetFigFont\undefined
\def\x#1#2#3#4#5#6#7\relax{\def\x{#1#2#3#4#5#6}}%
\expandafter\x\fmtname xxxxxx\relax \def\y{splain}%
\ifx\x\y   
\gdef\SetFigFont#1#2#3{%
  \ifnum #1<17\tiny\else \ifnum #1<20\small\else
  \ifnum #1<24\normalsize\else \ifnum #1<29\large\else
  \ifnum #1<34\Large\else \ifnum #1<41\LARGE\else
     \huge\fi\fi\fi\fi\fi\fi
  \csname #3\endcsname}%
\else
\gdef\SetFigFont#1#2#3{\begingroup
  \count@#1\relax \ifnum 25<\count@\count@25\fi
  \def\x{\endgroup\@setsize\SetFigFont{#2pt}}%
  \expandafter\x
    \csname \romannumeral\the\count@ pt\expandafter\endcsname
    \csname @\romannumeral\the\count@ pt\endcsname
  \csname #3\endcsname}%
\fi
\fi\endgroup
\begin{picture}(1553,1552)(439,-758)
\put(742,-521){\makebox(0,0)[lb]{\smash{
\SetFigFont{7}{8.4}{rm}\psve
}}}
\end{picture}

%% file: fig_4enulbis.pstex_t
\begin{picture}(0,0)%
\epsfig{file=fig_4enulbis.pstex}%
\end{picture}%
\setlength{\unitlength}{0.00087500in}%
\begingroup\makeatletter\ifx\SetFigFont\undefined
\def\x#1#2#3#4#5#6#7\relax{\def\x{#1#2#3#4#5#6}}%
\expandafter\x\fmtname xxxxxx\relax \def\y{splain}%
\ifx\x\y   
\gdef\SetFigFont#1#2#3{%
  \ifnum #1<17\tiny\else \ifnum #1<20\small\else
  \ifnum #1<24\normalsize\else \ifnum #1<29\large\else
  \ifnum #1<34\Large\else \ifnum #1<41\LARGE\else
     \huge\fi\fi\fi\fi\fi\fi
  \csname #3\endcsname}%
\else
\gdef\SetFigFont#1#2#3{\begingroup
  \count@#1\relax \ifnum 25<\count@\count@25\fi
  \def\x{\endgroup\@setsize\SetFigFont{#2pt}}%
  \expandafter\x
    \csname \romannumeral\the\count@ pt\expandafter\endcsname
    \csname @\romannumeral\the\count@ pt\endcsname
  \csname #3\endcsname}%
\fi
\fi\endgroup
\begin{picture}(1576,1577)(-11,-739)
\end{picture}

%% file: fig_4enegbis.pstex_t
\begin{picture}(0,0)%
\epsfig{file=fig_4enegbis.pstex}%
\end{picture}%
\setlength{\unitlength}{0.00087500in}%
\begingroup\makeatletter\ifx\SetFigFont\undefined
\def\x#1#2#3#4#5#6#7\relax{\def\x{#1#2#3#4#5#6}}%
\expandafter\x\fmtname xxxxxx\relax \def\y{splain}%
\ifx\x\y   
\gdef\SetFigFont#1#2#3{%
  \ifnum #1<17\tiny\else \ifnum #1<20\small\else
  \ifnum #1<24\normalsize\else \ifnum #1<29\large\else
  \ifnum #1<34\Large\else \ifnum #1<41\LARGE\else
     \huge\fi\fi\fi\fi\fi\fi
  \csname #3\endcsname}%
\else
\gdef\SetFigFont#1#2#3{\begingroup
  \count@#1\relax \ifnum 25<\count@\count@25\fi
  \def\x{\endgroup\@setsize\SetFigFont{#2pt}}%
  \expandafter\x
    \csname \romannumeral\the\count@ pt\expandafter\endcsname
    \csname @\romannumeral\the\count@ pt\endcsname
  \csname #3\endcsname}%
\fi
\fi\endgroup
\begin{picture}(1613,1615)(444,-777)
\put(1309,-747){\makebox(0,0)[lb]{\smash{\SetFigFont{7}{8.4}{rm}\psve}}}
\end{picture}

%% file: fig_5epos.pstex_t
\begin{picture}(0,0)%
\epsfig{file=fig_5epos.pstex}%
\end{picture}%
\setlength{\unitlength}{0.00087500in}%
\begingroup\makeatletter\ifx\SetFigFont\undefined
\def\x#1#2#3#4#5#6#7\relax{\def\x{#1#2#3#4#5#6}}%
\expandafter\x\fmtname xxxxxx\relax \def\y{splain}%
\ifx\x\y   
\gdef\SetFigFont#1#2#3{%
  \ifnum #1<17\tiny\else \ifnum #1<20\small\else
  \ifnum #1<24\normalsize\else \ifnum #1<29\large\else
  \ifnum #1<34\Large\else \ifnum #1<41\LARGE\else
     \huge\fi\fi\fi\fi\fi\fi
  \csname #3\endcsname}%
\else
\gdef\SetFigFont#1#2#3{\begingroup
  \count@#1\relax \ifnum 25<\count@\count@25\fi
  \def\x{\endgroup\@setsize\SetFigFont{#2pt}}%
  \expandafter\x
    \csname \romannumeral\the\count@ pt\expandafter\endcsname
    \csname @\romannumeral\the\count@ pt\endcsname
  \csname #3\endcsname}%
\fi
\fi\endgroup
\begin{picture}(1842,2010)(192,-1145)
\put(629,-1082){\makebox(0,0)[lb]{\smash{
\SetFigFont{7}{8.4}{rm}\pvek
}}}
\put(192,216){\makebox(0,0)[lb]{\smash{\SetFigFont{7}{8.4}{rm}\psve}}}
\end{picture}

%% file: fig_5enul.pstex_t
\begin{picture}(0,0)%
\epsfig{file=fig_5enul.pstex}%
\end{picture}%
\setlength{\unitlength}{0.00087500in}%
\begingroup\makeatletter\ifx\SetFigFont\undefined
\def\x#1#2#3#4#5#6#7\relax{\def\x{#1#2#3#4#5#6}}%
\expandafter\x\fmtname xxxxxx\relax \def\y{splain}%
\ifx\x\y   
\gdef\SetFigFont#1#2#3{%
  \ifnum #1<17\tiny\else \ifnum #1<20\small\else
  \ifnum #1<24\normalsize\else \ifnum #1<29\large\else
  \ifnum #1<34\Large\else \ifnum #1<41\LARGE\else
     \huge\fi\fi\fi\fi\fi\fi
  \csname #3\endcsname}%
\else
\gdef\SetFigFont#1#2#3{\begingroup
  \count@#1\relax \ifnum 25<\count@\count@25\fi
  \def\x{\endgroup\@setsize\SetFigFont{#2pt}}%
  \expandafter\x
    \csname \romannumeral\the\count@ pt\expandafter\endcsname
    \csname @\romannumeral\the\count@ pt\endcsname
  \csname #3\endcsname}%
\fi
\fi\endgroup
\begin{picture}(1546,1544)(40,-742)
\end{picture}

%% file: fig_5eneg.pstex_t
\begin{picture}(0,0)%
\epsfig{file=fig_5eneg.pstex}%
\end{picture}%
\setlength{\unitlength}{0.00087500in}%
\begingroup\makeatletter\ifx\SetFigFont\undefined
\def\x#1#2#3#4#5#6#7\relax{\def\x{#1#2#3#4#5#6}}%
\expandafter\x\fmtname xxxxxx\relax \def\y{splain}%
\ifx\x\y   
\gdef\SetFigFont#1#2#3{%
  \ifnum #1<17\tiny\else \ifnum #1<20\small\else
  \ifnum #1<24\normalsize\else \ifnum #1<29\large\else
  \ifnum #1<34\Large\else \ifnum #1<41\LARGE\else
     \huge\fi\fi\fi\fi\fi\fi
  \csname #3\endcsname}%
\else
\gdef\SetFigFont#1#2#3{\begingroup
  \count@#1\relax \ifnum 25<\count@\count@25\fi
  \def\x{\endgroup\@setsize\SetFigFont{#2pt}}%
  \expandafter\x
    \csname \romannumeral\the\count@ pt\expandafter\endcsname
    \csname @\romannumeral\the\count@ pt\endcsname
  \csname #3\endcsname}%
\fi
\fi\endgroup
\begin{picture}(1591,1588)(-3,-747)
\end{picture}

%% file: articledef.bbl
\begin{thebibliography}{10}
\newcommand{\enquote}[1]{``#1''}

\bibitem{Abraham/Marsden85a}
\textsc{R.~Abraham and J.~E. Marsden}, \enquote{Foundations of Mechanics},
  {A}ddison-{W}esley {P}ublishing {C}ompany {I}nc., {N}ew {Y}ork,  1985,
  (second edition).

\bibitem{Arnold88a}
\textsc{V.~I. Arnold}, \enquote{Geometrical methods in the theory of ordinary
  differential equations}, vol. 250 of \textit{Die grundlehren der
  mathematischen wissenschaften}, {S}pringer-{V}erlag, {N}ew {Y}ork,  1988,
  (2nd edition).

\bibitem{Arnold/Avez67a}
\textsc{V.~I. Arnold and A.~Avez}, \enquote{Probl{\`e}mes ergodique de la
  M{\'e}canique classique}, vol.~9 of \textit{Monographies internationales de
  math{\'e}matiques modernes}, {G}authier-{V}illars, Paris,  1967, English
  translation: {B}enjamin ({N}ew {Y}ork).

\bibitem{Balian/Bloch74b}
\textsc{R.~Balian and C.~Bloch}, \textit{Ann. Physics} \textbf{85} (1974), 514.

\bibitem{Berry78a}
\textsc{M.~V. Berry}, in J.~{S}iebe (ed.), \textit{Topics in Nonlinear Dynamics
  --- A Tribute to Sir {E}dward {B}ullard}, vol.~46,  (1978), 1978 pp. 16--120,
  Reprinted in~\cite{Mackay/Meiss87a}.

\bibitem{Berry/Howls91a}
\textsc{M.~V. Berry and C.~J. Howls}, \textit{Proc. Roy. Soc. London Ser. A}
  \textbf{434} (1991), 657.

\bibitem{Berry/Howls94a}
\textsc{M.~V. Berry and C.~J. Howls}, \textit{Proc. Roy. Soc. London Ser. A}
  \textbf{447} (1994), 527.

\bibitem{Birkhoff27a}
\textsc{G.~D. Birkhoff}, \enquote{Dynamical systems}, vol.~9 of
  \textit{American Mathematical Society Colloquium Publications}, American
  Mathematical Society, {N}ew {Y}ork,  1927, reprinted in~\cite[vol
  II]{Birkhoff50a}.

\bibitem{Birkhoff50a}
\textsc{G.~D. Birkhoff}, \enquote{{G}eorge {D}avid {B}irkhoff, collected
  mathematical papers}, American Mathematical Society, {N}ew {Y}ork,  1950,
  vol. I, II, III.

\bibitem{Bogomolny92a}
\textsc{E.~B. Bogomolny}, \textit{Nonlinearity} \textbf{5} (1992), 805.

\bibitem{Bogomolny+80a}
\textsc{E.~B. Bogomolny, V.~A. Fateev and L.~N. Lipatov}, \textit{Soviet Sc.
  Rev. (Physical Reviews)} \textbf{2} (1980), 247.

\bibitem{Bohigas+93a}
\textsc{O.~Bohigas, D.~Boos{\'e}, R.~Egydio~de Carvalho and V.~Marvulle},
  \textit{Nuclear Phys. A} \textbf{560} (1993), 197.

\bibitem{Bruno70a}
\textsc{A.~Bruno}, \textit{Math. USSR-Sb.} \textbf{12}(2) (1970), 271, (russian
  original paper: {\em Mat. Sbornik}, {\bf 83}(125), No. 2 (1970)
  pp.~273--312).

\bibitem{Carruthers/Nieto68a}
\textsc{P.~Carruthers and M.~M. Nieto}, \textit{Rev. Modern Phys.}
  \textbf{40}(2) (1968), 411.

\bibitem{Cartier/Voros88a}
\textsc{P.~Cartier and A.~Voros}, \textit{C. R. Acad. Sci. Paris S{\'e}r. I
  Math.} \textbf{307} (1988), 143, (in french).

\bibitem{Cherry27b}
\textsc{T.~M. Cherry}, \textit{Proc. London Math. Soc. (2nd ser.)} \textbf{27}
  (1927), 151.

\bibitem{Cherry27a}
\textsc{T.~M. Cherry}, \textit{Proc. London Math. Soc. (2nd ser.)} \textbf{26}
  (1927), 211.

\bibitem{Cherry28a}
\textsc{T.~M. Cherry}, \textit{Philos. Trans. Roy. Soc. London Ser. A}
  \textbf{227}(5) (1928), 137.

\bibitem{Creagh/Whelan96a}
\textsc{S.~C. Creagh and N.~D. Whelan}, \enquote{Complex Periodic Orbits and
  Tunnelling in Chaotic Potentials,} submitted to Phys. Rev. Lett.

\bibitem{Cushman+83a}
\textsc{R.~Cushman, A.~Deprit and R.~Mosak}, \textit{J. Math. Phys.}
  \textbf{24}(8) (1983), 2102.

\bibitem{deAguiar+87a}
\textsc{M.~A.~M. de~Aguiar, C.~P. Malta, M.~Baranger and K.~T.~R. Davies},
  \textit{Ann. Physics} \textbf{180} (1987), 167.

\bibitem{Doron/Frischat95a}
\textsc{E.~Doron and S.~D. Frischat}, \textit{Phys. Rev. Lett.} \textbf{75}(20)
  (1995), 3661.

\bibitem{Floquet1883a}
\textsc{G.~Floquet}, \textit{Ann. de l'{\'E}c. Normale, {$2^{\mbox{\small e}}$}
  S{\'e}rie} \textbf{12} (1883), 47, (in french).

\bibitem{Frischat/Doron98a}
\textsc{S.~Frischat and E.~Doron}, \textit{Phys. Rev. E} \textbf{57} (1998),
  1421.

\bibitem{Gustavson66a}
\textsc{F.~G. Gustavson}, \textit{Astron. J.} \textbf{71} (1966), 670.

\bibitem{Gutzwiller90a}
\textsc{M.~C. Gutzwiller}, \enquote{Chaos in Classical and Quantum Mechanics},
  vol.~1 of \textit{Interdisciplinary Applied Mathematics},
  {S}pringer-{V}erlag, {N}ew {Y}ork,  1990.

\bibitem{Henrard70a}
\textsc{J.~Henrard}, \textit{Celestial Mech.} \textbf{3} (1970), 107.

\bibitem{Howls91a}
\textsc{C.~J. Howls}, \enquote{Exponential Asymptotics}, Ph.D. thesis,
  University of {B}ristol, Bristol,  (1991).

\bibitem{Itzykson/Zuber80a}
\textsc{C.~Itzykson and J.-B. Zuber}, \enquote{Quantum Field Theory},
  International Series in Pure and Applied Physics, {McGraw-Hill}, {N}ew
  {Y}ork,  1980.

\bibitem{Kus+93a}
\textsc{M.~Ku{\'s}, F.~Haake and D.~Delande}, \textit{Phys. Rev. Lett.}
  \textbf{71}(14) (1993), 2167.

\bibitem{Leboeuf/Mouchet94a}
\textsc{P.~Leb{\oe}uf and A.~Mouchet}, \textit{Phys. Rev. Lett.}
  \textbf{73}(10) (1994), 1360.

\bibitem{Lin/Ballentine90a}
\textsc{W.~Lin and L.~Ballentine}, \textit{Phys. Rev. Lett.} \textbf{65}
  (1990), 2927.

\bibitem{Mackay/Meiss87a}
\textsc{R.~S. Mackay and J.~D. Meiss}, \enquote{Hamiltonian Dynamical Systems},
  {A}dam {H}ilger, Bristol and Philadelphia,  1987.

\bibitem{Meyer70a}
\textsc{K.~R. Meyer}, \textit{Trans. Amer. Math. Soc.} \textbf{149} (1970), 95,
  reprinted in \cite{Mackay/Meiss87a}.

\bibitem{Meyer71a}
\textsc{K.~R. Meyer}, \textit{Trans. Amer. Math. Soc.} \textbf{154} (1971),
  227.

\bibitem{Meyer86a}
\textsc{K.~R. Meyer}, \textit{Contemporary Mathematics} \textbf{56} (1986),
  373.

\bibitem{Meyer/Hall92a}
\textsc{K.~R. Meyer and G.~H. Hall}, \enquote{Introduction to Hamiltonian
  Dynamical Systems and the N-Body Problem}, vol.~90 of \textit{Applied
  Mathematical Sciences}, {S}pringer-{V}erlag, {N}ew {Y}ork,  1992.

\bibitem{Miller70a}
\textsc{W.~H. Miller}, \textit{J. Chem. Phys.} \textbf{53}(5) (1970), 1949.

\bibitem{Moser56a}
\textsc{J.~k. Moser}, \textit{Comm. Pure Appl. Math.} \textbf{9} (1956), 673.

\bibitem{Ozorio/Hannay87a}
\textsc{A.~M. Ozorio~de Almeida and J.~H. Hannay}, \textit{J. Phys. A}
  \textbf{20} (1987), 5873.

\bibitem{Poincare57a}
\textsc{H.~Poincar{\'e}}, \enquote{Les m{\'e}thodes nouvelles de la
  m{\'e}canique c{\'e}leste, 3 vol.}, {D}over {P}ublications, {I}nc., {N}ew
  {Y}ork,  1957, (First edition {G}authier-{V}illars, Paris,
  {\oldstyle{1892}}).

\bibitem{Poston/Stewart78a}
\textsc{T.~Poston and I.~Stewart}, \enquote{Catastrophe Theory and its
  Applications}, Pitman, London,  1978.

\bibitem{Prosen95a}
\textsc{T.~Prosen}, \textit{J. Phys. A} \textbf{28} (1995), 4133.

\bibitem{Schomerus/Sieber97a}
\textsc{H.~Schomerus and M.~Sieber}, \textit{J. Phys. A} \textbf{30} (1997),
  4537.

\bibitem{Schulman81a}
\textsc{L.~S. Schulman}, \enquote{Techniques and Applications of Path
  Integration}, {J}ohn {W}iley and sons, Inc., {N}ew {Y}ork,  1981.

\bibitem{Sieber96a}
\textsc{M.~Sieber}, \textit{J. Phys. A} \textbf{29} (1996), 4715.

\bibitem{Sieber97a}
\textsc{M.~Sieber}, \textit{J. Phys. A} \textbf{30} (1997), 4563.

\bibitem{Tabor83a}
\textsc{M.~Tabor}, \textit{Physica D} \textbf{6} (1983), 195.

\bibitem{Tomsovic+95a}
\textsc{S.~Tomsovic, M.~Grinberg and D.~Ullmo}, \textit{Phys. Rev. Lett.}
  \textbf{75}(25) (1995), 4346.

\bibitem{Tomsovic/Ullmo94a}
\textsc{S.~Tomsovic and D.~Ullmo}, \textit{Phys. Rev. E} \textbf{50}(1) (1994),
  145.

\bibitem{Turnbull/Aitken61a}
\textsc{H.~W. Turnbull and A.~C. Aitken}, \enquote{An Introduction to the
  Theory of Canonical Matrices}, {D}over publications, {N}ew {Y}ork,  1961.

\bibitem{Voros83a}
\textsc{A.~Voros}, \textit{Ann. Inst. H. Poincar{\'e}. Phys. Th{\'e}or.}
  \textbf{39}(3) (1983), 211.

\bibitem{Voros94a}
\textsc{A.~Voros}, \textit{Progr. Theoret. Phys. Suppl.} \textbf{116} (1994),
  17.

\bibitem{Voros94b}
\textsc{A.~Voros}, \textit{J. Phys. A} \textbf{27} (1994), 4653.

\bibitem{Whittaker64a}
\textsc{E.~T. Whittaker}, \enquote{A treatise on the analytical dynamics of
  particles and rigid bodies}, Cambridge University Press, Cambridge,  1964.

\bibitem{Williamson36a}
\textsc{J.~Williamson}, \textit{Amer. J. Math.} \textbf{58}(1) (1936), 141.

\end{thebibliography}
